\documentclass[10pt,aps,prl,twocolumn,superscriptaddress,floatfix,longbibliography]{revtex4-1}

\usepackage{graphicx}
\usepackage{amsfonts}
\usepackage{amssymb}
\usepackage{amsmath}
\usepackage{txfonts}
\usepackage{lipsum}
\usepackage[dvipsnames]{xcolor}
\usepackage{wasysym}
\usepackage[colorlinks=true, allcolors=blue]{hyperref}
\usepackage{bbold}
\usepackage{etoolbox} % for additional refs as footnotes
\usepackage[normalem]{ulem}
\usepackage{mathtools} % enhanced math capabilities
\usepackage{multirow} % for the tables
\usepackage{hhline}
\usepackage{mathrsfs} % curly letters by \mathscr{}
\usepackage{soul}
\usepackage{array}

\usepackage{letltxmacro}
\LetLtxMacro{\oldsqrt}{\sqrt}
\renewcommand{\sqrt}[2][\mkern8mu]{\mkern-6mu\mathop{}\oldsqrt[#1]{#2}}

\begin{document}

\title{Orbital-Selective Diffuse Magnetic Fluctuations in Sr$_2$RuO$_4$: a Unified Theoretical Picture}

\author{Maria Chatzieleftheriou}
\altaffiliation[Present address: ]{chatzieleftheriou@itp.uni-frankfurt.de}
\affiliation{CPHT, CNRS, {\'E}cole polytechnique, Institut Polytechnique de Paris, 91120 Palaiseau, France}

\author{Alexander N. Rudenko}
\affiliation{\mbox{Radboud University, Institute for Molecules and Materials, Heijendaalseweg 135, 6525AJ Nijmegen, The Netherlands}}

\author{Yvan Sidis}
\affiliation{Universit{\'e} Paris-Saclay, CNRS, CEA, Laboratoire L{\'e}on Brillouin, 91191, Gif-sur-Yvette, France}

\author{Silke Biermann}
\affiliation{CPHT, CNRS, {\'E}cole polytechnique, Institut Polytechnique de Paris, 91120 Palaiseau, France}
\affiliation{Coll{\`e}ge de France, Universit{\'e} PSL, 11 place Marcelin Berthelot, 75005 Paris, France}
\affiliation{European Theoretical Spectroscopy Facility, 91128 Palaiseau, France}

\author{Evgeny A. Stepanov}
\affiliation{CPHT, CNRS, {\'E}cole polytechnique, Institut Polytechnique de Paris, 91120 Palaiseau, France}
\affiliation{Coll{\`e}ge de France, Universit{\'e} PSL, 11 place Marcelin Berthelot, 75005 Paris, France}

\begin{abstract}
The quasi-two-dimensional material Sr$_2$RuO$_4$ is a paradigmatic example of a correlated system that exhibits unconventional superconductivity and intriguing magnetic properties.
The interplay between these two effects and the resulting strength and nature of spin fluctuations and their role for the properties of the compound have sparked significant debates.
Here, elaborating a theory that self-consistently incorporates spatial magnetic fluctuations into a realistic many-body description, we show that these fluctuations significantly reduce many-body correlations in the system, thereby preventing magnetic ordering in Sr$_2$RuO$_4$, in agreement with experimental observations.
Our conclusion is supported by a theoretical calculation of the spin susceptibility that closely matches the experimental results.
We obtain finite peaks at the incommensurate wave vectors, a broad dome-shaped structure centered around the $\Gamma$ point and a diminished magnetic response at the edges of the BZ. 
We identify the orbital character of the unusual dome structure as resulting predominantly from the 2D-like $xy$-orbital, which is believed to be responsible for the superconductivity.
\end{abstract}

\maketitle

In materials with strong electronic correlations, typically containing partially filled electronic $d$- or $f$-shells, Coulomb interactions between the electrons play a dominant role in determining their properties. 
The behavior of electrons cannot be described independently from one another, as is mostly the case for conventional metals. Instead, complex collective phenomena emerge, including high-temperature superconductivity, Mott insulating states, or exotic magnetic phases. 

The layered ruthenate Sr$_2$RuO$_4$ has attracted significant attention due to its structural similarity to high-temperature superconducting cuprates~\cite{keimer2015quantum} and its
unconventional superconductivity below ${T_c=1.5}$\,K~\cite{maeno1994superconductivity, RevModPhys.75.657, doi:10.1143/JPSJ.81.011009}. 
Currently, ongoing efforts focus on understanding the enhancement of $T_c$ under uniaxial strain~\cite{doi:10.1126/science.adf3348, doi:10.7566/JPSJ.93.062001, doi:10.1126/science.1248292, doi:10.1073/pnas.2020492118}.
Despite extensive research, however, the precise nature of the superconducting state 
in Sr$_2$RuO$_4$ remains unresolved.
Conflicting experimental evidence suggests scenarios ranging from a single $d_{x^2-y^2}$~\cite{doi:10.1073/pnas.2025313118, li2022elastocaloric, jerzembeck2022superconductivity, PhysRevB.108.094516} to broken time-reversal symmetry~\cite{PhysRevLett.97.167002, luke1998time}, and a two-component~\cite{doi:10.1143/JPSJ.81.011009, kivelson2020proposal, doi:10.7566/JPSJ.93.062001} order parameter. 
Numerous theoretical studies have tried to resolve this issue~\cite{PhysRevLett.123.247001, PhysRevB.105.155101, PhysRevResearch.4.033011, profe2024competition}, but no consensus has been reached so far.

At higher temperatures, significant magnetic fluctuations are found, and they are believed to be the source of the superconducting pairing mechanism in Sr$_2$RuO$_4$, calling urgently for their accurate theoretical description.
As seen in inelastic neutron scattering (INS) measurements, the leading spin excitations in Sr$_2$RuO$_4$ correspond to the incommensurate wave vector ${Q=(3\pi/5,3\pi/5,0)}$~\cite{PhysRevLett.83.3320, PhysRevB.66.064522, PhysRevLett.92.097402, PhysRevLett.122.047004, PhysRevB.103.104511}.
Remarkably, although spin excitations in this material are exceptionally strong, different experimental studies confirm the absence of magnetic ordering in the parent compound.
Nevertheless, Sr$_2$RuO$_4$ resides in close proximity to an ordered state, which can be induced already by a small concentration of impurities~\cite{PhysRevB.63.180504, PhysRevLett.88.197002,PhysRevB.89.045119}, or by applying strain~\cite{grinenko2021split,li2022elastocaloric}.

The description of spin excitations in Sr$_2$RuO$_4$ has sparked debates among theorists.
This material is a paradigmatic correlated metal that requires an accurate non-perturbative treatment of electronic interactions. 
Conventional theoretical methods correctly capture the leading spin fluctuations associated with the incommensurate wave vector $Q$~\cite{PhysRevLett.82.4324, Boehnke_2018, acharya2019evening, PhysRevB.100.125120}.
However, they significantly overestimate the strength of magnetic fluctuations, predicting a transition to a magnetically ordered state that is not realized in nature.
The final missing component in {\it state-of-the-art} theoretical calculations, which could potentially reconcile the discrepancies between theory and experiment, is a self-consistent incorporation of spatial magnetic fluctuations into the theory. 
This effect has not been addressed yet due to the lack of appropriate theoretical tools.

In this Letter, we resolve this long-standing problem by applying a recently developed advanced many-body approach that enables a self-consistent treatment of non-local collective electronic fluctuations within a realistic multi-orbital framework.
We demonstrate that a self-consistent treatment of spatial spin excitations suppresses their strength and reduces the electronic correlations in Sr$_2$RuO$_4$, ultimately eliminating the magnetic ordering predicted by {\it state-of-the-art} approaches.
Moreover, our results accurately reproduce the form of the spin susceptibility as deduced from experimental INS measurements, allowing us to detect an intriguing orbital dependence of the
latter.

\begin{figure*}[t!]
\includegraphics[width=1\linewidth]{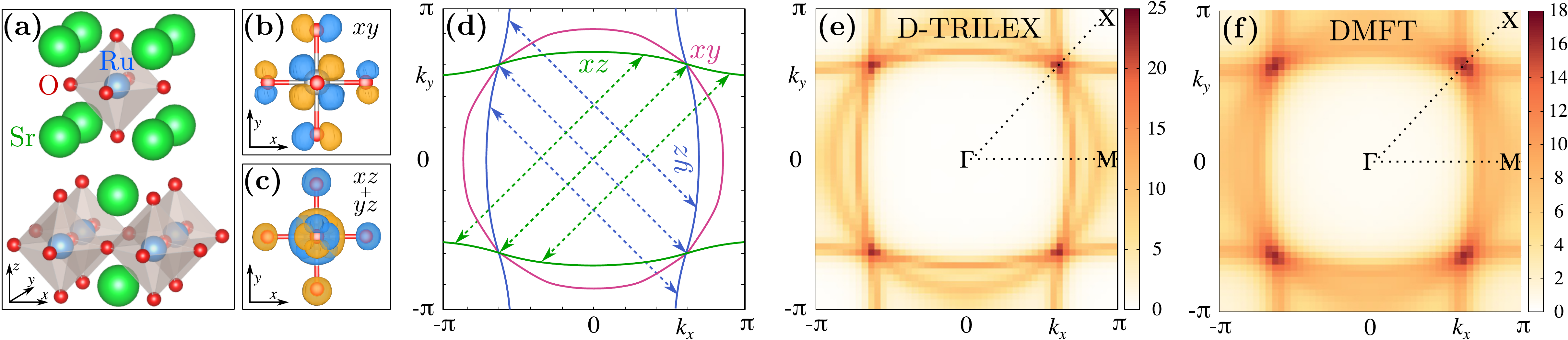}
\caption{\textbf{(a)} Crystal structure of Sr$_2$RuO$_4$ and the three Wannier orbitals ($xy$ in \textbf{(b)} and $xz+yz$ in \textbf{(c)}) on the $x-y$ plane. \textbf{(d)} Sketch of the non-interacting Fermi surface (FS) of Sr$_2$RuO$_4$ in the ${(k_x, k_y, 0)}$ plane, consisting of three sheets originating from the one-dimensional $xz$ and $yz$ orbitals and from the two-dimensional $xy$ one. The nesting vectors are shown with dashed lines. \textbf{(e)} \mbox{D-TRILEX} and \textbf{(f)} DMFT calculations of the FS performed at ${T=145}$\,K. 
In DMFT, the Fermi surface sheets appear broad in momentum, while in D-TRILEX, they are significantly sharper, suggesting reduced electronic correlations.
\label{fig:FS_susc}}
\end{figure*}

The Fermi surface (FS) of Sr$_2$RuO$_4$, shown in Fig.~\ref{fig:FS_susc}, consists of three Fermi sheets, originating from the three $4d$ orbitals: the quasi-one-dimensional (1D) $xz$ and $yz$, and the quasi-two-dimensional (2D) $xy$~\cite{PhysRevLett.76.3786,PhysRevLett.84.2662,PhysRevLett.85.5194,PhysRevX.9.021048}. 
Angle-resolved photoemission spectroscopy (ARPES) and quantum oscillations measurements reveal that Sr$_2$RuO$_4$ is a correlated metal~\cite{PhysRevB.72.205114,PhysRevLett.85.5194,PhysRevLett.76.3786}. Although a significant redistribution of the spectral weight and an enhancement of it around the Fermi energy are observed~\cite{PhysRevLett.76.4845,PhysRevLett.76.3009}, the band structure and FS remain relatively close~\cite{PhysRevLett.79.733,PhysRevB.74.035115} to those predicted by density functional theory (DFT)~\cite{PhysRev.140.A1133}. 
From the theoretical perspective, the compound is considered to be a paradigmatic correlated Fermi liquid, with strong signatures of Hund's metal physics~\cite{PhysRevLett.106.096401, annurev:/content/journals/10.1146/annurev-conmatphys-020911-125045,PhysRevLett.124.016401, PhysRevLett.131.236502, PhysRevResearch.6.023124,suzuki2023distinct}. Therefore, a number of studies have been conducted using this material as a testbed for realistic many-body calculations. 

Early electronic structure calculations using DFT combined with dynamical mean-field theory (DMFT)~\cite{RevModPhys.68.13}, that accounts for local correlation effects, have nearly 
succeeded to reproduce the experimental spectral function~\cite{PhysRevLett.84.1591}.
Since then, several DMFT studies have allowed for many important advances in our 
understanding of the physics of Sr$_2$RuO$_4$~\cite{PhysRevLett.106.096401, PhysRevLett.131.236502, PhysRevLett.116.106402}. 
Recent studies~\cite{PhysRevX.9.021048, PhysRevLett.131.236502} showed that the self-energy, derived from photoemission measurements at the {\bf k}-points of the Brillouin Zone (BZ) corresponding to the FS, is predominantly local for the $xz/yz$ orbitals. 
In contrast, the $xy$ orbital exhibits some momentum dependence, which becomes apparent at frequencies above ${\simeq10}$\,meV~\cite{PhysRevX.9.021048}. 
These findings suggest that, overall, the compound's single-particle properties are not particularly remarkable and are adequately captured by DMFT.

It is therefore very surprising that the material's two-particle quantities, particularly spin excitations, are in contrast not well understood. 
Within DFT calculations, the FS nesting arising from the compound's crystal structure leads to an ordered spin-density wave (SDW) state with the incommensurate wave vectors $Q$~\cite{PhysRevLett.82.4324}.
DMFT calculations of the spin susceptibility, incorporating local vertex corrections, have also reproduced the peaks at the $Q$ vectors~\cite{Boehnke_2018, acharya2019evening, PhysRevB.100.125120}. 
However, DMFT strongly overestimates the strength of magnetic fluctuations, compared to experiments, predicting a transition to a SDW ordered state at a finite temperature ${T \simeq 123}$\,K~\cite{PhysRevB.100.125120}.
While it has been shown that spin-orbit coupling (SOC) can suppress this magnetic transition in DMFT, achieving this requires a rather large coupling strength~\cite{PhysRevB.100.125120}.
Additionally, the SOC is not expected to significantly influence the spin susceptibility at the relatively high temperatures at which the DMFT calculations have been conducted~\cite{PhysRevB.65.220502, PhysRevLett.92.097402}. 

The key contribution of DMFT to calculating the spin susceptibility is its ability to reveal the suppression of the magnetic signal at the edges of the BZ, particularly at the ${\text{X}=(\pi, \pi, 0)}$ point, in agreement with experiments.
Within DMFT, spin excitations, aside from the $Q$ peaks, are found to be quasi-local, or nearly constant in momentum space.
However, this result does not fully align with experimental findings, where the magnetic response, in addition to the $Q$ peaks, exhibits a relatively broad dome centered at the ${\Gamma = (0,0,0)}$ point~\cite{PhysRevLett.122.047004, PhysRevB.103.104511, PhysRevB.103.104511}, rather than a quasi-local background signal.

To accurately describe magnetic fluctuations in Sr$_2$RuO$_4$, we consider an effective three-band model (see the Supplemental Material (SM)~\cite{SM}) corresponding to maximally localized $\{xz, yz, xy\}$ orbitals derived from DFT~\footnote{
The DFT calculations were performed within the projected augmented wave formalism~\cite{PhysRevB.50.17953, PhysRevB.59.1758} as implemented in the Vienna ab initio simulation package (VASP)~\cite{KRESSE199615, PhysRevB.54.11169}. 
The exchange-correlation effects were considered within the generalized-gradient approximation functional in the PBE parametrization~\cite{PhysRevLett.77.3865}. 
We used the standard pseudopotentials that include 10, 16, and 6 valence electrons for Sr, Ru, and O, respectively. 
An energy cut-off of 400\,eV for the plane-waves and a convergence threshold of ${10^{-7}}$\,eV were used in the calculations. 
The Wannier functions and the effective tight-binding Hamiltonian were constructed within the scheme of maximal localization~\cite{PhysRevB.56.12847, RevModPhys.84.1419} using the wannier90 package~\cite{MOSTOFI2008685}.
}.
We account for the on-site electronic interaction that is parametrized in the Kanamori form~\cite{10.1143/PTP.30.275}. 
The value of the intra-orbital Coulomb repulsion, ${U=2.56}$\,eV, is chosen based on the constrained random-phase approximation (cRPA) analysis conducted in~\cite{PhysRevB.86.165105}. 
A Hund's exchange coupling $J$, crucial for Sr$_2$RuO$_4$, is selected by evaluating results across different values of $J$. 
Specifically, we estimate the mass enhancement and spin susceptibility obtained for each $J$, as detailed in the SM~\cite{SM}. 
We find that the correct mass enhancement is achieved for ${J\simeq0.35-0.40}$\,eV, and the accurate spin susceptibility for ${J\simeq0.30-0.35}$\,eV, leading to an optimal choice of ${J=0.35}$\,eV.
We do not include spin-orbit coupling (SOC) in our calculations because it is prohibitively expensive computationally. 
This approximation is supported by experimental evidences, suggesting that SOC effects become less significant at higher temperatures, due to thermal effects.

In order to obtain a reliable description of the feedback of electronic correlations on the spectral and magnetic properties of the compound, a self-consistent method is required. 
We employ the dual triply irreducible local expansion (\mbox{D-TRILEX}) approach~\cite{PhysRevB.100.205115, PhysRevB.103.245123, 10.21468/SciPostPhys.13.2.036}, which is a diagrammatic extension of DMFT~\cite{RevModPhys.90.025003, Lyakhova_review}.
The decisive advantage of this method over other DMFT extensions is its computational efficiency, which enables calculations for multi-orbital systems~\cite{PhysRevLett.127.207205, PhysRevResearch.5.L022016, PhysRevLett.129.096404, PhysRevLett.132.226501, stepanov2023charge}, such as Sr$_2$RuO$_4$. 
In the \mbox{D-TRILEX} framework, local electronic correlations are treated non-perturbatively via a DMFT impurity problem~\cite{RevModPhys.68.13}, which is solved using the \textsc{w2dynamics} package~\cite{WALLERBERGER2019388}. 
The latter further serves as a reference system for the diagrammatic expansion that self-consistently incorporates non-local electronic correlations, including magnetic fluctuations. 
This approach allows the mutual influence of collective electronic fluctuations on single-particle quantities and {\it vice versa}~\cite{stepanov2021coexisting, vandelli2024doping, PhysRevLett.132.236504, PhysRevB.110.L161106}, yielding reliable results for both single- and two-particle observables in a broad regime of system's parameters~\cite{PhysRevB.103.245123, 10.21468/SciPostPhys.13.2.036, vandelli2022quantum}.

\begin{figure}[t!]
\includegraphics[width=1\linewidth]{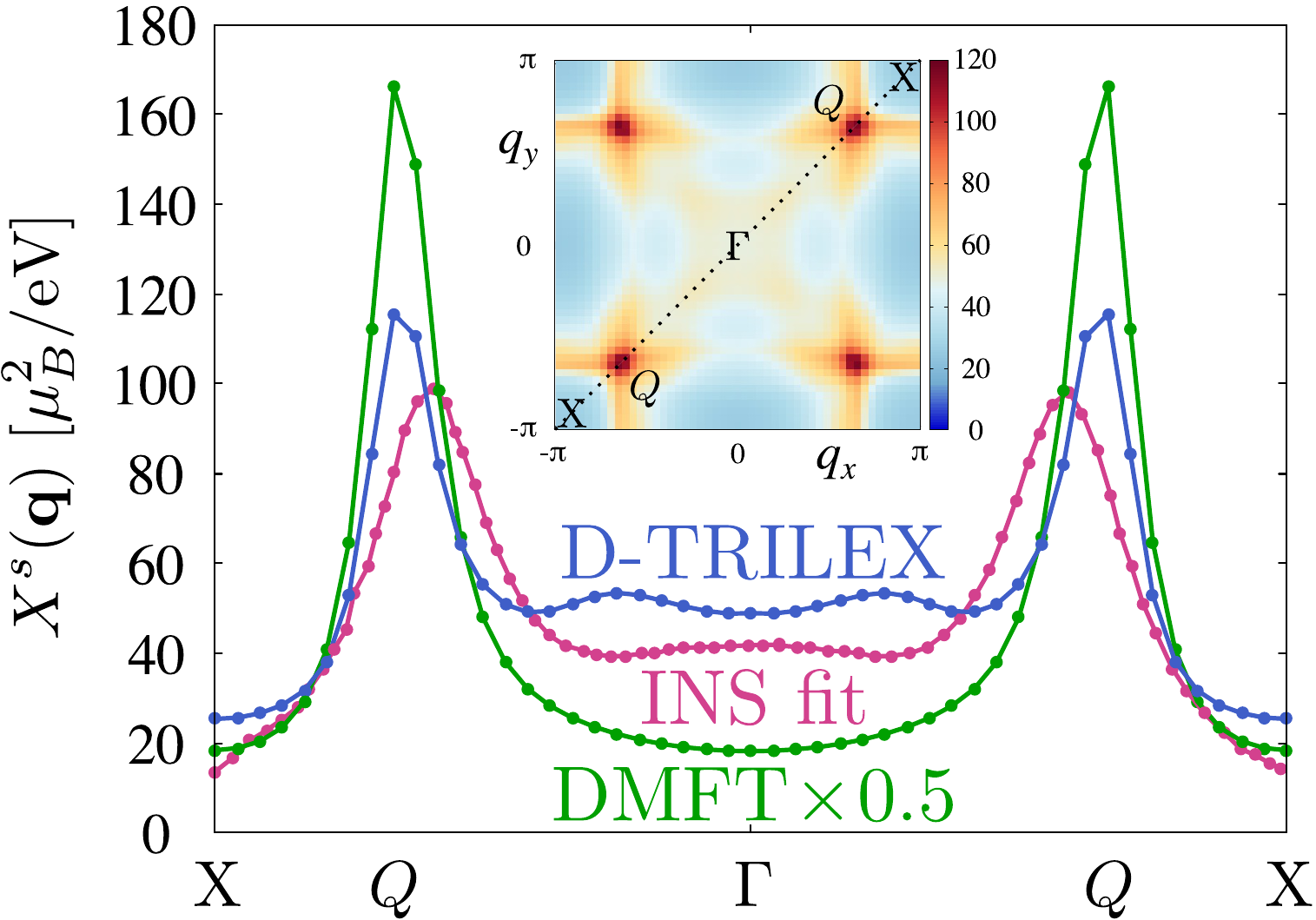}
\caption{The real part of the static spin susceptibility $X^{s}({\bf q})$ along the high symmetry path X-$\Gamma$-X (dashed line in the inset) of the first BZ.
The result is calculated using \mbox{D-TRILEX} (blue, ${T=145}$\,K) and is obtained by fitting the experimental INS data~\cite{PhysRevLett.122.047004} (magenta, ${T=150}$\,K). 
The result of the DMFT calculation, multiplied by 0.5, is also shown for comparison (green, ${T=193}$\,K). 
The inset shows the \mbox{D-TRILEX} susceptibility in the ${(q_x,q_y,0)}$ plane. 
The momentum-space structure of the magnetic susceptibility exhibits peaks at the incommensurate wave vector ${Q=(3\pi/5,3\pi/5,0)}$, the dome-like background signal centered at $\Gamma$ and minima around the $M$ and $X$ points. 
\label{fig:spin_susc}}
\end{figure}

We perform the \mbox{D-TRILEX} calculations at a relatively low temperature ${T=145}$\,K, where experimental data are available for comparison. 
The inset in Fig.~\ref{fig:spin_susc} shows the real part of the static magnetic (${\omega=0}$) susceptibility ${X^{s}({\bf q})}$ obtained for the first BZ using \mbox{D-TRILEX}. 
The main part of Fig.~\ref{fig:spin_susc} displays a cut of the susceptibility along the X-$\Gamma$-X diagonal of the BZ (dashed line in the inset).
The results are calculated numerically using \mbox{D-TRILEX} (blue) and DMFT (green), and are compared to the INS result (magenta)~\cite{PhysRevLett.122.047004}.
In the INS study, performed at ${T=150}$\,K, the real part of the static spin susceptibility is deduced from the fit of the low energy part of the experimentally measured spin excitation spectrum, with a set of single relaxors as described in Ref.~\cite{PhysRevLett.122.047004}. 
At this temperature the magnetic fluctuations are already very strong. 
In fact, for the considered model DMFT predicts a SDW ordered state already at ${T\simeq145}$\,K~\footnote{\label{note1}To be precise, our DMFT susceptibility is obtained through a single-shot \mbox{D-TRILEX} calculation, which, in practice, gives a very similar result to a direct evaluation of the local vertex corrections, as done in Ref.~\cite{PhysRevB.100.125120}. In that work the transition temperature is found to be lower at $T\simeq123K$, but this is related to the different interaction parameters used in the current study.}.
For this reason, the DMFT result in Fig.~\ref{fig:spin_susc} is shown for a bit higher temperature ${T=193}$\,K.
The spin susceptibility of DMFT, calculated in the vicinity of the SDW transition, features large peaks at the incommensurate $Q$ vectors, in agreement with previous works~\cite{PhysRevB.100.125120}.
For easier comparison, the DMFT result in Fig.~\ref{fig:spin_susc} is multiplied by $0.5$.
We find, that a self-consistent inclusion of the magnetic fluctuations beyond DMFT, using \mbox{D-TRILEX}, leads to a strong suppression of the SDW $Q$ peaks in a good agreement with the INS result, and no ordering is observed. 

\begin{figure}[t!]
\includegraphics[width=1\linewidth]{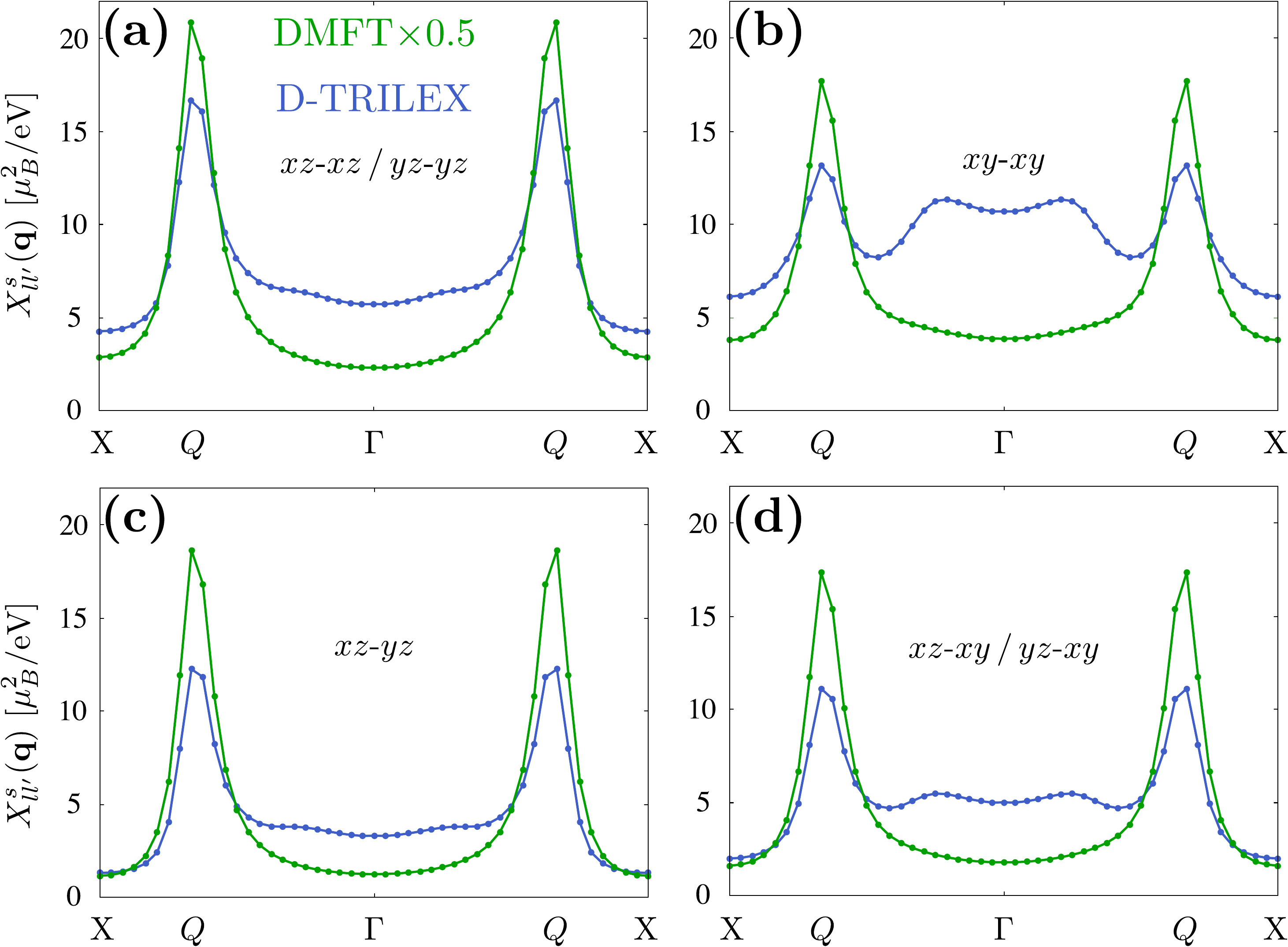} 
\caption{Orbitally resolved static spin susceptibility $X^{s}_{ll'}({\bf q})$ calculated along the X-$\Gamma$-X path in the first BZ using \mbox{D-TRILEX} (blue, ${T=145}$\,K) and DMFT (green, ${T=193}$\,K).
All, intra-(top row) and inter-orbital (bottom row) susceptibilities exhibit SDW peaks at the incommensurate $Q$ vectors.
The large response observed in \mbox{D-TRILEX} around the $\Gamma$ point, and also measured experimentally, is found to be related to the intra-orbital magnetic fluctuations within the $xy$ orbital.
This dome-like signal is completely absent in the DMFT result, where all orbital components the spin susceptibility exhibit a momentum-independent background signal.}
\label{fig:susc_perorb}
\end{figure}

The second important outcome of our results is related to the overall behavior of the spin susceptibility across the BZ. 
As discussed above, according to experimental measurements the magnetic signal can be decomposed into the sum of the SDW $Q$ peaks and a broad dome structure centered around the $\Gamma$ point, while DMFT calculations instead find a quasi-constant background signal besides the $Q$ peaks.
The \mbox{D-TRILEX} calculations reveal a significantly diminished spin susceptibility at the edges of the BZ, with a ``cross''-like structure in momentum space of higher intensity appearing at the center of the zone, visible in the inset of Fig.~\ref{fig:spin_susc}. 
The overall structure of the susceptibility agrees very well with the experimental results~\cite{PhysRevLett.122.047004}.
While early-unpolarized INS studies reported features of the spin excitation spectrum around $\Gamma$~\cite{PhysRevB.66.064522}, the most recent polarized INS data~\cite{PhysRevLett.122.047004,PhysRevB.103.104511} unambiguously confirmed the existence of such fluctuations, but cannot resolve the actual structure around $\Gamma$. 
Nevertheless, strong spin excitations around $\Gamma$ are found in metamagnetic Ca$_{2-x}$Sr$_x$RuO$_4$~\cite{PhysRevLett.93.147404,PhysRevLett.99.217402,PhysRevB.83.054429}, where the isostructural Ca/Sr substitution is, however, accompanied by a rotation and deformation of the RuO$_6$ octahedra~\cite{PhysRevB.63.174432,PhysRevLett.95.267403}. 

\begin{figure*}[t!]
\includegraphics[width=1\linewidth]{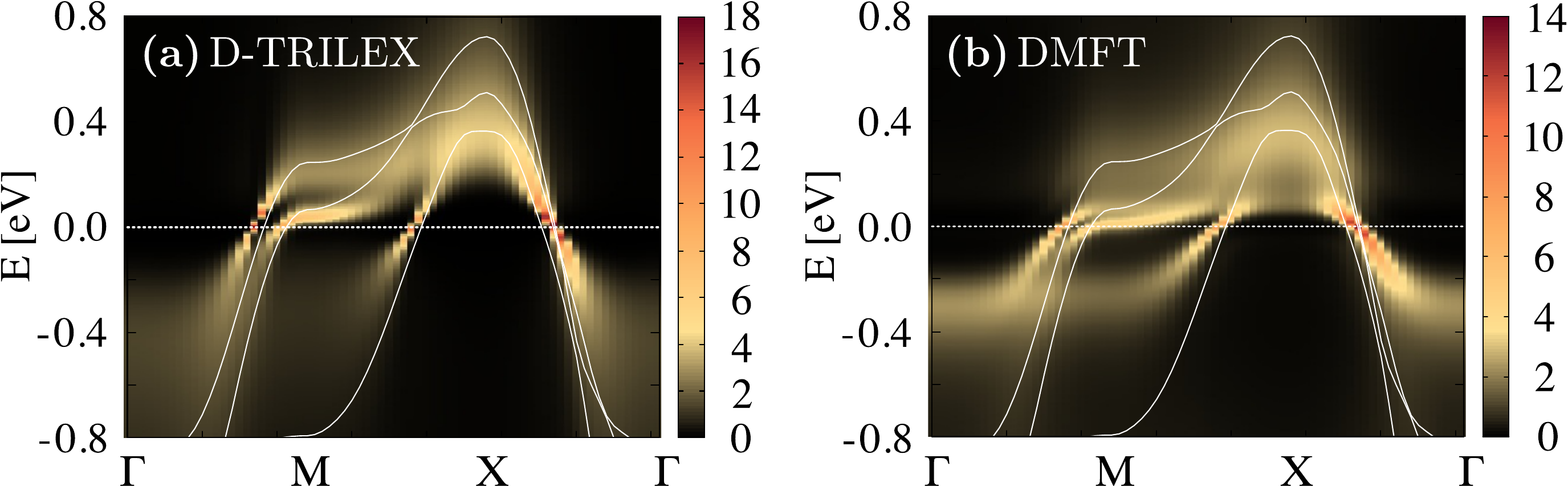}
\caption{
Momentum-resolved electronic spectral function along the high-symmetry path $\Gamma$-M-X-$\Gamma$ of the BZ, calculated with \mbox{D-TRILEX} (a) and DMFT (b) at ${T=145}$\,K. 
The white lines correspond to the bare DFT band-structure. Within DMFT a large band renormalization is observed, as well as large broadening of the bands. Within the \mbox{D-TRILEX} framework, however, the bands come closer to the DFT result and the overall spectral function is much sharper compared to the DMFT picture. 
\label{fig:spectr_func}}
\end{figure*}

In Fig.~\ref{fig:susc_perorb} we show the orbitally resolved static spin susceptibility ${X^{s}_{ll'}({\bf q})}$, where ${l^{(\prime)}\in\{xz, yz, xy\}}$.
The results are calculated using \mbox{D-TRILEX} (blue) and DMFT (green) and plotted along the X-${\Gamma}$-X path.
Although the incommensurate SDW $Q$ peaks are present in all orbital components due to mixing of different of orbital contributions to the susceptibility, the value of $X^{s}({\bf q}=Q)$ is the largest for the intra-orbital $xz$-$xz$ and $yz$-$yz$ components (a). 
This is expected, since the SDW $Q$ peaks originate from the nesting of the FS of these 1D-like orbitals.
Interestingly, the background dome-shaped magnetic signal, found experimentally and reproduced by \mbox{D-TRILEX}, appears to originate from magnetic fluctuations within the 2D-like $xy$ orbital (b).
Instead, the DMFT results do not exhibit any significant signal around the $\Gamma$ point, and for all the components a quasi-local response is obtained besides the $Q$ peaks. 
This result may have significant implications for the symmetry of the superconducting order parameter, as recent studies suggest that magnetic fluctuations within the $xy$ orbital are a key ingredient for the electron pairing~\cite{PhysRevLett.123.247001, profe2024competition}. 
In particular, the presence of spin excitations at more than one wave-vector allows for a competition between different order parameters of the superconducting state~\cite{ghosh2021thermodynamic, benhabib2021ultrasound}.

The breakdown of the SDW ordering predicted by DMFT, upon self-consistent inclusion of magnetic fluctuations in \mbox{D-TRILEX}, can be understood as follows.
In a metal, strong spin fluctuations induce a large electronic self-energy, which in turn diminishes the electronic Green's function. 
This renders the system less correlated, thereby decreasing the electronic polarizability, which 
renormalizes the spin excitations and thus suppresses them.  
The reduction of the many-body effects by magnetic fluctuations in Sr$_2$RuO$_4$ is clearly reflected in its single-particle properties.
In Fig.~\ref{fig:FS_susc}\,(e,\,f) we plot the imaginary part of the electronic Green's function 
${-\sum_{l}{\rm Im}\,G_{ll}({\bf k},\nu_0)}$
at the zeroth Matsubara frequency $\nu_0$, which approximates the FS.
The result is calculated using \mbox{D-TRILEX} (e) and DMFT (f) at ${T=145}$\,K. 
Both methods reveal three FS sheets originating from the 1D $xz$/$yz$ and the 2D $xy$ orbitals. 
However, one immediately finds that DMFT predicts a large broadening of the FS, while \mbox{D-TRILEX} shows instead very sharp FS sheets. 
In Fig.~\ref{fig:spectr_func} we plot the momentum-resolved spectral function along the high-symmetry path $\Gamma$-M-X-$\Gamma$ in the BZ. 
One immediately notes that DMFT (b) significantly renormalizes the non-interacting DFT band structure (white lines), particularly the bandwidth, and pins the van Hove singularity to the Fermi level. 
The latter is a well known effect of strong electronic correlations seen in various systems~\cite{PhysRevB.54.12505, PhysRevLett.89.076401, PhysRevB.72.205121, PhysRevB.84.245107, PhysRevB.82.155126, stepanov2023charge}. 
The renormalization predicted by DMFT broadens the bands and increases the density of states around the Fermi energy, strongly enhancing the leading nesting-driven spin excitations and ultimately resulting in a phase transition to a magnetically ordered state.
Essentially, DMFT appears to overestimate the degree of correlation of the system. 
Instead, the self-consistent inclusion of spatial magnetic fluctuations within \mbox{D-TRILEX} reduces the electronic correlations, resulting in sharper, less renormalized bands (a) that are rather close to the DFT picture.
In particular, we find that the van Hove singularity is no longer pinned to the Fermi level but appears above the Fermi energy, as in DFT, indicating that electronic correlations within the \mbox{D-TRILEX} framework are weaker than those in DMFT. 

The similarity between the \mbox{D-TRILEX} and DFT band structures also explains the emergence of the dome-shaped $\Gamma$-point structure in the spin susceptibility of \mbox{D-TRILEX}, as a similar signal has been observed in RPA calculations for the DFT band structure under specific conditions~\cite{PhysRevLett.123.247001}. 
The proximity of the spectral function of Sr$_2$RuO$_4$ to the DFT one is also reflected in the orbital polarization. 
The latter is found to be decreased in DMFT, yielding the values $n_{xz/yz}=0.67, n_{xy}=0.66$, compared to the DFT values $n_{xz/yz}=0.69, n_{xy}=0.62$. 
This significant polarization between the 1D and 2D orbitals is also observed in \mbox{D-TRILEX} ($n_{xz/yz}=0.70, n_{xy}=0.60$), and it is another signature of reduced electronic correlations. 

By analyzing the electron self-energy we find that, although it is not local and exhibits a noticeable momentum dependence perpendicularly to the FS, it has a surprisingly small momentum-dependence along the FS, as discussed in the SM~\cite{SM}. 
Along the FS the $xz$/$yz$ orbitals appear to have a practically constant self-energy, while the $xy$ orbital exhibits a finite but small momentum-modulation.
This result is in a qualitative agreement with recent ARPES data presented in Fig.~4\,(d) of Ref.~\cite{PhysRevX.9.021048}, and explains why DMFT was particularly successful in reproducing the FS of the material.

In conclusion, we have studied the effect of magnetic fluctuations on the electronic correlations and the spin susceptibility of Sr$_2$RuO$_4$.
These excitations are found to be significant in this material, such that DMFT calculations predict an ordered SDW state that is not observed experimentally. 
We demonstrate that the self-consistent inclusion of spatial magnetic fluctuations suppresses their strength by reducing many-body correlations to the extent that no magnetic ordering is realized in Sr$_2$RuO$_4$. 
The overall behavior of the spin susceptibility in momentum space deduced from INS measurements is well reproduced by our calculations. 
We obtain finite peaks at the incommensurate SDW $Q$ vectors, a broad dome-shaped structure centered around the $\Gamma$ point and a diminished magnetic response at the edges of the BZ. 
We identify the orbital character of the unusual dome structure as resulting predominantly from the 2D-like $xy$-orbital. This observation resonates with recent claims that it is the $xy$-orbital that
is mainly responsible for the superconductivity.
\\

\begin{acknowledgments}
We thank Antoine Georges, Alexander Lichtenstein, Hugo Strand and Alaska Subedi for 
fruitful discussions. 
We acknowledge support from IDRIS/GENCI under grant number 091393, and are thankful to the
CPHT computer support team for their help.
\end{acknowledgments}

\bibliography{Ref}

\begin{thebibliography}{93}%
\makeatletter
\providecommand \@ifxundefined [1]{%
 \@ifx{#1\undefined}
}%
\providecommand \@ifnum [1]{%
 \ifnum #1\expandafter \@firstoftwo
 \else \expandafter \@secondoftwo
 \fi
}%
\providecommand \@ifx [1]{%
 \ifx #1\expandafter \@firstoftwo
 \else \expandafter \@secondoftwo
 \fi
}%
\providecommand \natexlab [1]{#1}%
\providecommand \enquote  [1]{``#1''}%
\providecommand \bibnamefont  [1]{#1}%
\providecommand \bibfnamefont [1]{#1}%
\providecommand \citenamefont [1]{#1}%
\providecommand \href@noop [0]{\@secondoftwo}%
\providecommand \href [0]{\begingroup \@sanitize@url \@href}%
\providecommand \@href[1]{\@@startlink{#1}\@@href}%
\providecommand \@@href[1]{\endgroup#1\@@endlink}%
\providecommand \@sanitize@url [0]{\catcode `\\12\catcode `\$12\catcode `\&12\catcode `\#12\catcode `\^12\catcode `\_12\catcode `\%12\relax}%
\providecommand \@@startlink[1]{}%
\providecommand \@@endlink[0]{}%
\providecommand \url  [0]{\begingroup\@sanitize@url \@url }%
\providecommand \@url [1]{\endgroup\@href {#1}{\urlprefix }}%
\providecommand \urlprefix  [0]{URL }%
\providecommand \Eprint [0]{\href }%
\providecommand \doibase [0]{http://dx.doi.org/}%
\providecommand \selectlanguage [0]{\@gobble}%
\providecommand \bibinfo  [0]{\@secondoftwo}%
\providecommand \bibfield  [0]{\@secondoftwo}%
\providecommand \translation [1]{[#1]}%
\providecommand \BibitemOpen [0]{}%
\providecommand \bibitemStop [0]{}%
\providecommand \bibitemNoStop [0]{.\EOS\space}%
\providecommand \EOS [0]{\spacefactor3000\relax}%
\providecommand \BibitemShut  [1]{\csname bibitem#1\endcsname}%
\let\auto@bib@innerbib\@empty
%</preamble>
\bibitem [{\citenamefont {Keimer}\ \emph {et~al.}(2015)\citenamefont {Keimer}, \citenamefont {Kivelson}, \citenamefont {Norman}, \citenamefont {Uchida},\ and\ \citenamefont {Zaanen}}]{keimer2015quantum}%
  \BibitemOpen
  \bibfield  {author} {\bibinfo {author} {\bibfnamefont {B.}~\bibnamefont {Keimer}}, \bibinfo {author} {\bibfnamefont {S.~A.}\ \bibnamefont {Kivelson}}, \bibinfo {author} {\bibfnamefont {M.~R.}\ \bibnamefont {Norman}}, \bibinfo {author} {\bibfnamefont {S.}~\bibnamefont {Uchida}}, \ and\ \bibinfo {author} {\bibfnamefont {J.}~\bibnamefont {Zaanen}},\ }\bibfield  {title} {\enquote {\bibinfo {title} {{From quantum matter to high-temperature superconductivity in copper oxides}},}\ }\href {\doibase 10.1038/nature14165} {\bibfield  {journal} {\bibinfo  {journal} {Nature}\ }\textbf {\bibinfo {volume} {518}},\ \bibinfo {pages} {179--186} (\bibinfo {year} {2015})}\BibitemShut {NoStop}%
\bibitem [{\citenamefont {Maeno}\ \emph {et~al.}(1994)\citenamefont {Maeno}, \citenamefont {Hashimoto}, \citenamefont {Yoshida}, \citenamefont {Nishizaki}, \citenamefont {Fujita}, \citenamefont {Bednorz},\ and\ \citenamefont {Lichtenberg}}]{maeno1994superconductivity}%
  \BibitemOpen
  \bibfield  {author} {\bibinfo {author} {\bibfnamefont {Y.}~\bibnamefont {Maeno}}, \bibinfo {author} {\bibfnamefont {H.}~\bibnamefont {Hashimoto}}, \bibinfo {author} {\bibfnamefont {K.}~\bibnamefont {Yoshida}}, \bibinfo {author} {\bibfnamefont {S.}~\bibnamefont {Nishizaki}}, \bibinfo {author} {\bibfnamefont {T.}~\bibnamefont {Fujita}}, \bibinfo {author} {\bibfnamefont {J.~G.}\ \bibnamefont {Bednorz}}, \ and\ \bibinfo {author} {\bibfnamefont {F.}~\bibnamefont {Lichtenberg}},\ }\bibfield  {title} {\enquote {\bibinfo {title} {{Superconductivity in a layered perovskite without copper}},}\ }\href {\doibase 10.1038/372532a0} {\bibfield  {journal} {\bibinfo  {journal} {Nature}\ }\textbf {\bibinfo {volume} {372}},\ \bibinfo {pages} {532--534} (\bibinfo {year} {1994})}\BibitemShut {NoStop}%
\bibitem [{\citenamefont {Mackenzie}\ and\ \citenamefont {Maeno}(2003)}]{RevModPhys.75.657}%
  \BibitemOpen
  \bibfield  {author} {\bibinfo {author} {\bibfnamefont {Andrew~Peter}\ \bibnamefont {Mackenzie}}\ and\ \bibinfo {author} {\bibfnamefont {Yoshiteru}\ \bibnamefont {Maeno}},\ }\bibfield  {title} {\enquote {\bibinfo {title} {{The superconductivity of ${\mathrm{Sr}}_{2}{\mathrm{RuO}}_{4}$ and the physics of spin-triplet pairing}},}\ }\href {\doibase 10.1103/RevModPhys.75.657} {\bibfield  {journal} {\bibinfo  {journal} {Rev. Mod. Phys.}\ }\textbf {\bibinfo {volume} {75}},\ \bibinfo {pages} {657--712} (\bibinfo {year} {2003})}\BibitemShut {NoStop}%
\bibitem [{\citenamefont {Maeno}\ \emph {et~al.}(2012)\citenamefont {Maeno}, \citenamefont {Kittaka}, \citenamefont {Nomura}, \citenamefont {Yonezawa},\ and\ \citenamefont {Ishida}}]{doi:10.1143/JPSJ.81.011009}%
  \BibitemOpen
  \bibfield  {author} {\bibinfo {author} {\bibfnamefont {Yoshiteru}\ \bibnamefont {Maeno}}, \bibinfo {author} {\bibfnamefont {Shunichiro}\ \bibnamefont {Kittaka}}, \bibinfo {author} {\bibfnamefont {Takuji}\ \bibnamefont {Nomura}}, \bibinfo {author} {\bibfnamefont {Shingo}\ \bibnamefont {Yonezawa}}, \ and\ \bibinfo {author} {\bibfnamefont {Kenji}\ \bibnamefont {Ishida}},\ }\bibfield  {title} {\enquote {\bibinfo {title} {{Evaluation of Spin-Triplet Superconductivity in Sr$_2$RuO$_4$}},}\ }\href {\doibase 10.1143/JPSJ.81.011009} {\bibfield  {journal} {\bibinfo  {journal} {J. Phys. Soc. Jpn.}\ }\textbf {\bibinfo {volume} {81}},\ \bibinfo {pages} {011009} (\bibinfo {year} {2012})}\BibitemShut {NoStop}%
\bibitem [{\citenamefont {{H. M. L. Noad and K. Ishida and Y.-S. Li and E. Gati and V. Stangier and N. Kikugawa and D. A. Sokolov and M. Nicklas and B. Kim and I. I. Mazin and M. Garst and J. Schmalian and A. P. Mackenzie and C. W. Hicks }}(2023)}]{doi:10.1126/science.adf3348}%
  \BibitemOpen
  \bibfield  {author} {\bibinfo {author} {\bibnamefont {{H. M. L. Noad and K. Ishida and Y.-S. Li and E. Gati and V. Stangier and N. Kikugawa and D. A. Sokolov and M. Nicklas and B. Kim and I. I. Mazin and M. Garst and J. Schmalian and A. P. Mackenzie and C. W. Hicks }}},\ }\bibfield  {title} {\enquote {\bibinfo {title} {{Giant lattice softening at a Lifshitz transition in Sr$_2$RuO$_4$}},}\ }\href {\doibase 10.1126/science.adf3348} {\bibfield  {journal} {\bibinfo  {journal} {Science}\ }\textbf {\bibinfo {volume} {382}},\ \bibinfo {pages} {447--450} (\bibinfo {year} {2023})}\BibitemShut {NoStop}%
\bibitem [{\citenamefont {Maeno}\ \emph {et~al.}(2024)\citenamefont {Maeno}, \citenamefont {Yonezawa},\ and\ \citenamefont {Ramires}}]{doi:10.7566/JPSJ.93.062001}%
  \BibitemOpen
  \bibfield  {author} {\bibinfo {author} {\bibfnamefont {Yoshiteru}\ \bibnamefont {Maeno}}, \bibinfo {author} {\bibfnamefont {Shingo}\ \bibnamefont {Yonezawa}}, \ and\ \bibinfo {author} {\bibfnamefont {Aline}\ \bibnamefont {Ramires}},\ }\bibfield  {title} {\enquote {\bibinfo {title} {{Still Mystery after All These Years —Unconventional Superconductivity of Sr$_2$RuO$_4$—}},}\ }\href {\doibase 10.7566/JPSJ.93.062001} {\bibfield  {journal} {\bibinfo  {journal} {J. Phys. Soc. Jpn.}\ }\textbf {\bibinfo {volume} {93}},\ \bibinfo {pages} {062001} (\bibinfo {year} {2024})}\BibitemShut {NoStop}%
\bibitem [{\citenamefont {{Clifford W. Hicks and Daniel O. Brodsky and Edward A. Yelland and Alexandra S. Gibbs and Jan A. N. Bruin and Mark E. Barber and Stephen D. Edkins and Keigo Nishimura and Shingo Yonezawa and Yoshiteru Maeno and Andrew P. Mackenzie }}(2014)}]{doi:10.1126/science.1248292}%
  \BibitemOpen
  \bibfield  {author} {\bibinfo {author} {\bibnamefont {{Clifford W. Hicks and Daniel O. Brodsky and Edward A. Yelland and Alexandra S. Gibbs and Jan A. N. Bruin and Mark E. Barber and Stephen D. Edkins and Keigo Nishimura and Shingo Yonezawa and Yoshiteru Maeno and Andrew P. Mackenzie }}},\ }\bibfield  {title} {\enquote {\bibinfo {title} {{Strong Increase of $T_c$ of Sr$_2$RuO$_4$ Under Both Tensile and Compressive Strain}},}\ }\href {\doibase 10.1126/science.1248292} {\bibfield  {journal} {\bibinfo  {journal} {Science}\ }\textbf {\bibinfo {volume} {344}},\ \bibinfo {pages} {283--285} (\bibinfo {year} {2014})}\BibitemShut {NoStop}%
\bibitem [{\citenamefont {{You-Sheng Li and Naoki Kikugawa and Dmitry A. Sokolov and Fabian Jerzembeck and Alexandra S. Gibbs and Yoshiteru Maeno and Clifford W. Hicks and Jörg Schmalian and Michael Nicklas and Andrew P. Mackenzie }}(2021)}]{doi:10.1073/pnas.2020492118}%
  \BibitemOpen
  \bibfield  {author} {\bibinfo {author} {\bibnamefont {{You-Sheng Li and Naoki Kikugawa and Dmitry A. Sokolov and Fabian Jerzembeck and Alexandra S. Gibbs and Yoshiteru Maeno and Clifford W. Hicks and Jörg Schmalian and Michael Nicklas and Andrew P. Mackenzie }}},\ }\bibfield  {title} {\enquote {\bibinfo {title} {{High-sensitivity heat-capacity measurements on Sr$_2$RuO$_4$ under uniaxial pressure}},}\ }\href {\doibase 10.1073/pnas.2020492118} {\bibfield  {journal} {\bibinfo  {journal} {Proc. Natl. Acad. Sci. U.S.A.}\ }\textbf {\bibinfo {volume} {118}},\ \bibinfo {pages} {e2020492118} (\bibinfo {year} {2021})}\BibitemShut {NoStop}%
\bibitem [{\citenamefont {Chronister}\ \emph {et~al.}(2021)\citenamefont {Chronister}, \citenamefont {Pustogow}, \citenamefont {Kikugawa}, \citenamefont {Sokolov}, \citenamefont {Jerzembeck}, \citenamefont {Hicks}, \citenamefont {Mackenzie}, \citenamefont {Bauer},\ and\ \citenamefont {Brown}}]{doi:10.1073/pnas.2025313118}%
  \BibitemOpen
  \bibfield  {author} {\bibinfo {author} {\bibfnamefont {Aaron}\ \bibnamefont {Chronister}}, \bibinfo {author} {\bibfnamefont {Andrej}\ \bibnamefont {Pustogow}}, \bibinfo {author} {\bibfnamefont {Naoki}\ \bibnamefont {Kikugawa}}, \bibinfo {author} {\bibfnamefont {Dmitry~A.}\ \bibnamefont {Sokolov}}, \bibinfo {author} {\bibfnamefont {Fabian}\ \bibnamefont {Jerzembeck}}, \bibinfo {author} {\bibfnamefont {Clifford~W.}\ \bibnamefont {Hicks}}, \bibinfo {author} {\bibfnamefont {Andrew~P.}\ \bibnamefont {Mackenzie}}, \bibinfo {author} {\bibfnamefont {Eric~D.}\ \bibnamefont {Bauer}}, \ and\ \bibinfo {author} {\bibfnamefont {Stuart~E.}\ \bibnamefont {Brown}},\ }\bibfield  {title} {\enquote {\bibinfo {title} {{Evidence for even parity unconventional superconductivity in Sr$_2$RuO$_4$}},}\ }\href {\doibase 10.1073/pnas.2025313118} {\bibfield  {journal} {\bibinfo  {journal} {Proc. Natl. Acad. Sci. U.S.A.}\ }\textbf {\bibinfo {volume} {118}},\ \bibinfo {pages} {e2025313118} (\bibinfo {year} {2021})}\BibitemShut {NoStop}%
\bibitem [{\citenamefont {Li}\ \emph {et~al.}(2022)\citenamefont {Li}, \citenamefont {Garst}, \citenamefont {Schmalian}, \citenamefont {Ghosh}, \citenamefont {Kikugawa}, \citenamefont {Sokolov}, \citenamefont {Hicks}, \citenamefont {Jerzembeck}, \citenamefont {Ikeda}, \citenamefont {Hu} \emph {et~al.}}]{li2022elastocaloric}%
  \BibitemOpen
  \bibfield  {author} {\bibinfo {author} {\bibfnamefont {You-Sheng}\ \bibnamefont {Li}}, \bibinfo {author} {\bibfnamefont {Markus}\ \bibnamefont {Garst}}, \bibinfo {author} {\bibfnamefont {J{\"o}rg}\ \bibnamefont {Schmalian}}, \bibinfo {author} {\bibfnamefont {Sayak}\ \bibnamefont {Ghosh}}, \bibinfo {author} {\bibfnamefont {Naoki}\ \bibnamefont {Kikugawa}}, \bibinfo {author} {\bibfnamefont {Dmitry~A}\ \bibnamefont {Sokolov}}, \bibinfo {author} {\bibfnamefont {Clifford~W}\ \bibnamefont {Hicks}}, \bibinfo {author} {\bibfnamefont {Fabian}\ \bibnamefont {Jerzembeck}}, \bibinfo {author} {\bibfnamefont {Matthias~S}\ \bibnamefont {Ikeda}}, \bibinfo {author} {\bibfnamefont {Zhenhai}\ \bibnamefont {Hu}},  \emph {et~al.},\ }\bibfield  {title} {\enquote {\bibinfo {title} {{Elastocaloric determination of the phase diagram of Sr$_2$RuO$_4$}},}\ }\href {\doibase 10.1038/s41586-022-04820-z} {\bibfield  {journal} {\bibinfo  {journal} {Nature}\ }\textbf {\bibinfo {volume} {607}},\ \bibinfo {pages} {276--280} (\bibinfo {year}
  {2022})}\BibitemShut {NoStop}%
\bibitem [{\citenamefont {Jerzembeck}\ \emph {et~al.}(2022)\citenamefont {Jerzembeck}, \citenamefont {Røising}, \citenamefont {Steppke}, \citenamefont {Rosner}, \citenamefont {Sokolov}, \citenamefont {Kikugawa}, \citenamefont {Scaffidi}, \citenamefont {Simon}, \citenamefont {Mackenzie},\ and\ \citenamefont {Hicks}}]{jerzembeck2022superconductivity}%
  \BibitemOpen
  \bibfield  {author} {\bibinfo {author} {\bibfnamefont {Fabian}\ \bibnamefont {Jerzembeck}}, \bibinfo {author} {\bibfnamefont {Henrik~S.}\ \bibnamefont {Røising}}, \bibinfo {author} {\bibfnamefont {Alexander}\ \bibnamefont {Steppke}}, \bibinfo {author} {\bibfnamefont {Helge}\ \bibnamefont {Rosner}}, \bibinfo {author} {\bibfnamefont {Dmitry~A.}\ \bibnamefont {Sokolov}}, \bibinfo {author} {\bibfnamefont {Naoki}\ \bibnamefont {Kikugawa}}, \bibinfo {author} {\bibfnamefont {Thomas}\ \bibnamefont {Scaffidi}}, \bibinfo {author} {\bibfnamefont {Steven~H.}\ \bibnamefont {Simon}}, \bibinfo {author} {\bibfnamefont {Andrew~P.}\ \bibnamefont {Mackenzie}}, \ and\ \bibinfo {author} {\bibfnamefont {Clifford~W.}\ \bibnamefont {Hicks}},\ }\bibfield  {title} {\enquote {\bibinfo {title} {{The superconductivity of Sr$_2$RuO$_4$ under c-axis uniaxial stress}},}\ }\href {\doibase 10.1038/s41467-022-32177-4} {\bibfield  {journal} {\bibinfo  {journal} {Nat. Commun.}\ }\textbf {\bibinfo {volume} {13}},\ \bibinfo {pages} {4596}
  (\bibinfo {year} {2022})}\BibitemShut {NoStop}%
\bibitem [{\citenamefont {Palle}\ \emph {et~al.}(2023)\citenamefont {Palle}, \citenamefont {Hicks}, \citenamefont {Valent\'{\i}}, \citenamefont {Hu}, \citenamefont {Li}, \citenamefont {Rost}, \citenamefont {Nicklas}, \citenamefont {Mackenzie},\ and\ \citenamefont {Schmalian}}]{PhysRevB.108.094516}%
  \BibitemOpen
  \bibfield  {author} {\bibinfo {author} {\bibfnamefont {Grgur}\ \bibnamefont {Palle}}, \bibinfo {author} {\bibfnamefont {Clifford}\ \bibnamefont {Hicks}}, \bibinfo {author} {\bibfnamefont {Roser}\ \bibnamefont {Valent\'{\i}}}, \bibinfo {author} {\bibfnamefont {Zhenhai}\ \bibnamefont {Hu}}, \bibinfo {author} {\bibfnamefont {You-Sheng}\ \bibnamefont {Li}}, \bibinfo {author} {\bibfnamefont {Andreas}\ \bibnamefont {Rost}}, \bibinfo {author} {\bibfnamefont {Michael}\ \bibnamefont {Nicklas}}, \bibinfo {author} {\bibfnamefont {Andrew~P.}\ \bibnamefont {Mackenzie}}, \ and\ \bibinfo {author} {\bibfnamefont {J\"org}\ \bibnamefont {Schmalian}},\ }\bibfield  {title} {\enquote {\bibinfo {title} {{Constraints on the superconducting state of ${\text{Sr}}_{2}{\text{RuO}}_{4}$ from elastocaloric measurements}},}\ }\href {\doibase 10.1103/PhysRevB.108.094516} {\bibfield  {journal} {\bibinfo  {journal} {Phys. Rev. B}\ }\textbf {\bibinfo {volume} {108}},\ \bibinfo {pages} {094516} (\bibinfo {year} {2023})}\BibitemShut {NoStop}%
\bibitem [{\citenamefont {Xia}\ \emph {et~al.}(2006)\citenamefont {Xia}, \citenamefont {Maeno}, \citenamefont {Beyersdorf}, \citenamefont {Fejer},\ and\ \citenamefont {Kapitulnik}}]{PhysRevLett.97.167002}%
  \BibitemOpen
  \bibfield  {author} {\bibinfo {author} {\bibfnamefont {Jing}\ \bibnamefont {Xia}}, \bibinfo {author} {\bibfnamefont {Yoshiteru}\ \bibnamefont {Maeno}}, \bibinfo {author} {\bibfnamefont {Peter~T.}\ \bibnamefont {Beyersdorf}}, \bibinfo {author} {\bibfnamefont {M.~M.}\ \bibnamefont {Fejer}}, \ and\ \bibinfo {author} {\bibfnamefont {Aharon}\ \bibnamefont {Kapitulnik}},\ }\bibfield  {title} {\enquote {\bibinfo {title} {{High Resolution Polar Kerr Effect Measurements of ${\mathrm{Sr}}_{2}{\mathrm{RuO}}_{4}$: Evidence for Broken Time-Reversal Symmetry in the Superconducting State}},}\ }\href {\doibase 10.1103/PhysRevLett.97.167002} {\bibfield  {journal} {\bibinfo  {journal} {Phys. Rev. Lett.}\ }\textbf {\bibinfo {volume} {97}},\ \bibinfo {pages} {167002} (\bibinfo {year} {2006})}\BibitemShut {NoStop}%
\bibitem [{\citenamefont {Luke}\ \emph {et~al.}(1998)\citenamefont {Luke}, \citenamefont {Fudamoto}, \citenamefont {Kojima}, \citenamefont {Larkin}, \citenamefont {Merrin}, \citenamefont {Nachumi}, \citenamefont {Uemura}, \citenamefont {Maeno}, \citenamefont {Mao}, \citenamefont {Mori}, \citenamefont {Nakamura},\ and\ \citenamefont {Sigrist}}]{luke1998time}%
  \BibitemOpen
  \bibfield  {author} {\bibinfo {author} {\bibfnamefont {G.~Ml.}\ \bibnamefont {Luke}}, \bibinfo {author} {\bibfnamefont {Y.}~\bibnamefont {Fudamoto}}, \bibinfo {author} {\bibfnamefont {K.~M.}\ \bibnamefont {Kojima}}, \bibinfo {author} {\bibfnamefont {M.~I.}\ \bibnamefont {Larkin}}, \bibinfo {author} {\bibfnamefont {J.}~\bibnamefont {Merrin}}, \bibinfo {author} {\bibfnamefont {B.}~\bibnamefont {Nachumi}}, \bibinfo {author} {\bibfnamefont {Y.~J.}\ \bibnamefont {Uemura}}, \bibinfo {author} {\bibfnamefont {Y.}~\bibnamefont {Maeno}}, \bibinfo {author} {\bibfnamefont {Z.~Q.}\ \bibnamefont {Mao}}, \bibinfo {author} {\bibfnamefont {Y.}~\bibnamefont {Mori}}, \bibinfo {author} {\bibfnamefont {H.}~\bibnamefont {Nakamura}}, \ and\ \bibinfo {author} {\bibfnamefont {M.}~\bibnamefont {Sigrist}},\ }\bibfield  {title} {\enquote {\bibinfo {title} {{Time-reversal symmetry-breaking superconductivity in Sr$_2$RuO$_4$}},}\ }\href {\doibase 10.1038/29038} {\bibfield  {journal} {\bibinfo  {journal} {Nature}\ }\textbf {\bibinfo
  {volume} {394}},\ \bibinfo {pages} {558--561} (\bibinfo {year} {1998})}\BibitemShut {NoStop}%
\bibitem [{\citenamefont {Kivelson}\ \emph {et~al.}(2020)\citenamefont {Kivelson}, \citenamefont {Yuan}, \citenamefont {Ramshaw},\ and\ \citenamefont {Thomale}}]{kivelson2020proposal}%
  \BibitemOpen
  \bibfield  {author} {\bibinfo {author} {\bibfnamefont {Steven~Allan}\ \bibnamefont {Kivelson}}, \bibinfo {author} {\bibfnamefont {Andrew~Chang}\ \bibnamefont {Yuan}}, \bibinfo {author} {\bibfnamefont {Brad}\ \bibnamefont {Ramshaw}}, \ and\ \bibinfo {author} {\bibfnamefont {Ronny}\ \bibnamefont {Thomale}},\ }\bibfield  {title} {\enquote {\bibinfo {title} {{A proposal for reconciling diverse experiments on the superconducting state in Sr$_2$RuO$_4$}},}\ }\href {\doibase 10.1038/s41535-020-0245-1} {\bibfield  {journal} {\bibinfo  {journal} {npj Quantum Mater.}\ }\textbf {\bibinfo {volume} {5}},\ \bibinfo {pages} {43} (\bibinfo {year} {2020})}\BibitemShut {NoStop}%
\bibitem [{\citenamefont {R\o{}mer}\ \emph {et~al.}(2019)\citenamefont {R\o{}mer}, \citenamefont {Scherer}, \citenamefont {Eremin}, \citenamefont {Hirschfeld},\ and\ \citenamefont {Andersen}}]{PhysRevLett.123.247001}%
  \BibitemOpen
  \bibfield  {author} {\bibinfo {author} {\bibfnamefont {A.~T.}\ \bibnamefont {R\o{}mer}}, \bibinfo {author} {\bibfnamefont {D.~D.}\ \bibnamefont {Scherer}}, \bibinfo {author} {\bibfnamefont {I.~M.}\ \bibnamefont {Eremin}}, \bibinfo {author} {\bibfnamefont {P.~J.}\ \bibnamefont {Hirschfeld}}, \ and\ \bibinfo {author} {\bibfnamefont {B.~M.}\ \bibnamefont {Andersen}},\ }\bibfield  {title} {\enquote {\bibinfo {title} {{Knight Shift and Leading Superconducting Instability from Spin Fluctuations in ${\mathrm{Sr}}_{2}{\mathrm{RuO}}_{4}$}},}\ }\href {\doibase 10.1103/PhysRevLett.123.247001} {\bibfield  {journal} {\bibinfo  {journal} {Phys. Rev. Lett.}\ }\textbf {\bibinfo {volume} {123}},\ \bibinfo {pages} {247001} (\bibinfo {year} {2019})}\BibitemShut {NoStop}%
\bibitem [{\citenamefont {K\"aser}\ \emph {et~al.}(2022)\citenamefont {K\"aser}, \citenamefont {Strand}, \citenamefont {Wentzell}, \citenamefont {Georges}, \citenamefont {Parcollet},\ and\ \citenamefont {Hansmann}}]{PhysRevB.105.155101}%
  \BibitemOpen
  \bibfield  {author} {\bibinfo {author} {\bibfnamefont {Stefan}\ \bibnamefont {K\"aser}}, \bibinfo {author} {\bibfnamefont {Hugo U.~R.}\ \bibnamefont {Strand}}, \bibinfo {author} {\bibfnamefont {Nils}\ \bibnamefont {Wentzell}}, \bibinfo {author} {\bibfnamefont {Antoine}\ \bibnamefont {Georges}}, \bibinfo {author} {\bibfnamefont {Olivier}\ \bibnamefont {Parcollet}}, \ and\ \bibinfo {author} {\bibfnamefont {Philipp}\ \bibnamefont {Hansmann}},\ }\bibfield  {title} {\enquote {\bibinfo {title} {{Interorbital singlet pairing in ${\mathrm{Sr}}_{2}{\mathrm{RuO}}_{4}$: A Hund's superconductor}},}\ }\href {\doibase 10.1103/PhysRevB.105.155101} {\bibfield  {journal} {\bibinfo  {journal} {Phys. Rev. B}\ }\textbf {\bibinfo {volume} {105}},\ \bibinfo {pages} {155101} (\bibinfo {year} {2022})}\BibitemShut {NoStop}%
\bibitem [{\citenamefont {R\o{}mer}\ \emph {et~al.}(2022)\citenamefont {R\o{}mer}, \citenamefont {Maier}, \citenamefont {Kreisel}, \citenamefont {Hirschfeld},\ and\ \citenamefont {Andersen}}]{PhysRevResearch.4.033011}%
  \BibitemOpen
  \bibfield  {author} {\bibinfo {author} {\bibfnamefont {Astrid~T.}\ \bibnamefont {R\o{}mer}}, \bibinfo {author} {\bibfnamefont {T.~A.}\ \bibnamefont {Maier}}, \bibinfo {author} {\bibfnamefont {Andreas}\ \bibnamefont {Kreisel}}, \bibinfo {author} {\bibfnamefont {P.~J.}\ \bibnamefont {Hirschfeld}}, \ and\ \bibinfo {author} {\bibfnamefont {Brian~M.}\ \bibnamefont {Andersen}},\ }\bibfield  {title} {\enquote {\bibinfo {title} {{Leading superconducting instabilities in three-dimensional models for ${\mathrm{Sr}}_{2}{\mathrm{RuO}}_{4}$}},}\ }\href {\doibase 10.1103/PhysRevResearch.4.033011} {\bibfield  {journal} {\bibinfo  {journal} {Phys. Rev. Res.}\ }\textbf {\bibinfo {volume} {4}},\ \bibinfo {pages} {033011} (\bibinfo {year} {2022})}\BibitemShut {NoStop}%
\bibitem [{\citenamefont {Profe}\ \emph {et~al.}(2024)\citenamefont {Profe}, \citenamefont {Beck}, \citenamefont {Kennes}, \citenamefont {Georges},\ and\ \citenamefont {Gingras}}]{profe2024competition}%
  \BibitemOpen
  \bibfield  {author} {\bibinfo {author} {\bibfnamefont {Jonas~B.}\ \bibnamefont {Profe}}, \bibinfo {author} {\bibfnamefont {Sophie}\ \bibnamefont {Beck}}, \bibinfo {author} {\bibfnamefont {Dante~M.}\ \bibnamefont {Kennes}}, \bibinfo {author} {\bibfnamefont {Antoine}\ \bibnamefont {Georges}}, \ and\ \bibinfo {author} {\bibfnamefont {Olivier}\ \bibnamefont {Gingras}},\ }\bibfield  {title} {\enquote {\bibinfo {title} {{Competition between d-wave superconductivity and magnetism in uniaxially strained Sr$_2$RuO$_4$}},}\ }\href {\doibase 10.1038/s41535-024-00661-3} {\bibfield  {journal} {\bibinfo  {journal} {npj Quantum Mater.}\ }\textbf {\bibinfo {volume} {9}},\ \bibinfo {pages} {53} (\bibinfo {year} {2024})}\BibitemShut {NoStop}%
\bibitem [{\citenamefont {Sidis}\ \emph {et~al.}(1999)\citenamefont {Sidis}, \citenamefont {Braden}, \citenamefont {Bourges}, \citenamefont {Hennion}, \citenamefont {NishiZaki}, \citenamefont {Maeno},\ and\ \citenamefont {Mori}}]{PhysRevLett.83.3320}%
  \BibitemOpen
  \bibfield  {author} {\bibinfo {author} {\bibfnamefont {Y.}~\bibnamefont {Sidis}}, \bibinfo {author} {\bibfnamefont {M.}~\bibnamefont {Braden}}, \bibinfo {author} {\bibfnamefont {P.}~\bibnamefont {Bourges}}, \bibinfo {author} {\bibfnamefont {B.}~\bibnamefont {Hennion}}, \bibinfo {author} {\bibfnamefont {S.}~\bibnamefont {NishiZaki}}, \bibinfo {author} {\bibfnamefont {Y.}~\bibnamefont {Maeno}}, \ and\ \bibinfo {author} {\bibfnamefont {Y.}~\bibnamefont {Mori}},\ }\bibfield  {title} {\enquote {\bibinfo {title} {{Evidence for Incommensurate Spin Fluctuations in ${\mathrm{Sr}}_{2}{\mathrm{RuO}}_{4}$}},}\ }\href {\doibase 10.1103/PhysRevLett.83.3320} {\bibfield  {journal} {\bibinfo  {journal} {Phys. Rev. Lett.}\ }\textbf {\bibinfo {volume} {83}},\ \bibinfo {pages} {3320--3323} (\bibinfo {year} {1999})}\BibitemShut {NoStop}%
\bibitem [{\citenamefont {Braden}\ \emph {et~al.}(2002{\natexlab{a}})\citenamefont {Braden}, \citenamefont {Sidis}, \citenamefont {Bourges}, \citenamefont {Pfeuty}, \citenamefont {Kulda}, \citenamefont {Mao},\ and\ \citenamefont {Maeno}}]{PhysRevB.66.064522}%
  \BibitemOpen
  \bibfield  {author} {\bibinfo {author} {\bibfnamefont {M.}~\bibnamefont {Braden}}, \bibinfo {author} {\bibfnamefont {Y.}~\bibnamefont {Sidis}}, \bibinfo {author} {\bibfnamefont {P.}~\bibnamefont {Bourges}}, \bibinfo {author} {\bibfnamefont {P.}~\bibnamefont {Pfeuty}}, \bibinfo {author} {\bibfnamefont {J.}~\bibnamefont {Kulda}}, \bibinfo {author} {\bibfnamefont {Z.}~\bibnamefont {Mao}}, \ and\ \bibinfo {author} {\bibfnamefont {Y.}~\bibnamefont {Maeno}},\ }\bibfield  {title} {\enquote {\bibinfo {title} {{Inelastic neutron scattering study of magnetic excitations in ${\mathrm{Sr}}_{2}{\mathrm{RuO}}_{4}$}},}\ }\href {\doibase 10.1103/PhysRevB.66.064522} {\bibfield  {journal} {\bibinfo  {journal} {Phys. Rev. B}\ }\textbf {\bibinfo {volume} {66}},\ \bibinfo {pages} {064522} (\bibinfo {year} {2002}{\natexlab{a}})}\BibitemShut {NoStop}%
\bibitem [{\citenamefont {Braden}\ \emph {et~al.}(2004)\citenamefont {Braden}, \citenamefont {Steffens}, \citenamefont {Sidis}, \citenamefont {Kulda}, \citenamefont {Bourges}, \citenamefont {Hayden}, \citenamefont {Kikugawa},\ and\ \citenamefont {Maeno}}]{PhysRevLett.92.097402}%
  \BibitemOpen
  \bibfield  {author} {\bibinfo {author} {\bibfnamefont {M.}~\bibnamefont {Braden}}, \bibinfo {author} {\bibfnamefont {P.}~\bibnamefont {Steffens}}, \bibinfo {author} {\bibfnamefont {Y.}~\bibnamefont {Sidis}}, \bibinfo {author} {\bibfnamefont {J.}~\bibnamefont {Kulda}}, \bibinfo {author} {\bibfnamefont {P.}~\bibnamefont {Bourges}}, \bibinfo {author} {\bibfnamefont {S.}~\bibnamefont {Hayden}}, \bibinfo {author} {\bibfnamefont {N.}~\bibnamefont {Kikugawa}}, \ and\ \bibinfo {author} {\bibfnamefont {Y.}~\bibnamefont {Maeno}},\ }\bibfield  {title} {\enquote {\bibinfo {title} {{Anisotropy of the Incommensurate Fluctuations in ${\mathrm{S}\mathrm{r}}_{2}{\mathrm{R}\mathrm{u}\mathrm{O}}_{4}$: A Study with Polarized Neutrons}},}\ }\href {\doibase 10.1103/PhysRevLett.92.097402} {\bibfield  {journal} {\bibinfo  {journal} {Phys. Rev. Lett.}\ }\textbf {\bibinfo {volume} {92}},\ \bibinfo {pages} {097402} (\bibinfo {year} {2004})}\BibitemShut {NoStop}%
\bibitem [{\citenamefont {Steffens}\ \emph {et~al.}(2019)\citenamefont {Steffens}, \citenamefont {Sidis}, \citenamefont {Kulda}, \citenamefont {Mao}, \citenamefont {Maeno}, \citenamefont {Mazin},\ and\ \citenamefont {Braden}}]{PhysRevLett.122.047004}%
  \BibitemOpen
  \bibfield  {author} {\bibinfo {author} {\bibfnamefont {P.}~\bibnamefont {Steffens}}, \bibinfo {author} {\bibfnamefont {Y.}~\bibnamefont {Sidis}}, \bibinfo {author} {\bibfnamefont {J.}~\bibnamefont {Kulda}}, \bibinfo {author} {\bibfnamefont {Z.~Q.}\ \bibnamefont {Mao}}, \bibinfo {author} {\bibfnamefont {Y.}~\bibnamefont {Maeno}}, \bibinfo {author} {\bibfnamefont {I.~I.}\ \bibnamefont {Mazin}}, \ and\ \bibinfo {author} {\bibfnamefont {M.}~\bibnamefont {Braden}},\ }\bibfield  {title} {\enquote {\bibinfo {title} {{Spin Fluctuations in ${\mathrm{Sr}}_{2}{\mathrm{RuO}}_{4}$ from Polarized Neutron Scattering: Implications for Superconductivity}},}\ }\href {\doibase 10.1103/PhysRevLett.122.047004} {\bibfield  {journal} {\bibinfo  {journal} {Phys. Rev. Lett.}\ }\textbf {\bibinfo {volume} {122}},\ \bibinfo {pages} {047004} (\bibinfo {year} {2019})}\BibitemShut {NoStop}%
\bibitem [{\citenamefont {Jenni}\ \emph {et~al.}(2021)\citenamefont {Jenni}, \citenamefont {Kunkem\"oller}, \citenamefont {Steffens}, \citenamefont {Sidis}, \citenamefont {Bewley}, \citenamefont {Mao}, \citenamefont {Maeno},\ and\ \citenamefont {Braden}}]{PhysRevB.103.104511}%
  \BibitemOpen
  \bibfield  {author} {\bibinfo {author} {\bibfnamefont {K.}~\bibnamefont {Jenni}}, \bibinfo {author} {\bibfnamefont {S.}~\bibnamefont {Kunkem\"oller}}, \bibinfo {author} {\bibfnamefont {P.}~\bibnamefont {Steffens}}, \bibinfo {author} {\bibfnamefont {Y.}~\bibnamefont {Sidis}}, \bibinfo {author} {\bibfnamefont {R.}~\bibnamefont {Bewley}}, \bibinfo {author} {\bibfnamefont {Z.~Q.}\ \bibnamefont {Mao}}, \bibinfo {author} {\bibfnamefont {Y.}~\bibnamefont {Maeno}}, \ and\ \bibinfo {author} {\bibfnamefont {M.}~\bibnamefont {Braden}},\ }\bibfield  {title} {\enquote {\bibinfo {title} {{Neutron scattering studies on spin fluctuations in ${\mathrm{Sr}}_{2}{\mathrm{RuO}}_{4}$}},}\ }\href {\doibase 10.1103/PhysRevB.103.104511} {\bibfield  {journal} {\bibinfo  {journal} {Phys. Rev. B}\ }\textbf {\bibinfo {volume} {103}},\ \bibinfo {pages} {104511} (\bibinfo {year} {2021})}\BibitemShut {NoStop}%
\bibitem [{\citenamefont {Minakata}\ and\ \citenamefont {Maeno}(2001)}]{PhysRevB.63.180504}%
  \BibitemOpen
  \bibfield  {author} {\bibinfo {author} {\bibfnamefont {M.}~\bibnamefont {Minakata}}\ and\ \bibinfo {author} {\bibfnamefont {Y.}~\bibnamefont {Maeno}},\ }\bibfield  {title} {\enquote {\bibinfo {title} {{Magnetic ordering in ${\mathrm{Sr}}_{2}{\mathrm{RuO}}_{4}$ induced by nonmagnetic impurities}},}\ }\href {\doibase 10.1103/PhysRevB.63.180504} {\bibfield  {journal} {\bibinfo  {journal} {Phys. Rev. B}\ }\textbf {\bibinfo {volume} {63}},\ \bibinfo {pages} {180504} (\bibinfo {year} {2001})}\BibitemShut {NoStop}%
\bibitem [{\citenamefont {Braden}\ \emph {et~al.}(2002{\natexlab{b}})\citenamefont {Braden}, \citenamefont {Friedt}, \citenamefont {Sidis}, \citenamefont {Bourges}, \citenamefont {Minakata},\ and\ \citenamefont {Maeno}}]{PhysRevLett.88.197002}%
  \BibitemOpen
  \bibfield  {author} {\bibinfo {author} {\bibfnamefont {M.}~\bibnamefont {Braden}}, \bibinfo {author} {\bibfnamefont {O.}~\bibnamefont {Friedt}}, \bibinfo {author} {\bibfnamefont {Y.}~\bibnamefont {Sidis}}, \bibinfo {author} {\bibfnamefont {P.}~\bibnamefont {Bourges}}, \bibinfo {author} {\bibfnamefont {M.}~\bibnamefont {Minakata}}, \ and\ \bibinfo {author} {\bibfnamefont {Y.}~\bibnamefont {Maeno}},\ }\bibfield  {title} {\enquote {\bibinfo {title} {{Incommensurate Magnetic Ordering in ${\mathrm{Sr}}_{2}{\mathrm{Ru}}_{1\ensuremath{-}\mathit{x}}{\mathrm{Ti}}_{\mathit{x}}{O}_{4}$}},}\ }\href {\doibase 10.1103/PhysRevLett.88.197002} {\bibfield  {journal} {\bibinfo  {journal} {Phys. Rev. Lett.}\ }\textbf {\bibinfo {volume} {88}},\ \bibinfo {pages} {197002} (\bibinfo {year} {2002}{\natexlab{b}})}\BibitemShut {NoStop}%
\bibitem [{\citenamefont {Kunkem\"oller}\ \emph {et~al.}(2014)\citenamefont {Kunkem\"oller}, \citenamefont {Nugroho}, \citenamefont {Sidis},\ and\ \citenamefont {Braden}}]{PhysRevB.89.045119}%
  \BibitemOpen
  \bibfield  {author} {\bibinfo {author} {\bibfnamefont {S.}~\bibnamefont {Kunkem\"oller}}, \bibinfo {author} {\bibfnamefont {A.~A.}\ \bibnamefont {Nugroho}}, \bibinfo {author} {\bibfnamefont {Y.}~\bibnamefont {Sidis}}, \ and\ \bibinfo {author} {\bibfnamefont {M.}~\bibnamefont {Braden}},\ }\bibfield  {title} {\enquote {\bibinfo {title} {{Spin-density-wave ordering in ${\mathrm{Ca}}_{0.5}$Sr${}_{1.5}$${\mathrm{RuO}}_{4}$ studied by neutron scattering}},}\ }\href {\doibase 10.1103/PhysRevB.89.045119} {\bibfield  {journal} {\bibinfo  {journal} {Phys. Rev. B}\ }\textbf {\bibinfo {volume} {89}},\ \bibinfo {pages} {045119} (\bibinfo {year} {2014})}\BibitemShut {NoStop}%
\bibitem [{\citenamefont {Grinenko}\ \emph {et~al.}(2021)\citenamefont {Grinenko}, \citenamefont {Ghosh}, \citenamefont {Sarkar}, \citenamefont {Orain}, \citenamefont {Nikitin}, \citenamefont {Elender}, \citenamefont {Das}, \citenamefont {Guguchia}, \citenamefont {Br{\"u}ckner}, \citenamefont {Barber} \emph {et~al.}}]{grinenko2021split}%
  \BibitemOpen
  \bibfield  {author} {\bibinfo {author} {\bibfnamefont {Vadim}\ \bibnamefont {Grinenko}}, \bibinfo {author} {\bibfnamefont {Shreenanda}\ \bibnamefont {Ghosh}}, \bibinfo {author} {\bibfnamefont {Rajib}\ \bibnamefont {Sarkar}}, \bibinfo {author} {\bibfnamefont {Jean-Christophe}\ \bibnamefont {Orain}}, \bibinfo {author} {\bibfnamefont {Artem}\ \bibnamefont {Nikitin}}, \bibinfo {author} {\bibfnamefont {Matthias}\ \bibnamefont {Elender}}, \bibinfo {author} {\bibfnamefont {Debarchan}\ \bibnamefont {Das}}, \bibinfo {author} {\bibfnamefont {Zurab}\ \bibnamefont {Guguchia}}, \bibinfo {author} {\bibfnamefont {Felix}\ \bibnamefont {Br{\"u}ckner}}, \bibinfo {author} {\bibfnamefont {Mark~E}\ \bibnamefont {Barber}},  \emph {et~al.},\ }\bibfield  {title} {\enquote {\bibinfo {title} {{Split superconducting and time-reversal symmetry-breaking transitions in Sr$_2$RuO$_4$ under stress}},}\ }\href {\doibase 10.1038/s41567-021-01182-7} {\bibfield  {journal} {\bibinfo  {journal} {Nat. Phys.}\ }\textbf {\bibinfo {volume} {17}},\
  \bibinfo {pages} {748--754} (\bibinfo {year} {2021})}\BibitemShut {NoStop}%
\bibitem [{\citenamefont {Mazin}\ and\ \citenamefont {Singh}(1999)}]{PhysRevLett.82.4324}%
  \BibitemOpen
  \bibfield  {author} {\bibinfo {author} {\bibfnamefont {I.~I.}\ \bibnamefont {Mazin}}\ and\ \bibinfo {author} {\bibfnamefont {D.~J.}\ \bibnamefont {Singh}},\ }\bibfield  {title} {\enquote {\bibinfo {title} {Competitions in layered ruthenates: Ferromagnetism versus antiferromagnetism and triplet versus singlet pairing},}\ }\href {\doibase 10.1103/PhysRevLett.82.4324} {\bibfield  {journal} {\bibinfo  {journal} {Phys. Rev. Lett.}\ }\textbf {\bibinfo {volume} {82}},\ \bibinfo {pages} {4324--4327} (\bibinfo {year} {1999})}\BibitemShut {NoStop}%
\bibitem [{\citenamefont {Boehnke}\ \emph {et~al.}(2018)\citenamefont {Boehnke}, \citenamefont {Werner},\ and\ \citenamefont {Lechermann}}]{Boehnke_2018}%
  \BibitemOpen
  \bibfield  {author} {\bibinfo {author} {\bibfnamefont {Lewin}\ \bibnamefont {Boehnke}}, \bibinfo {author} {\bibfnamefont {Philipp}\ \bibnamefont {Werner}}, \ and\ \bibinfo {author} {\bibfnamefont {Frank}\ \bibnamefont {Lechermann}},\ }\bibfield  {title} {\enquote {\bibinfo {title} {{Multi-orbital nature of the spin fluctuations in Sr$_2$RuO$_4$}},}\ }\href {\doibase 10.1209/0295-5075/122/57001} {\bibfield  {journal} {\bibinfo  {journal} {EPL}\ }\textbf {\bibinfo {volume} {122}},\ \bibinfo {pages} {57001} (\bibinfo {year} {2018})}\BibitemShut {NoStop}%
\bibitem [{\citenamefont {Acharya}\ \emph {et~al.}(2019)\citenamefont {Acharya}, \citenamefont {Pashov}, \citenamefont {Weber}, \citenamefont {Park}, \citenamefont {Sponza},\ and\ \citenamefont {Schilfgaarde}}]{acharya2019evening}%
  \BibitemOpen
  \bibfield  {author} {\bibinfo {author} {\bibfnamefont {Swagata}\ \bibnamefont {Acharya}}, \bibinfo {author} {\bibfnamefont {Dimitar}\ \bibnamefont {Pashov}}, \bibinfo {author} {\bibfnamefont {C{\'e}dric}\ \bibnamefont {Weber}}, \bibinfo {author} {\bibfnamefont {Hyowon}\ \bibnamefont {Park}}, \bibinfo {author} {\bibfnamefont {Lorenzo}\ \bibnamefont {Sponza}}, \ and\ \bibinfo {author} {\bibfnamefont {Mark~Van}\ \bibnamefont {Schilfgaarde}},\ }\bibfield  {title} {\enquote {\bibinfo {title} {{Evening out the spin and charge parity to increase $T_c$ in Sr$_2$RuO$_4$}},}\ }\href {\doibase 10.1038/s42005-019-0254-1} {\bibfield  {journal} {\bibinfo  {journal} {Commun. Phys.}\ }\textbf {\bibinfo {volume} {2}},\ \bibinfo {pages} {163} (\bibinfo {year} {2019})}\BibitemShut {NoStop}%
\bibitem [{\citenamefont {Strand}\ \emph {et~al.}(2019)\citenamefont {Strand}, \citenamefont {Zingl}, \citenamefont {Wentzell}, \citenamefont {Parcollet},\ and\ \citenamefont {Georges}}]{PhysRevB.100.125120}%
  \BibitemOpen
  \bibfield  {author} {\bibinfo {author} {\bibfnamefont {Hugo U.~R.}\ \bibnamefont {Strand}}, \bibinfo {author} {\bibfnamefont {Manuel}\ \bibnamefont {Zingl}}, \bibinfo {author} {\bibfnamefont {Nils}\ \bibnamefont {Wentzell}}, \bibinfo {author} {\bibfnamefont {Olivier}\ \bibnamefont {Parcollet}}, \ and\ \bibinfo {author} {\bibfnamefont {Antoine}\ \bibnamefont {Georges}},\ }\bibfield  {title} {\enquote {\bibinfo {title} {{Magnetic response of ${\mathrm{Sr}}_{2}{\mathrm{RuO}}_{4}$: Quasi-local spin fluctuations due to Hund's coupling}},}\ }\href {\doibase 10.1103/PhysRevB.100.125120} {\bibfield  {journal} {\bibinfo  {journal} {Phys. Rev. B}\ }\textbf {\bibinfo {volume} {100}},\ \bibinfo {pages} {125120} (\bibinfo {year} {2019})}\BibitemShut {NoStop}%
\bibitem [{\citenamefont {Mackenzie}\ \emph {et~al.}(1996)\citenamefont {Mackenzie}, \citenamefont {Julian}, \citenamefont {Diver}, \citenamefont {McMullan}, \citenamefont {Ray}, \citenamefont {Lonzarich}, \citenamefont {Maeno}, \citenamefont {Nishizaki},\ and\ \citenamefont {Fujita}}]{PhysRevLett.76.3786}%
  \BibitemOpen
  \bibfield  {author} {\bibinfo {author} {\bibfnamefont {A.~P.}\ \bibnamefont {Mackenzie}}, \bibinfo {author} {\bibfnamefont {S.~R.}\ \bibnamefont {Julian}}, \bibinfo {author} {\bibfnamefont {A.~J.}\ \bibnamefont {Diver}}, \bibinfo {author} {\bibfnamefont {G.~J.}\ \bibnamefont {McMullan}}, \bibinfo {author} {\bibfnamefont {M.~P.}\ \bibnamefont {Ray}}, \bibinfo {author} {\bibfnamefont {G.~G.}\ \bibnamefont {Lonzarich}}, \bibinfo {author} {\bibfnamefont {Y.}~\bibnamefont {Maeno}}, \bibinfo {author} {\bibfnamefont {S.}~\bibnamefont {Nishizaki}}, \ and\ \bibinfo {author} {\bibfnamefont {T.}~\bibnamefont {Fujita}},\ }\bibfield  {title} {\enquote {\bibinfo {title} {{Quantum Oscillations in the Layered Perovskite Superconductor S${\mathrm{r}}_{2}$Ru${\mathrm{O}}_{4}$}},}\ }\href {\doibase 10.1103/PhysRevLett.76.3786} {\bibfield  {journal} {\bibinfo  {journal} {Phys. Rev. Lett.}\ }\textbf {\bibinfo {volume} {76}},\ \bibinfo {pages} {3786--3789} (\bibinfo {year} {1996})}\BibitemShut {NoStop}%
\bibitem [{\citenamefont {Bergemann}\ \emph {et~al.}(2000)\citenamefont {Bergemann}, \citenamefont {Julian}, \citenamefont {Mackenzie}, \citenamefont {NishiZaki},\ and\ \citenamefont {Maeno}}]{PhysRevLett.84.2662}%
  \BibitemOpen
  \bibfield  {author} {\bibinfo {author} {\bibfnamefont {C.}~\bibnamefont {Bergemann}}, \bibinfo {author} {\bibfnamefont {S.~R.}\ \bibnamefont {Julian}}, \bibinfo {author} {\bibfnamefont {A.~P.}\ \bibnamefont {Mackenzie}}, \bibinfo {author} {\bibfnamefont {S.}~\bibnamefont {NishiZaki}}, \ and\ \bibinfo {author} {\bibfnamefont {Y.}~\bibnamefont {Maeno}},\ }\bibfield  {title} {\enquote {\bibinfo {title} {{Detailed Topography of the Fermi Surface of ${\mathrm{Sr}}_{2}{\mathrm{RuO}}_{4}$}},}\ }\href {\doibase 10.1103/PhysRevLett.84.2662} {\bibfield  {journal} {\bibinfo  {journal} {Phys. Rev. Lett.}\ }\textbf {\bibinfo {volume} {84}},\ \bibinfo {pages} {2662--2665} (\bibinfo {year} {2000})}\BibitemShut {NoStop}%
\bibitem [{\citenamefont {Damascelli}\ \emph {et~al.}(2000)\citenamefont {Damascelli}, \citenamefont {Lu}, \citenamefont {Shen}, \citenamefont {Armitage}, \citenamefont {Ronning}, \citenamefont {Feng}, \citenamefont {Kim}, \citenamefont {Shen}, \citenamefont {Kimura}, \citenamefont {Tokura}, \citenamefont {Mao},\ and\ \citenamefont {Maeno}}]{PhysRevLett.85.5194}%
  \BibitemOpen
  \bibfield  {author} {\bibinfo {author} {\bibfnamefont {A.}~\bibnamefont {Damascelli}}, \bibinfo {author} {\bibfnamefont {D.~H.}\ \bibnamefont {Lu}}, \bibinfo {author} {\bibfnamefont {K.~M.}\ \bibnamefont {Shen}}, \bibinfo {author} {\bibfnamefont {N.~P.}\ \bibnamefont {Armitage}}, \bibinfo {author} {\bibfnamefont {F.}~\bibnamefont {Ronning}}, \bibinfo {author} {\bibfnamefont {D.~L.}\ \bibnamefont {Feng}}, \bibinfo {author} {\bibfnamefont {C.}~\bibnamefont {Kim}}, \bibinfo {author} {\bibfnamefont {Z.-X.}\ \bibnamefont {Shen}}, \bibinfo {author} {\bibfnamefont {T.}~\bibnamefont {Kimura}}, \bibinfo {author} {\bibfnamefont {Y.}~\bibnamefont {Tokura}}, \bibinfo {author} {\bibfnamefont {Z.~Q.}\ \bibnamefont {Mao}}, \ and\ \bibinfo {author} {\bibfnamefont {Y.}~\bibnamefont {Maeno}},\ }\bibfield  {title} {\enquote {\bibinfo {title} {{Fermi Surface, Surface States, and Surface Reconstruction in ${\mathrm{Sr}}_{2}{\mathrm{RuO}}_{4}$}},}\ }\href {\doibase 10.1103/PhysRevLett.85.5194} {\bibfield  {journal} {\bibinfo
  {journal} {Phys. Rev. Lett.}\ }\textbf {\bibinfo {volume} {85}},\ \bibinfo {pages} {5194--5197} (\bibinfo {year} {2000})}\BibitemShut {NoStop}%
\bibitem [{\citenamefont {Tamai}\ \emph {et~al.}(2019)\citenamefont {Tamai}, \citenamefont {Zingl}, \citenamefont {Rozbicki}, \citenamefont {Cappelli}, \citenamefont {Ricc\`o}, \citenamefont {de~la Torre}, \citenamefont {McKeown~Walker}, \citenamefont {Bruno}, \citenamefont {King}, \citenamefont {Meevasana}, \citenamefont {Shi}, \citenamefont {Radovi\ifmmode~\acute{c}\else \'{c}\fi{}}, \citenamefont {Plumb}, \citenamefont {Gibbs}, \citenamefont {Mackenzie}, \citenamefont {Berthod}, \citenamefont {Strand}, \citenamefont {Kim}, \citenamefont {Georges},\ and\ \citenamefont {Baumberger}}]{PhysRevX.9.021048}%
  \BibitemOpen
  \bibfield  {author} {\bibinfo {author} {\bibfnamefont {A.}~\bibnamefont {Tamai}}, \bibinfo {author} {\bibfnamefont {M.}~\bibnamefont {Zingl}}, \bibinfo {author} {\bibfnamefont {E.}~\bibnamefont {Rozbicki}}, \bibinfo {author} {\bibfnamefont {E.}~\bibnamefont {Cappelli}}, \bibinfo {author} {\bibfnamefont {S.}~\bibnamefont {Ricc\`o}}, \bibinfo {author} {\bibfnamefont {A.}~\bibnamefont {de~la Torre}}, \bibinfo {author} {\bibfnamefont {S.}~\bibnamefont {McKeown~Walker}}, \bibinfo {author} {\bibfnamefont {F.~Y.}\ \bibnamefont {Bruno}}, \bibinfo {author} {\bibfnamefont {P.~D.~C.}\ \bibnamefont {King}}, \bibinfo {author} {\bibfnamefont {W.}~\bibnamefont {Meevasana}}, \bibinfo {author} {\bibfnamefont {M.}~\bibnamefont {Shi}}, \bibinfo {author} {\bibfnamefont {M.}~\bibnamefont {Radovi\ifmmode~\acute{c}\else \'{c}\fi{}}}, \bibinfo {author} {\bibfnamefont {N.~C.}\ \bibnamefont {Plumb}}, \bibinfo {author} {\bibfnamefont {A.~S.}\ \bibnamefont {Gibbs}}, \bibinfo {author} {\bibfnamefont {A.~P.}\ \bibnamefont {Mackenzie}},
  \bibinfo {author} {\bibfnamefont {C.}~\bibnamefont {Berthod}}, \bibinfo {author} {\bibfnamefont {H.~U.~R.}\ \bibnamefont {Strand}}, \bibinfo {author} {\bibfnamefont {M.}~\bibnamefont {Kim}}, \bibinfo {author} {\bibfnamefont {A.}~\bibnamefont {Georges}}, \ and\ \bibinfo {author} {\bibfnamefont {F.}~\bibnamefont {Baumberger}},\ }\bibfield  {title} {\enquote {\bibinfo {title} {{High-Resolution Photoemission on ${\mathrm{Sr}}_{2}{\mathrm{RuO}}_{4}$ Reveals Correlation-Enhanced Effective Spin-Orbit Coupling and Dominantly Local Self-Energies}},}\ }\href {\doibase 10.1103/PhysRevX.9.021048} {\bibfield  {journal} {\bibinfo  {journal} {Phys. Rev. X}\ }\textbf {\bibinfo {volume} {9}},\ \bibinfo {pages} {021048} (\bibinfo {year} {2019})}\BibitemShut {NoStop}%
\bibitem [{\citenamefont {Ingle}\ \emph {et~al.}(2005)\citenamefont {Ingle}, \citenamefont {Shen}, \citenamefont {Baumberger}, \citenamefont {Meevasana}, \citenamefont {Lu}, \citenamefont {Shen}, \citenamefont {Damascelli}, \citenamefont {Nakatsuji}, \citenamefont {Mao}, \citenamefont {Maeno}, \citenamefont {Kimura},\ and\ \citenamefont {Tokura}}]{PhysRevB.72.205114}%
  \BibitemOpen
  \bibfield  {author} {\bibinfo {author} {\bibfnamefont {N.~J.~C.}\ \bibnamefont {Ingle}}, \bibinfo {author} {\bibfnamefont {K.~M.}\ \bibnamefont {Shen}}, \bibinfo {author} {\bibfnamefont {F.}~\bibnamefont {Baumberger}}, \bibinfo {author} {\bibfnamefont {W.}~\bibnamefont {Meevasana}}, \bibinfo {author} {\bibfnamefont {D.~H.}\ \bibnamefont {Lu}}, \bibinfo {author} {\bibfnamefont {Z.-X.}\ \bibnamefont {Shen}}, \bibinfo {author} {\bibfnamefont {A.}~\bibnamefont {Damascelli}}, \bibinfo {author} {\bibfnamefont {S.}~\bibnamefont {Nakatsuji}}, \bibinfo {author} {\bibfnamefont {Z.~Q.}\ \bibnamefont {Mao}}, \bibinfo {author} {\bibfnamefont {Y.}~\bibnamefont {Maeno}}, \bibinfo {author} {\bibfnamefont {T.}~\bibnamefont {Kimura}}, \ and\ \bibinfo {author} {\bibfnamefont {Y.}~\bibnamefont {Tokura}},\ }\bibfield  {title} {\enquote {\bibinfo {title} {{Quantitative analysis of ${\mathrm{Sr}}_{2}{\mathrm{RuO}}_{4}$ angle-resolved photoemission spectra: Many-body interactions in a model Fermi liquid}},}\ }\href {\doibase
  10.1103/PhysRevB.72.205114} {\bibfield  {journal} {\bibinfo  {journal} {Phys. Rev. B}\ }\textbf {\bibinfo {volume} {72}},\ \bibinfo {pages} {205114} (\bibinfo {year} {2005})}\BibitemShut {NoStop}%
\bibitem [{\citenamefont {Lu}\ \emph {et~al.}(1996)\citenamefont {Lu}, \citenamefont {Schmidt}, \citenamefont {Cummins}, \citenamefont {Schuppler}, \citenamefont {Lichtenberg},\ and\ \citenamefont {Bednorz}}]{PhysRevLett.76.4845}%
  \BibitemOpen
  \bibfield  {author} {\bibinfo {author} {\bibfnamefont {D.~H.}\ \bibnamefont {Lu}}, \bibinfo {author} {\bibfnamefont {M.}~\bibnamefont {Schmidt}}, \bibinfo {author} {\bibfnamefont {T.~R.}\ \bibnamefont {Cummins}}, \bibinfo {author} {\bibfnamefont {S.}~\bibnamefont {Schuppler}}, \bibinfo {author} {\bibfnamefont {F.}~\bibnamefont {Lichtenberg}}, \ and\ \bibinfo {author} {\bibfnamefont {J.~G.}\ \bibnamefont {Bednorz}},\ }\bibfield  {title} {\enquote {\bibinfo {title} {{Fermi Surface and Extended van Hove Singularity in the Noncuprate Superconductor S${\mathrm{r}}_{2}$Ru${\mathrm{O}}_{4}$}},}\ }\href {\doibase 10.1103/PhysRevLett.76.4845} {\bibfield  {journal} {\bibinfo  {journal} {Phys. Rev. Lett.}\ }\textbf {\bibinfo {volume} {76}},\ \bibinfo {pages} {4845--4848} (\bibinfo {year} {1996})}\BibitemShut {NoStop}%
\bibitem [{\citenamefont {Yokoya}\ \emph {et~al.}(1996)\citenamefont {Yokoya}, \citenamefont {Chainani}, \citenamefont {Takahashi}, \citenamefont {Katayama-Yoshida}, \citenamefont {Kasai},\ and\ \citenamefont {Tokura}}]{PhysRevLett.76.3009}%
  \BibitemOpen
  \bibfield  {author} {\bibinfo {author} {\bibfnamefont {T.}~\bibnamefont {Yokoya}}, \bibinfo {author} {\bibfnamefont {A.}~\bibnamefont {Chainani}}, \bibinfo {author} {\bibfnamefont {T.}~\bibnamefont {Takahashi}}, \bibinfo {author} {\bibfnamefont {H.}~\bibnamefont {Katayama-Yoshida}}, \bibinfo {author} {\bibfnamefont {M.}~\bibnamefont {Kasai}}, \ and\ \bibinfo {author} {\bibfnamefont {Y.}~\bibnamefont {Tokura}},\ }\bibfield  {title} {\enquote {\bibinfo {title} {{Extended Van Hove Singularity in a Noncuprate Layered Superconductor S${\mathrm{r}}_{2}$Ru${\mathrm{O}}_{4}$}},}\ }\href {\doibase 10.1103/PhysRevLett.76.3009} {\bibfield  {journal} {\bibinfo  {journal} {Phys. Rev. Lett.}\ }\textbf {\bibinfo {volume} {76}},\ \bibinfo {pages} {3009--3012} (\bibinfo {year} {1996})}\BibitemShut {NoStop}%
\bibitem [{\citenamefont {Mazin}\ and\ \citenamefont {Singh}(1997)}]{PhysRevLett.79.733}%
  \BibitemOpen
  \bibfield  {author} {\bibinfo {author} {\bibfnamefont {I.~I.}\ \bibnamefont {Mazin}}\ and\ \bibinfo {author} {\bibfnamefont {David~J.}\ \bibnamefont {Singh}},\ }\bibfield  {title} {\enquote {\bibinfo {title} {{Ferromagnetic Spin Fluctuation Induced Superconductivity in ${\mathrm{Sr}}_{2}{\mathrm{RuO}}_{4}$}},}\ }\href {\doibase 10.1103/PhysRevLett.79.733} {\bibfield  {journal} {\bibinfo  {journal} {Phys. Rev. Lett.}\ }\textbf {\bibinfo {volume} {79}},\ \bibinfo {pages} {733--736} (\bibinfo {year} {1997})}\BibitemShut {NoStop}%
\bibitem [{\citenamefont {Pavarini}\ and\ \citenamefont {Mazin}(2006)}]{PhysRevB.74.035115}%
  \BibitemOpen
  \bibfield  {author} {\bibinfo {author} {\bibfnamefont {E.}~\bibnamefont {Pavarini}}\ and\ \bibinfo {author} {\bibfnamefont {I.~I.}\ \bibnamefont {Mazin}},\ }\bibfield  {title} {\enquote {\bibinfo {title} {{First-principles study of spin-orbit effects and NMR in ${\mathrm{Sr}}_{2}\mathrm{Ru}{\mathrm{O}}_{4}$}},}\ }\href {\doibase 10.1103/PhysRevB.74.035115} {\bibfield  {journal} {\bibinfo  {journal} {Phys. Rev. B}\ }\textbf {\bibinfo {volume} {74}},\ \bibinfo {pages} {035115} (\bibinfo {year} {2006})}\BibitemShut {NoStop}%
\bibitem [{\citenamefont {Kohn}\ and\ \citenamefont {Sham}(1965)}]{PhysRev.140.A1133}%
  \BibitemOpen
  \bibfield  {author} {\bibinfo {author} {\bibfnamefont {W.}~\bibnamefont {Kohn}}\ and\ \bibinfo {author} {\bibfnamefont {L.~J.}\ \bibnamefont {Sham}},\ }\bibfield  {title} {\enquote {\bibinfo {title} {Self-consistent equations including exchange and correlation effects},}\ }\href {\doibase 10.1103/PhysRev.140.A1133} {\bibfield  {journal} {\bibinfo  {journal} {Phys. Rev.}\ }\textbf {\bibinfo {volume} {140}},\ \bibinfo {pages} {A1133--A1138} (\bibinfo {year} {1965})}\BibitemShut {NoStop}%
\bibitem [{\citenamefont {Mravlje}\ \emph {et~al.}(2011)\citenamefont {Mravlje}, \citenamefont {Aichhorn}, \citenamefont {Miyake}, \citenamefont {Haule}, \citenamefont {Kotliar},\ and\ \citenamefont {Georges}}]{PhysRevLett.106.096401}%
  \BibitemOpen
  \bibfield  {author} {\bibinfo {author} {\bibfnamefont {Jernej}\ \bibnamefont {Mravlje}}, \bibinfo {author} {\bibfnamefont {Markus}\ \bibnamefont {Aichhorn}}, \bibinfo {author} {\bibfnamefont {Takashi}\ \bibnamefont {Miyake}}, \bibinfo {author} {\bibfnamefont {Kristjan}\ \bibnamefont {Haule}}, \bibinfo {author} {\bibfnamefont {Gabriel}\ \bibnamefont {Kotliar}}, \ and\ \bibinfo {author} {\bibfnamefont {Antoine}\ \bibnamefont {Georges}},\ }\bibfield  {title} {\enquote {\bibinfo {title} {{Coherence-Incoherence Crossover and the Mass-Renormalization Puzzles in ${\mathrm{Sr}}_{2}{\mathrm{RuO}}_{4}$}},}\ }\href {\doibase 10.1103/PhysRevLett.106.096401} {\bibfield  {journal} {\bibinfo  {journal} {Phys. Rev. Lett.}\ }\textbf {\bibinfo {volume} {106}},\ \bibinfo {pages} {096401} (\bibinfo {year} {2011})}\BibitemShut {NoStop}%
\bibitem [{\citenamefont {Georges}\ \emph {et~al.}(2013)\citenamefont {Georges}, \citenamefont {Medici},\ and\ \citenamefont {Mravlje}}]{annurev:/content/journals/10.1146/annurev-conmatphys-020911-125045}%
  \BibitemOpen
  \bibfield  {author} {\bibinfo {author} {\bibfnamefont {Antoine}\ \bibnamefont {Georges}}, \bibinfo {author} {\bibfnamefont {Luca~de'}\ \bibnamefont {Medici}}, \ and\ \bibinfo {author} {\bibfnamefont {Jernej}\ \bibnamefont {Mravlje}},\ }\bibfield  {title} {\enquote {\bibinfo {title} {{Strong Correlations from Hund’s Coupling}},}\ }\href {\doibase https://doi.org/10.1146/annurev-conmatphys-020911-125045} {\bibfield  {journal} {\bibinfo  {journal} {Annu. Rev. Condens. Matter Phys.}\ }\textbf {\bibinfo {volume} {4}},\ \bibinfo {pages} {137--178} (\bibinfo {year} {2013})}\BibitemShut {NoStop}%
\bibitem [{\citenamefont {Kugler}\ \emph {et~al.}(2020)\citenamefont {Kugler}, \citenamefont {Zingl}, \citenamefont {Strand}, \citenamefont {Lee}, \citenamefont {von Delft},\ and\ \citenamefont {Georges}}]{PhysRevLett.124.016401}%
  \BibitemOpen
  \bibfield  {author} {\bibinfo {author} {\bibfnamefont {Fabian~B.}\ \bibnamefont {Kugler}}, \bibinfo {author} {\bibfnamefont {Manuel}\ \bibnamefont {Zingl}}, \bibinfo {author} {\bibfnamefont {Hugo U.~R.}\ \bibnamefont {Strand}}, \bibinfo {author} {\bibfnamefont {Seung-Sup~B.}\ \bibnamefont {Lee}}, \bibinfo {author} {\bibfnamefont {Jan}\ \bibnamefont {von Delft}}, \ and\ \bibinfo {author} {\bibfnamefont {Antoine}\ \bibnamefont {Georges}},\ }\bibfield  {title} {\enquote {\bibinfo {title} {{Strongly Correlated Materials from a Numerical Renormalization Group Perspective: How the Fermi-Liquid State of ${\mathrm{Sr}}_{2}{\mathrm{RuO}}_{4}$ Emerges}},}\ }\href {\doibase 10.1103/PhysRevLett.124.016401} {\bibfield  {journal} {\bibinfo  {journal} {Phys. Rev. Lett.}\ }\textbf {\bibinfo {volume} {124}},\ \bibinfo {pages} {016401} (\bibinfo {year} {2020})}\BibitemShut {NoStop}%
\bibitem [{\citenamefont {Hunter}\ \emph {et~al.}(2023)\citenamefont {Hunter}, \citenamefont {Beck}, \citenamefont {Cappelli}, \citenamefont {Margot}, \citenamefont {Straub}, \citenamefont {Alexanian}, \citenamefont {Gatti}, \citenamefont {Watson}, \citenamefont {Kim}, \citenamefont {Cacho}, \citenamefont {Plumb}, \citenamefont {Shi}, \citenamefont {Radovi\ifmmode~\acute{c}\else \'{c}\fi{}}, \citenamefont {Sokolov}, \citenamefont {Mackenzie}, \citenamefont {Zingl}, \citenamefont {Mravlje}, \citenamefont {Georges}, \citenamefont {Baumberger},\ and\ \citenamefont {Tamai}}]{PhysRevLett.131.236502}%
  \BibitemOpen
  \bibfield  {author} {\bibinfo {author} {\bibfnamefont {A.}~\bibnamefont {Hunter}}, \bibinfo {author} {\bibfnamefont {S.}~\bibnamefont {Beck}}, \bibinfo {author} {\bibfnamefont {E.}~\bibnamefont {Cappelli}}, \bibinfo {author} {\bibfnamefont {F.}~\bibnamefont {Margot}}, \bibinfo {author} {\bibfnamefont {M.}~\bibnamefont {Straub}}, \bibinfo {author} {\bibfnamefont {Y.}~\bibnamefont {Alexanian}}, \bibinfo {author} {\bibfnamefont {G.}~\bibnamefont {Gatti}}, \bibinfo {author} {\bibfnamefont {M.~D.}\ \bibnamefont {Watson}}, \bibinfo {author} {\bibfnamefont {T.~K.}\ \bibnamefont {Kim}}, \bibinfo {author} {\bibfnamefont {C.}~\bibnamefont {Cacho}}, \bibinfo {author} {\bibfnamefont {N.~C.}\ \bibnamefont {Plumb}}, \bibinfo {author} {\bibfnamefont {M.}~\bibnamefont {Shi}}, \bibinfo {author} {\bibfnamefont {M.}~\bibnamefont {Radovi\ifmmode~\acute{c}\else \'{c}\fi{}}}, \bibinfo {author} {\bibfnamefont {D.~A.}\ \bibnamefont {Sokolov}}, \bibinfo {author} {\bibfnamefont {A.~P.}\ \bibnamefont {Mackenzie}}, \bibinfo {author}
  {\bibfnamefont {M.}~\bibnamefont {Zingl}}, \bibinfo {author} {\bibfnamefont {J.}~\bibnamefont {Mravlje}}, \bibinfo {author} {\bibfnamefont {A.}~\bibnamefont {Georges}}, \bibinfo {author} {\bibfnamefont {F.}~\bibnamefont {Baumberger}}, \ and\ \bibinfo {author} {\bibfnamefont {A.}~\bibnamefont {Tamai}},\ }\bibfield  {title} {\enquote {\bibinfo {title} {{Fate of Quasiparticles at High Temperature in the Correlated Metal Sr$_{2}$RuO$_{4}$}},}\ }\href {\doibase 10.1103/PhysRevLett.131.236502} {\bibfield  {journal} {\bibinfo  {journal} {Phys. Rev. Lett.}\ }\textbf {\bibinfo {volume} {131}},\ \bibinfo {pages} {236502} (\bibinfo {year} {2023})}\BibitemShut {NoStop}%
\bibitem [{\citenamefont {{Blesio, Germ\'an and Beck, Sophie and Gingras, Olivier and Georges, Antoine and Mravlje, Jernej}}(2024)}]{PhysRevResearch.6.023124}%
  \BibitemOpen
  \bibfield  {author} {\bibinfo {author} {\bibnamefont {{Blesio, Germ\'an and Beck, Sophie and Gingras, Olivier and Georges, Antoine and Mravlje, Jernej}}},\ }\bibfield  {title} {\enquote {\bibinfo {title} {{Signatures of Hund metal and finite-frequency nesting in ${\mathrm{Sr}}_{2}{\mathrm{RuO}}_{4}$ revealed by electronic Raman scattering}},}\ }\href {\doibase 10.1103/PhysRevResearch.6.023124} {\bibfield  {journal} {\bibinfo  {journal} {Phys. Rev. Res.}\ }\textbf {\bibinfo {volume} {6}},\ \bibinfo {pages} {023124} (\bibinfo {year} {2024})}\BibitemShut {NoStop}%
\bibitem [{\citenamefont {Suzuki}\ \emph {et~al.}(2023)\citenamefont {Suzuki}, \citenamefont {Wang}, \citenamefont {Bertinshaw}, \citenamefont {Strand}, \citenamefont {K{\"a}ser}, \citenamefont {Krautloher}, \citenamefont {Yang}, \citenamefont {Wentzell}, \citenamefont {Parcollet}, \citenamefont {Jerzembeck}, \citenamefont {Kikugawa}, \citenamefont {Mackenzie}, \citenamefont {Georges}, \citenamefont {Hansmann}, \citenamefont {Gretarsson},\ and\ \citenamefont {Keimer}}]{suzuki2023distinct}%
  \BibitemOpen
  \bibfield  {author} {\bibinfo {author} {\bibfnamefont {H.}~\bibnamefont {Suzuki}}, \bibinfo {author} {\bibfnamefont {L.}~\bibnamefont {Wang}}, \bibinfo {author} {\bibfnamefont {J.}~\bibnamefont {Bertinshaw}}, \bibinfo {author} {\bibfnamefont {H.~U.~R.}\ \bibnamefont {Strand}}, \bibinfo {author} {\bibfnamefont {S.}~\bibnamefont {K{\"a}ser}}, \bibinfo {author} {\bibfnamefont {M.}~\bibnamefont {Krautloher}}, \bibinfo {author} {\bibfnamefont {Z.}~\bibnamefont {Yang}}, \bibinfo {author} {\bibfnamefont {N.}~\bibnamefont {Wentzell}}, \bibinfo {author} {\bibfnamefont {O.}~\bibnamefont {Parcollet}}, \bibinfo {author} {\bibfnamefont {F.}~\bibnamefont {Jerzembeck}}, \bibinfo {author} {\bibfnamefont {N.}~\bibnamefont {Kikugawa}}, \bibinfo {author} {\bibfnamefont {A.~P.}\ \bibnamefont {Mackenzie}}, \bibinfo {author} {\bibfnamefont {A.}~\bibnamefont {Georges}}, \bibinfo {author} {\bibfnamefont {P.}~\bibnamefont {Hansmann}}, \bibinfo {author} {\bibfnamefont {H.}~\bibnamefont {Gretarsson}}, \ and\ \bibinfo {author}
  {\bibfnamefont {B.}~\bibnamefont {Keimer}},\ }\bibfield  {title} {\enquote {\bibinfo {title} {{Distinct spin and orbital dynamics in Sr$_2$RuO$_4$}},}\ }\href {\doibase 10.1038/s41467-023-42804-3} {\bibfield  {journal} {\bibinfo  {journal} {Nat. Commun.}\ }\textbf {\bibinfo {volume} {14}},\ \bibinfo {pages} {7042} (\bibinfo {year} {2023})}\BibitemShut {NoStop}%
\bibitem [{\citenamefont {Georges}\ \emph {et~al.}(1996)\citenamefont {Georges}, \citenamefont {Kotliar}, \citenamefont {Krauth},\ and\ \citenamefont {Rozenberg}}]{RevModPhys.68.13}%
  \BibitemOpen
  \bibfield  {author} {\bibinfo {author} {\bibfnamefont {Antoine}\ \bibnamefont {Georges}}, \bibinfo {author} {\bibfnamefont {Gabriel}\ \bibnamefont {Kotliar}}, \bibinfo {author} {\bibfnamefont {Werner}\ \bibnamefont {Krauth}}, \ and\ \bibinfo {author} {\bibfnamefont {Marcelo~J.}\ \bibnamefont {Rozenberg}},\ }\bibfield  {title} {\enquote {\bibinfo {title} {{Dynamical mean-field theory of strongly correlated fermion systems and the limit of infinite dimensions}},}\ }\href {\doibase 10.1103/RevModPhys.68.13} {\bibfield  {journal} {\bibinfo  {journal} {Rev. Mod. Phys.}\ }\textbf {\bibinfo {volume} {68}},\ \bibinfo {pages} {13--125} (\bibinfo {year} {1996})}\BibitemShut {NoStop}%
\bibitem [{\citenamefont {Liebsch}\ and\ \citenamefont {Lichtenstein}(2000)}]{PhysRevLett.84.1591}%
  \BibitemOpen
  \bibfield  {author} {\bibinfo {author} {\bibfnamefont {A.}~\bibnamefont {Liebsch}}\ and\ \bibinfo {author} {\bibfnamefont {A.}~\bibnamefont {Lichtenstein}},\ }\bibfield  {title} {\enquote {\bibinfo {title} {{Photoemission Quasiparticle Spectra of Sr$_{2}$RuO$_{4}$}},}\ }\href {\doibase 10.1103/PhysRevLett.84.1591} {\bibfield  {journal} {\bibinfo  {journal} {Phys. Rev. Lett.}\ }\textbf {\bibinfo {volume} {84}},\ \bibinfo {pages} {1591--1594} (\bibinfo {year} {2000})}\BibitemShut {NoStop}%
\bibitem [{\citenamefont {Zhang}\ \emph {et~al.}(2016)\citenamefont {Zhang}, \citenamefont {Gorelov}, \citenamefont {Sarvestani},\ and\ \citenamefont {Pavarini}}]{PhysRevLett.116.106402}%
  \BibitemOpen
  \bibfield  {author} {\bibinfo {author} {\bibfnamefont {Guoren}\ \bibnamefont {Zhang}}, \bibinfo {author} {\bibfnamefont {Evgeny}\ \bibnamefont {Gorelov}}, \bibinfo {author} {\bibfnamefont {Esmaeel}\ \bibnamefont {Sarvestani}}, \ and\ \bibinfo {author} {\bibfnamefont {Eva}\ \bibnamefont {Pavarini}},\ }\bibfield  {title} {\enquote {\bibinfo {title} {{Fermi Surface of ${\mathrm{Sr}}_{2}{\mathrm{RuO}}_{4}$: Spin-Orbit and Anisotropic Coulomb Interaction Effects}},}\ }\href {\doibase 10.1103/PhysRevLett.116.106402} {\bibfield  {journal} {\bibinfo  {journal} {Phys. Rev. Lett.}\ }\textbf {\bibinfo {volume} {116}},\ \bibinfo {pages} {106402} (\bibinfo {year} {2016})}\BibitemShut {NoStop}%
\bibitem [{\citenamefont {Eremin}\ \emph {et~al.}(2002)\citenamefont {Eremin}, \citenamefont {Manske},\ and\ \citenamefont {Bennemann}}]{PhysRevB.65.220502}%
  \BibitemOpen
  \bibfield  {author} {\bibinfo {author} {\bibfnamefont {I.}~\bibnamefont {Eremin}}, \bibinfo {author} {\bibfnamefont {D.}~\bibnamefont {Manske}}, \ and\ \bibinfo {author} {\bibfnamefont {K.~H.}\ \bibnamefont {Bennemann}},\ }\bibfield  {title} {\enquote {\bibinfo {title} {{Electronic theory for the normal-state spin dynamics in ${\mathrm{Sr}}_{2}{\mathrm{RuO}}_{4}:$ Anisotropy due to spin-orbit coupling}},}\ }\href {\doibase 10.1103/PhysRevB.65.220502} {\bibfield  {journal} {\bibinfo  {journal} {Phys. Rev. B}\ }\textbf {\bibinfo {volume} {65}},\ \bibinfo {pages} {220502} (\bibinfo {year} {2002})}\BibitemShut {NoStop}%
\bibitem [{SM()}]{SM}%
  \BibitemOpen
  \bibinfo {note} {See Supplemental Material for an analysis of the effect of Hund's exchange coupling and of temperature on the system.}\BibitemShut {Stop}%
\bibitem [{Note1()}]{Note1}%
  \BibitemOpen
  \bibinfo {note} {The DFT calculations were performed within the projected augmented wave formalism~\cite {PhysRevB.50.17953, PhysRevB.59.1758} as implemented in the Vienna ab initio simulation package (VASP)~\cite {KRESSE199615, PhysRevB.54.11169}. The exchange-correlation effects were considered within the generalized-gradient approximation functional in the PBE parametrization~\cite {PhysRevLett.77.3865}. We used the standard pseudopotentials that include 10, 16, and 6 valence electrons for Sr, Ru, and O, respectively. An energy cut-off of 400\protect \,eV for the plane-waves and a convergence threshold of ${10^{-7}}$\protect \,eV were used in the calculations. The Wannier functions and the effective tight-binding Hamiltonian were constructed within the scheme of maximal localization~\cite {PhysRevB.56.12847, RevModPhys.84.1419} using the wannier90 package~\cite {MOSTOFI2008685}.}\BibitemShut {Stop}%
\bibitem [{\citenamefont {Kanamori}(1963)}]{10.1143/PTP.30.275}%
  \BibitemOpen
  \bibfield  {author} {\bibinfo {author} {\bibfnamefont {Junjiro}\ \bibnamefont {Kanamori}},\ }\bibfield  {title} {\enquote {\bibinfo {title} {{Electron Correlation and Ferromagnetism of Transition Metals}},}\ }\href {\doibase 10.1143/PTP.30.275} {\bibfield  {journal} {\bibinfo  {journal} {Prog. Theor. Phys.}\ }\textbf {\bibinfo {volume} {30}},\ \bibinfo {pages} {275--289} (\bibinfo {year} {1963})}\BibitemShut {NoStop}%
\bibitem [{\citenamefont {Vaugier}\ \emph {et~al.}(2012)\citenamefont {Vaugier}, \citenamefont {Jiang},\ and\ \citenamefont {Biermann}}]{PhysRevB.86.165105}%
  \BibitemOpen
  \bibfield  {author} {\bibinfo {author} {\bibfnamefont {Lo\"{\i}g}\ \bibnamefont {Vaugier}}, \bibinfo {author} {\bibfnamefont {Hong}\ \bibnamefont {Jiang}}, \ and\ \bibinfo {author} {\bibfnamefont {Silke}\ \bibnamefont {Biermann}},\ }\bibfield  {title} {\enquote {\bibinfo {title} {{Hubbard $U$ and Hund exchange $J$ in transition metal oxides: Screening versus localization trends from constrained random phase approximation}},}\ }\href {\doibase 10.1103/PhysRevB.86.165105} {\bibfield  {journal} {\bibinfo  {journal} {Phys. Rev. B}\ }\textbf {\bibinfo {volume} {86}},\ \bibinfo {pages} {165105} (\bibinfo {year} {2012})}\BibitemShut {NoStop}%
\bibitem [{\citenamefont {Stepanov}\ \emph {et~al.}(2019)\citenamefont {Stepanov}, \citenamefont {Harkov},\ and\ \citenamefont {Lichtenstein}}]{PhysRevB.100.205115}%
  \BibitemOpen
  \bibfield  {author} {\bibinfo {author} {\bibfnamefont {E.~A.}\ \bibnamefont {Stepanov}}, \bibinfo {author} {\bibfnamefont {V.}~\bibnamefont {Harkov}}, \ and\ \bibinfo {author} {\bibfnamefont {A.~I.}\ \bibnamefont {Lichtenstein}},\ }\bibfield  {title} {\enquote {\bibinfo {title} {{Consistent partial bosonization of the extended Hubbard model}},}\ }\href {\doibase 10.1103/PhysRevB.100.205115} {\bibfield  {journal} {\bibinfo  {journal} {Phys. Rev. B}\ }\textbf {\bibinfo {volume} {100}},\ \bibinfo {pages} {205115} (\bibinfo {year} {2019})}\BibitemShut {NoStop}%
\bibitem [{\citenamefont {Harkov}\ \emph {et~al.}(2021)\citenamefont {Harkov}, \citenamefont {Vandelli}, \citenamefont {Brener}, \citenamefont {Lichtenstein},\ and\ \citenamefont {Stepanov}}]{PhysRevB.103.245123}%
  \BibitemOpen
  \bibfield  {author} {\bibinfo {author} {\bibfnamefont {V.}~\bibnamefont {Harkov}}, \bibinfo {author} {\bibfnamefont {M.}~\bibnamefont {Vandelli}}, \bibinfo {author} {\bibfnamefont {S.}~\bibnamefont {Brener}}, \bibinfo {author} {\bibfnamefont {A.~I.}\ \bibnamefont {Lichtenstein}}, \ and\ \bibinfo {author} {\bibfnamefont {E.~A.}\ \bibnamefont {Stepanov}},\ }\bibfield  {title} {\enquote {\bibinfo {title} {{Impact of partially bosonized collective fluctuations on electronic degrees of freedom}},}\ }\href {\doibase 10.1103/PhysRevB.103.245123} {\bibfield  {journal} {\bibinfo  {journal} {Phys. Rev. B}\ }\textbf {\bibinfo {volume} {103}},\ \bibinfo {pages} {245123} (\bibinfo {year} {2021})}\BibitemShut {NoStop}%
\bibitem [{\citenamefont {Vandelli}\ \emph {et~al.}(2022)\citenamefont {Vandelli}, \citenamefont {Kaufmann}, \citenamefont {El-Nabulsi}, \citenamefont {Harkov}, \citenamefont {Lichtenstein},\ and\ \citenamefont {Stepanov}}]{10.21468/SciPostPhys.13.2.036}%
  \BibitemOpen
  \bibfield  {author} {\bibinfo {author} {\bibfnamefont {Matteo}\ \bibnamefont {Vandelli}}, \bibinfo {author} {\bibfnamefont {Josef}\ \bibnamefont {Kaufmann}}, \bibinfo {author} {\bibfnamefont {Mohammed}\ \bibnamefont {El-Nabulsi}}, \bibinfo {author} {\bibfnamefont {Viktor}\ \bibnamefont {Harkov}}, \bibinfo {author} {\bibfnamefont {Alexander~I.}\ \bibnamefont {Lichtenstein}}, \ and\ \bibinfo {author} {\bibfnamefont {Evgeny~A.}\ \bibnamefont {Stepanov}},\ }\bibfield  {title} {\enquote {\bibinfo {title} {{Multi-band D-TRILEX approach to materials with strong electronic correlations}},}\ }\href {\doibase 10.21468/SciPostPhys.13.2.036} {\bibfield  {journal} {\bibinfo  {journal} {SciPost Phys.}\ }\textbf {\bibinfo {volume} {13}},\ \bibinfo {pages} {036} (\bibinfo {year} {2022})}\BibitemShut {NoStop}%
\bibitem [{\citenamefont {Rohringer}\ \emph {et~al.}(2018)\citenamefont {Rohringer}, \citenamefont {Hafermann}, \citenamefont {Toschi}, \citenamefont {Katanin}, \citenamefont {Antipov}, \citenamefont {Katsnelson}, \citenamefont {Lichtenstein}, \citenamefont {Rubtsov},\ and\ \citenamefont {Held}}]{RevModPhys.90.025003}%
  \BibitemOpen
  \bibfield  {author} {\bibinfo {author} {\bibfnamefont {G.}~\bibnamefont {Rohringer}}, \bibinfo {author} {\bibfnamefont {H.}~\bibnamefont {Hafermann}}, \bibinfo {author} {\bibfnamefont {A.}~\bibnamefont {Toschi}}, \bibinfo {author} {\bibfnamefont {A.~A.}\ \bibnamefont {Katanin}}, \bibinfo {author} {\bibfnamefont {A.~E.}\ \bibnamefont {Antipov}}, \bibinfo {author} {\bibfnamefont {M.~I.}\ \bibnamefont {Katsnelson}}, \bibinfo {author} {\bibfnamefont {A.~I.}\ \bibnamefont {Lichtenstein}}, \bibinfo {author} {\bibfnamefont {A.~N.}\ \bibnamefont {Rubtsov}}, \ and\ \bibinfo {author} {\bibfnamefont {K.}~\bibnamefont {Held}},\ }\bibfield  {title} {\enquote {\bibinfo {title} {{Diagrammatic routes to nonlocal correlations beyond dynamical mean field theory}},}\ }\href {\doibase 10.1103/RevModPhys.90.025003} {\bibfield  {journal} {\bibinfo  {journal} {Rev. Mod. Phys.}\ }\textbf {\bibinfo {volume} {90}},\ \bibinfo {pages} {025003} (\bibinfo {year} {2018})}\BibitemShut {NoStop}%
\bibitem [{\citenamefont {Lyakhova}\ \emph {et~al.}(2023)\citenamefont {Lyakhova}, \citenamefont {Astretsov},\ and\ \citenamefont {Rubtsov}}]{Lyakhova_review}%
  \BibitemOpen
  \bibfield  {author} {\bibinfo {author} {\bibfnamefont {Ya.~S.}\ \bibnamefont {Lyakhova}}, \bibinfo {author} {\bibfnamefont {G.~V.}\ \bibnamefont {Astretsov}}, \ and\ \bibinfo {author} {\bibfnamefont {A.~N.}\ \bibnamefont {Rubtsov}},\ }\bibfield  {title} {\enquote {\bibinfo {title} {{The mean-field concept and post-DMFT methods in the contemporary theory of correlated systems}},}\ }\href {\doibase 10.3367/UFNe.2022.09.039231} {\bibfield  {journal} {\bibinfo  {journal} {Phys. Usp.}\ }\textbf {\bibinfo {volume} {66}},\ \bibinfo {pages} {775–793} (\bibinfo {year} {2023})}\BibitemShut {NoStop}%
\bibitem [{\citenamefont {Stepanov}\ \emph {et~al.}(2021)\citenamefont {Stepanov}, \citenamefont {Nomura}, \citenamefont {Lichtenstein},\ and\ \citenamefont {Biermann}}]{PhysRevLett.127.207205}%
  \BibitemOpen
  \bibfield  {author} {\bibinfo {author} {\bibfnamefont {Evgeny~A.}\ \bibnamefont {Stepanov}}, \bibinfo {author} {\bibfnamefont {Yusuke}\ \bibnamefont {Nomura}}, \bibinfo {author} {\bibfnamefont {Alexander~I.}\ \bibnamefont {Lichtenstein}}, \ and\ \bibinfo {author} {\bibfnamefont {Silke}\ \bibnamefont {Biermann}},\ }\bibfield  {title} {\enquote {\bibinfo {title} {{Orbital Isotropy of Magnetic Fluctuations in Correlated Electron Materials Induced by Hund's Exchange Coupling}},}\ }\href {\doibase 10.1103/PhysRevLett.127.207205} {\bibfield  {journal} {\bibinfo  {journal} {Phys. Rev. Lett.}\ }\textbf {\bibinfo {volume} {127}},\ \bibinfo {pages} {207205} (\bibinfo {year} {2021})}\BibitemShut {NoStop}%
\bibitem [{\citenamefont {Vandelli}\ \emph {et~al.}(2023)\citenamefont {Vandelli}, \citenamefont {Kaufmann}, \citenamefont {Harkov}, \citenamefont {Lichtenstein}, \citenamefont {Held},\ and\ \citenamefont {Stepanov}}]{PhysRevResearch.5.L022016}%
  \BibitemOpen
  \bibfield  {author} {\bibinfo {author} {\bibfnamefont {M.}~\bibnamefont {Vandelli}}, \bibinfo {author} {\bibfnamefont {J.}~\bibnamefont {Kaufmann}}, \bibinfo {author} {\bibfnamefont {V.}~\bibnamefont {Harkov}}, \bibinfo {author} {\bibfnamefont {A.~I.}\ \bibnamefont {Lichtenstein}}, \bibinfo {author} {\bibfnamefont {K.}~\bibnamefont {Held}}, \ and\ \bibinfo {author} {\bibfnamefont {E.~A.}\ \bibnamefont {Stepanov}},\ }\bibfield  {title} {\enquote {\bibinfo {title} {{Extended regime of metastable metallic and insulating phases in a two-orbital electronic system}},}\ }\href {\doibase 10.1103/PhysRevResearch.5.L022016} {\bibfield  {journal} {\bibinfo  {journal} {Phys. Rev. Res.}\ }\textbf {\bibinfo {volume} {5}},\ \bibinfo {pages} {L022016} (\bibinfo {year} {2023})}\BibitemShut {NoStop}%
\bibitem [{\citenamefont {Stepanov}(2022)}]{PhysRevLett.129.096404}%
  \BibitemOpen
  \bibfield  {author} {\bibinfo {author} {\bibfnamefont {Evgeny~A.}\ \bibnamefont {Stepanov}},\ }\bibfield  {title} {\enquote {\bibinfo {title} {{Eliminating Orbital Selectivity from the Metal-Insulator Transition by Strong Magnetic Fluctuations}},}\ }\href {\doibase 10.1103/PhysRevLett.129.096404} {\bibfield  {journal} {\bibinfo  {journal} {Phys. Rev. Lett.}\ }\textbf {\bibinfo {volume} {129}},\ \bibinfo {pages} {096404} (\bibinfo {year} {2022})}\BibitemShut {NoStop}%
\bibitem [{\citenamefont {Stepanov}\ and\ \citenamefont {Biermann}(2024)}]{PhysRevLett.132.226501}%
  \BibitemOpen
  \bibfield  {author} {\bibinfo {author} {\bibfnamefont {Evgeny~A.}\ \bibnamefont {Stepanov}}\ and\ \bibinfo {author} {\bibfnamefont {Silke}\ \bibnamefont {Biermann}},\ }\bibfield  {title} {\enquote {\bibinfo {title} {{Can Orbital-Selective N\'eel Transitions Survive Strong Nonlocal Electronic Correlations?}}}\ }\href {\doibase 10.1103/PhysRevLett.132.226501} {\bibfield  {journal} {\bibinfo  {journal} {Phys. Rev. Lett.}\ }\textbf {\bibinfo {volume} {132}},\ \bibinfo {pages} {226501} (\bibinfo {year} {2024})}\BibitemShut {NoStop}%
\bibitem [{\citenamefont {Stepanov}\ \emph {et~al.}(2024{\natexlab{a}})\citenamefont {Stepanov}, \citenamefont {Vandelli}, \citenamefont {Lichtenstein},\ and\ \citenamefont {Lechermann}}]{stepanov2023charge}%
  \BibitemOpen
  \bibfield  {author} {\bibinfo {author} {\bibfnamefont {Evgeny~A.}\ \bibnamefont {Stepanov}}, \bibinfo {author} {\bibfnamefont {Matteo}\ \bibnamefont {Vandelli}}, \bibinfo {author} {\bibfnamefont {Alexander~I.}\ \bibnamefont {Lichtenstein}}, \ and\ \bibinfo {author} {\bibfnamefont {Frank}\ \bibnamefont {Lechermann}},\ }\bibfield  {title} {\enquote {\bibinfo {title} {{Charge Density Wave Ordering in NdNiO$_2$: Effects of Multiorbital Nonlocal Correlations}},}\ }\href {\doibase 10.1038/s41524-024-01298-3} {\bibfield  {journal} {\bibinfo  {journal} {npj Comput. Mater.}\ }\textbf {\bibinfo {volume} {10}},\ \bibinfo {pages} {108} (\bibinfo {year} {2024}{\natexlab{a}})}\BibitemShut {NoStop}%
\bibitem [{\citenamefont {Wallerberger}\ \emph {et~al.}(2019)\citenamefont {Wallerberger}, \citenamefont {Hausoel}, \citenamefont {Gunacker}, \citenamefont {Kowalski}, \citenamefont {Parragh}, \citenamefont {Goth}, \citenamefont {Held},\ and\ \citenamefont {Sangiovanni}}]{WALLERBERGER2019388}%
  \BibitemOpen
  \bibfield  {author} {\bibinfo {author} {\bibfnamefont {Markus}\ \bibnamefont {Wallerberger}}, \bibinfo {author} {\bibfnamefont {Andreas}\ \bibnamefont {Hausoel}}, \bibinfo {author} {\bibfnamefont {Patrik}\ \bibnamefont {Gunacker}}, \bibinfo {author} {\bibfnamefont {Alexander}\ \bibnamefont {Kowalski}}, \bibinfo {author} {\bibfnamefont {Nicolaus}\ \bibnamefont {Parragh}}, \bibinfo {author} {\bibfnamefont {Florian}\ \bibnamefont {Goth}}, \bibinfo {author} {\bibfnamefont {Karsten}\ \bibnamefont {Held}}, \ and\ \bibinfo {author} {\bibfnamefont {Giorgio}\ \bibnamefont {Sangiovanni}},\ }\bibfield  {title} {\enquote {\bibinfo {title} {{w2dynamics: Local one- and two-particle quantities from dynamical mean field theory}},}\ }\href {\doibase https://doi.org/10.1016/j.cpc.2018.09.007} {\bibfield  {journal} {\bibinfo  {journal} {Comput. Phys. Commun.}\ }\textbf {\bibinfo {volume} {235}},\ \bibinfo {pages} {388--399} (\bibinfo {year} {2019})}\BibitemShut {NoStop}%
\bibitem [{\citenamefont {Stepanov}\ \emph {et~al.}(2022)\citenamefont {Stepanov}, \citenamefont {Harkov}, \citenamefont {R\"osner}, \citenamefont {Lichtenstein}, \citenamefont {Katsnelson},\ and\ \citenamefont {Rudenko}}]{stepanov2021coexisting}%
  \BibitemOpen
  \bibfield  {author} {\bibinfo {author} {\bibfnamefont {E.~A.}\ \bibnamefont {Stepanov}}, \bibinfo {author} {\bibfnamefont {V.}~\bibnamefont {Harkov}}, \bibinfo {author} {\bibfnamefont {M.}~\bibnamefont {R\"osner}}, \bibinfo {author} {\bibfnamefont {A.~I.}\ \bibnamefont {Lichtenstein}}, \bibinfo {author} {\bibfnamefont {M.~I.}\ \bibnamefont {Katsnelson}}, \ and\ \bibinfo {author} {\bibfnamefont {A.~N.}\ \bibnamefont {Rudenko}},\ }\bibfield  {title} {\enquote {\bibinfo {title} {{Coexisting charge density wave and ferromagnetic instabilities in monolayer InSe}},}\ }\href {\doibase 10.1038/s41524-022-00798-4} {\bibfield  {journal} {\bibinfo  {journal} {npj Comput. Mater.}\ }\textbf {\bibinfo {volume} {8}},\ \bibinfo {pages} {118} (\bibinfo {year} {2022})}\BibitemShut {NoStop}%
\bibitem [{\citenamefont {Vandelli}\ \emph {et~al.}(2024)\citenamefont {Vandelli}, \citenamefont {Galler}, \citenamefont {Rubio}, \citenamefont {Lichtenstein}, \citenamefont {Biermann},\ and\ \citenamefont {Stepanov}}]{vandelli2024doping}%
  \BibitemOpen
  \bibfield  {author} {\bibinfo {author} {\bibfnamefont {M.}~\bibnamefont {Vandelli}}, \bibinfo {author} {\bibfnamefont {A.}~\bibnamefont {Galler}}, \bibinfo {author} {\bibfnamefont {A.}~\bibnamefont {Rubio}}, \bibinfo {author} {\bibfnamefont {A.~I.}\ \bibnamefont {Lichtenstein}}, \bibinfo {author} {\bibfnamefont {S.}~\bibnamefont {Biermann}}, \ and\ \bibinfo {author} {\bibfnamefont {E.~A.}\ \bibnamefont {Stepanov}},\ }\bibfield  {title} {\enquote {\bibinfo {title} {{Doping-dependent charge- and spin-density wave orderings in a monolayer of Pb adatoms on Si(111)}},}\ }\href {\doibase 10.1038/s41535-024-00630-w} {\bibfield  {journal} {\bibinfo  {journal} {npj Quantum Mater.}\ }\textbf {\bibinfo {volume} {9}},\ \bibinfo {pages} {19} (\bibinfo {year} {2024})}\BibitemShut {NoStop}%
\bibitem [{\citenamefont {Chatzieleftheriou}\ \emph {et~al.}(2024)\citenamefont {Chatzieleftheriou}, \citenamefont {Biermann},\ and\ \citenamefont {Stepanov}}]{PhysRevLett.132.236504}%
  \BibitemOpen
  \bibfield  {author} {\bibinfo {author} {\bibfnamefont {Maria}\ \bibnamefont {Chatzieleftheriou}}, \bibinfo {author} {\bibfnamefont {Silke}\ \bibnamefont {Biermann}}, \ and\ \bibinfo {author} {\bibfnamefont {Evgeny~A.}\ \bibnamefont {Stepanov}},\ }\bibfield  {title} {\enquote {\bibinfo {title} {{Local and Nonlocal Electronic Correlations at the Metal-Insulator Transition in the Two-Dimensional Hubbard Model}},}\ }\href {\doibase 10.1103/PhysRevLett.132.236504} {\bibfield  {journal} {\bibinfo  {journal} {Phys. Rev. Lett.}\ }\textbf {\bibinfo {volume} {132}},\ \bibinfo {pages} {236504} (\bibinfo {year} {2024})}\BibitemShut {NoStop}%
\bibitem [{\citenamefont {Stepanov}\ \emph {et~al.}(2024{\natexlab{b}})\citenamefont {Stepanov}, \citenamefont {Chatzieleftheriou}, \citenamefont {Wagner},\ and\ \citenamefont {Sangiovanni}}]{PhysRevB.110.L161106}%
  \BibitemOpen
  \bibfield  {author} {\bibinfo {author} {\bibfnamefont {Evgeny~A.}\ \bibnamefont {Stepanov}}, \bibinfo {author} {\bibfnamefont {Maria}\ \bibnamefont {Chatzieleftheriou}}, \bibinfo {author} {\bibfnamefont {Niklas}\ \bibnamefont {Wagner}}, \ and\ \bibinfo {author} {\bibfnamefont {Giorgio}\ \bibnamefont {Sangiovanni}},\ }\bibfield  {title} {\enquote {\bibinfo {title} {{Interconnected renormalization of Hubbard bands and Green's function zeros in Mott insulators induced by strong magnetic fluctuations}},}\ }\href {\doibase 10.1103/PhysRevB.110.L161106} {\bibfield  {journal} {\bibinfo  {journal} {Phys. Rev. B}\ }\textbf {\bibinfo {volume} {110}},\ \bibinfo {pages} {L161106} (\bibinfo {year} {2024}{\natexlab{b}})}\BibitemShut {NoStop}%
\bibitem [{\citenamefont {Vandelli}(2022)}]{vandelli2022quantum}%
  \BibitemOpen
  \bibfield  {author} {\bibinfo {author} {\bibfnamefont {Matteo}\ \bibnamefont {Vandelli}},\ }\emph {\bibinfo {title} {Quantum embedding methods in dual space for strongly interacting electronic systems}},\ \href@noop {} {Ph.D. thesis},\ \bibinfo  {school} {Universit{\"a}t Hamburg Hamburg} (\bibinfo {year} {2022})\BibitemShut {NoStop}%
\bibitem [{Note2()}]{Note2}%
  \BibitemOpen
  \bibinfo {note} {\label {note1}To be precise, our DMFT susceptibility is obtained through a single-shot \protect \mbox {D-TRILEX} calculation, which, in practice, gives a very similar result to a direct evaluation of the local vertex corrections, as done in Ref.~\cite {PhysRevB.100.125120}. In that work the transition temperature is found to be lower at $T\simeq 123K$, but this is related to the different interaction parameters used in the current study.}\BibitemShut {Stop}%
\bibitem [{\citenamefont {Friedt}\ \emph {et~al.}(2004)\citenamefont {Friedt}, \citenamefont {Steffens}, \citenamefont {Braden}, \citenamefont {Sidis}, \citenamefont {Nakatsuji},\ and\ \citenamefont {Maeno}}]{PhysRevLett.93.147404}%
  \BibitemOpen
  \bibfield  {author} {\bibinfo {author} {\bibfnamefont {O.}~\bibnamefont {Friedt}}, \bibinfo {author} {\bibfnamefont {P.}~\bibnamefont {Steffens}}, \bibinfo {author} {\bibfnamefont {M.}~\bibnamefont {Braden}}, \bibinfo {author} {\bibfnamefont {Y.}~\bibnamefont {Sidis}}, \bibinfo {author} {\bibfnamefont {S.}~\bibnamefont {Nakatsuji}}, \ and\ \bibinfo {author} {\bibfnamefont {Y.}~\bibnamefont {Maeno}},\ }\bibfield  {title} {\enquote {\bibinfo {title} {{Strongly Enhanced Magnetic Fluctuations in a Large-Mass Layered Ruthenate}},}\ }\href {\doibase 10.1103/PhysRevLett.93.147404} {\bibfield  {journal} {\bibinfo  {journal} {Phys. Rev. Lett.}\ }\textbf {\bibinfo {volume} {93}},\ \bibinfo {pages} {147404} (\bibinfo {year} {2004})}\BibitemShut {NoStop}%
\bibitem [{\citenamefont {Steffens}\ \emph {et~al.}(2007)\citenamefont {Steffens}, \citenamefont {Sidis}, \citenamefont {Link}, \citenamefont {Schmalzl}, \citenamefont {Nakatsuji}, \citenamefont {Maeno},\ and\ \citenamefont {Braden}}]{PhysRevLett.99.217402}%
  \BibitemOpen
  \bibfield  {author} {\bibinfo {author} {\bibfnamefont {P.}~\bibnamefont {Steffens}}, \bibinfo {author} {\bibfnamefont {Y.}~\bibnamefont {Sidis}}, \bibinfo {author} {\bibfnamefont {P.}~\bibnamefont {Link}}, \bibinfo {author} {\bibfnamefont {K.}~\bibnamefont {Schmalzl}}, \bibinfo {author} {\bibfnamefont {S.}~\bibnamefont {Nakatsuji}}, \bibinfo {author} {\bibfnamefont {Y.}~\bibnamefont {Maeno}}, \ and\ \bibinfo {author} {\bibfnamefont {M.}~\bibnamefont {Braden}},\ }\bibfield  {title} {\enquote {\bibinfo {title} {{Field-Induced Paramagnons at the Metamagnetic Transition of ${\mathrm{Ca}}_{1.8}{\mathrm{Sr}}_{0.2}{\mathrm{RuO}}_{4}$}},}\ }\href {\doibase 10.1103/PhysRevLett.99.217402} {\bibfield  {journal} {\bibinfo  {journal} {Phys. Rev. Lett.}\ }\textbf {\bibinfo {volume} {99}},\ \bibinfo {pages} {217402} (\bibinfo {year} {2007})}\BibitemShut {NoStop}%
\bibitem [{\citenamefont {Steffens}\ \emph {et~al.}(2011)\citenamefont {Steffens}, \citenamefont {Friedt}, \citenamefont {Sidis}, \citenamefont {Link}, \citenamefont {Kulda}, \citenamefont {Schmalzl}, \citenamefont {Nakatsuji},\ and\ \citenamefont {Braden}}]{PhysRevB.83.054429}%
  \BibitemOpen
  \bibfield  {author} {\bibinfo {author} {\bibfnamefont {P.}~\bibnamefont {Steffens}}, \bibinfo {author} {\bibfnamefont {O.}~\bibnamefont {Friedt}}, \bibinfo {author} {\bibfnamefont {Y.}~\bibnamefont {Sidis}}, \bibinfo {author} {\bibfnamefont {P.}~\bibnamefont {Link}}, \bibinfo {author} {\bibfnamefont {J.}~\bibnamefont {Kulda}}, \bibinfo {author} {\bibfnamefont {K.}~\bibnamefont {Schmalzl}}, \bibinfo {author} {\bibfnamefont {S.}~\bibnamefont {Nakatsuji}}, \ and\ \bibinfo {author} {\bibfnamefont {M.}~\bibnamefont {Braden}},\ }\bibfield  {title} {\enquote {\bibinfo {title} {{Magnetic excitations in the metallic single-layer ruthenates Ca${}_{2\ensuremath{-}x}$Sr${}_{x}$RuO${}_{4}$ studied by inelastic neutron scattering}},}\ }\href {\doibase 10.1103/PhysRevB.83.054429} {\bibfield  {journal} {\bibinfo  {journal} {Phys. Rev. B}\ }\textbf {\bibinfo {volume} {83}},\ \bibinfo {pages} {054429} (\bibinfo {year} {2011})}\BibitemShut {NoStop}%
\bibitem [{\citenamefont {Friedt}\ \emph {et~al.}(2001)\citenamefont {Friedt}, \citenamefont {Braden}, \citenamefont {Andr\'e}, \citenamefont {Adelmann}, \citenamefont {Nakatsuji},\ and\ \citenamefont {Maeno}}]{PhysRevB.63.174432}%
  \BibitemOpen
  \bibfield  {author} {\bibinfo {author} {\bibfnamefont {O.}~\bibnamefont {Friedt}}, \bibinfo {author} {\bibfnamefont {M.}~\bibnamefont {Braden}}, \bibinfo {author} {\bibfnamefont {G.}~\bibnamefont {Andr\'e}}, \bibinfo {author} {\bibfnamefont {P.}~\bibnamefont {Adelmann}}, \bibinfo {author} {\bibfnamefont {S.}~\bibnamefont {Nakatsuji}}, \ and\ \bibinfo {author} {\bibfnamefont {Y.}~\bibnamefont {Maeno}},\ }\bibfield  {title} {\enquote {\bibinfo {title} {{Structural and magnetic aspects of the metal-insulator transition in ${\mathrm{Ca}}_{2\ensuremath{-}x}{\mathrm{Sr}}_{x}{\mathrm{RuO}}_{4}$}},}\ }\href {\doibase 10.1103/PhysRevB.63.174432} {\bibfield  {journal} {\bibinfo  {journal} {Phys. Rev. B}\ }\textbf {\bibinfo {volume} {63}},\ \bibinfo {pages} {174432} (\bibinfo {year} {2001})}\BibitemShut {NoStop}%
\bibitem [{\citenamefont {Kriener}\ \emph {et~al.}(2005)\citenamefont {Kriener}, \citenamefont {Steffens}, \citenamefont {Baier}, \citenamefont {Schumann}, \citenamefont {Zabel}, \citenamefont {Lorenz}, \citenamefont {Friedt}, \citenamefont {M\"uller}, \citenamefont {Gukasov}, \citenamefont {Radaelli}, \citenamefont {Reutler}, \citenamefont {Revcolevschi}, \citenamefont {Nakatsuji}, \citenamefont {Maeno},\ and\ \citenamefont {Braden}}]{PhysRevLett.95.267403}%
  \BibitemOpen
  \bibfield  {author} {\bibinfo {author} {\bibfnamefont {M.}~\bibnamefont {Kriener}}, \bibinfo {author} {\bibfnamefont {P.}~\bibnamefont {Steffens}}, \bibinfo {author} {\bibfnamefont {J.}~\bibnamefont {Baier}}, \bibinfo {author} {\bibfnamefont {O.}~\bibnamefont {Schumann}}, \bibinfo {author} {\bibfnamefont {T.}~\bibnamefont {Zabel}}, \bibinfo {author} {\bibfnamefont {T.}~\bibnamefont {Lorenz}}, \bibinfo {author} {\bibfnamefont {O.}~\bibnamefont {Friedt}}, \bibinfo {author} {\bibfnamefont {R.}~\bibnamefont {M\"uller}}, \bibinfo {author} {\bibfnamefont {A.}~\bibnamefont {Gukasov}}, \bibinfo {author} {\bibfnamefont {P.~G.}\ \bibnamefont {Radaelli}}, \bibinfo {author} {\bibfnamefont {P.}~\bibnamefont {Reutler}}, \bibinfo {author} {\bibfnamefont {A.}~\bibnamefont {Revcolevschi}}, \bibinfo {author} {\bibfnamefont {S.}~\bibnamefont {Nakatsuji}}, \bibinfo {author} {\bibfnamefont {Y.}~\bibnamefont {Maeno}}, \ and\ \bibinfo {author} {\bibfnamefont {M.}~\bibnamefont {Braden}},\ }\bibfield  {title} {\enquote {\bibinfo
  {title} {{Structural Aspects of Metamagnetism in ${\mathrm{Ca}}_{2\ensuremath{-}x}{\mathrm{Sr}}_{x}{\mathrm{RuO}}_{4}$: Evidence for Field Tuning of Orbital Occupation}},}\ }\href {\doibase 10.1103/PhysRevLett.95.267403} {\bibfield  {journal} {\bibinfo  {journal} {Phys. Rev. Lett.}\ }\textbf {\bibinfo {volume} {95}},\ \bibinfo {pages} {267403} (\bibinfo {year} {2005})}\BibitemShut {NoStop}%
\bibitem [{\citenamefont {Ghosh}\ \emph {et~al.}(2021)\citenamefont {Ghosh}, \citenamefont {Shekhter}, \citenamefont {Jerzembeck}, \citenamefont {Kikugawa}, \citenamefont {Sokolov}, \citenamefont {Brando}, \citenamefont {Mackenzie}, \citenamefont {Hicks},\ and\ \citenamefont {Ramshaw}}]{ghosh2021thermodynamic}%
  \BibitemOpen
  \bibfield  {author} {\bibinfo {author} {\bibfnamefont {Sayak}\ \bibnamefont {Ghosh}}, \bibinfo {author} {\bibfnamefont {Arkady}\ \bibnamefont {Shekhter}}, \bibinfo {author} {\bibfnamefont {F.}~\bibnamefont {Jerzembeck}}, \bibinfo {author} {\bibfnamefont {N.}~\bibnamefont {Kikugawa}}, \bibinfo {author} {\bibfnamefont {Dmitry~A.}\ \bibnamefont {Sokolov}}, \bibinfo {author} {\bibfnamefont {Manuel}\ \bibnamefont {Brando}}, \bibinfo {author} {\bibfnamefont {A.P.}\ \bibnamefont {Mackenzie}}, \bibinfo {author} {\bibfnamefont {Clifford~W.}\ \bibnamefont {Hicks}}, \ and\ \bibinfo {author} {\bibfnamefont {B.~J.}\ \bibnamefont {Ramshaw}},\ }\bibfield  {title} {\enquote {\bibinfo {title} {{Thermodynamic evidence for a two-component superconducting order parameter in Sr$_2$RuO$_4$}},}\ }\href {\doibase 10.1038/s41567-020-1032-4} {\bibfield  {journal} {\bibinfo  {journal} {Nat. Phys.}\ }\textbf {\bibinfo {volume} {17}},\ \bibinfo {pages} {199--204} (\bibinfo {year} {2021})}\BibitemShut {NoStop}%
\bibitem [{\citenamefont {Benhabib}\ \emph {et~al.}(2021)\citenamefont {Benhabib}, \citenamefont {Lupien}, \citenamefont {Paul}, \citenamefont {Berges}, \citenamefont {Dion}, \citenamefont {Nardone}, \citenamefont {Zitouni}, \citenamefont {Mao}, \citenamefont {Maeno}, \citenamefont {Georges}, \citenamefont {Taillefer},\ and\ \citenamefont {Proust}}]{benhabib2021ultrasound}%
  \BibitemOpen
  \bibfield  {author} {\bibinfo {author} {\bibfnamefont {Siham}\ \bibnamefont {Benhabib}}, \bibinfo {author} {\bibfnamefont {C.}~\bibnamefont {Lupien}}, \bibinfo {author} {\bibfnamefont {I.}~\bibnamefont {Paul}}, \bibinfo {author} {\bibfnamefont {L.}~\bibnamefont {Berges}}, \bibinfo {author} {\bibfnamefont {M.}~\bibnamefont {Dion}}, \bibinfo {author} {\bibfnamefont {Marc}\ \bibnamefont {Nardone}}, \bibinfo {author} {\bibfnamefont {Abdelaziz}\ \bibnamefont {Zitouni}}, \bibinfo {author} {\bibfnamefont {Z.~Q.}\ \bibnamefont {Mao}}, \bibinfo {author} {\bibfnamefont {Y.}~\bibnamefont {Maeno}}, \bibinfo {author} {\bibfnamefont {A.}~\bibnamefont {Georges}}, \bibinfo {author} {\bibfnamefont {L.}~\bibnamefont {Taillefer}}, \ and\ \bibinfo {author} {\bibfnamefont {C.}~\bibnamefont {Proust}},\ }\bibfield  {title} {\enquote {\bibinfo {title} {{Ultrasound evidence for a two-component superconducting order parameter in Sr$_2$RuO$_4$}},}\ }\href {\doibase 10.1038/s41567-020-1033-3} {\bibfield  {journal} {\bibinfo  {journal}
  {Nat. Phys.}\ }\textbf {\bibinfo {volume} {17}},\ \bibinfo {pages} {194--198} (\bibinfo {year} {2021})}\BibitemShut {NoStop}%
\bibitem [{\citenamefont {Liechtenstein}\ \emph {et~al.}(1996)\citenamefont {Liechtenstein}, \citenamefont {Gunnarsson}, \citenamefont {Andersen},\ and\ \citenamefont {Martin}}]{PhysRevB.54.12505}%
  \BibitemOpen
  \bibfield  {author} {\bibinfo {author} {\bibfnamefont {A.~I.}\ \bibnamefont {Liechtenstein}}, \bibinfo {author} {\bibfnamefont {O.}~\bibnamefont {Gunnarsson}}, \bibinfo {author} {\bibfnamefont {O.~K.}\ \bibnamefont {Andersen}}, \ and\ \bibinfo {author} {\bibfnamefont {R.~M.}\ \bibnamefont {Martin}},\ }\bibfield  {title} {\enquote {\bibinfo {title} {Quasiparticle bands and superconductivity in bilayer cuprates},}\ }\href {\doibase 10.1103/PhysRevB.54.12505} {\bibfield  {journal} {\bibinfo  {journal} {Phys. Rev. B}\ }\textbf {\bibinfo {volume} {54}},\ \bibinfo {pages} {12505--12508} (\bibinfo {year} {1996})}\BibitemShut {NoStop}%
\bibitem [{\citenamefont {Irkhin}\ \emph {et~al.}(2002)\citenamefont {Irkhin}, \citenamefont {Katanin},\ and\ \citenamefont {Katsnelson}}]{PhysRevLett.89.076401}%
  \BibitemOpen
  \bibfield  {author} {\bibinfo {author} {\bibfnamefont {V.~Yu.}\ \bibnamefont {Irkhin}}, \bibinfo {author} {\bibfnamefont {A.~A.}\ \bibnamefont {Katanin}}, \ and\ \bibinfo {author} {\bibfnamefont {M.~I.}\ \bibnamefont {Katsnelson}},\ }\bibfield  {title} {\enquote {\bibinfo {title} {{Robustness of the Van Hove Scenario for High-${T}_{c}$ Superconductors}},}\ }\href {\doibase 10.1103/PhysRevLett.89.076401} {\bibfield  {journal} {\bibinfo  {journal} {Phys. Rev. Lett.}\ }\textbf {\bibinfo {volume} {89}},\ \bibinfo {pages} {076401} (\bibinfo {year} {2002})}\BibitemShut {NoStop}%
\bibitem [{\citenamefont {Odashima}\ \emph {et~al.}(2005)\citenamefont {Odashima}, \citenamefont {Avella},\ and\ \citenamefont {Mancini}}]{PhysRevB.72.205121}%
  \BibitemOpen
  \bibfield  {author} {\bibinfo {author} {\bibfnamefont {Satoru}\ \bibnamefont {Odashima}}, \bibinfo {author} {\bibfnamefont {Adolfo}\ \bibnamefont {Avella}}, \ and\ \bibinfo {author} {\bibfnamefont {Ferdinando}\ \bibnamefont {Mancini}},\ }\bibfield  {title} {\enquote {\bibinfo {title} {{High-order correlation effects in the two-dimensional Hubbard model}},}\ }\href {\doibase 10.1103/PhysRevB.72.205121} {\bibfield  {journal} {\bibinfo  {journal} {Phys. Rev. B}\ }\textbf {\bibinfo {volume} {72}},\ \bibinfo {pages} {205121} (\bibinfo {year} {2005})}\BibitemShut {NoStop}%
\bibitem [{\citenamefont {Chen}\ \emph {et~al.}(2011)\citenamefont {Chen}, \citenamefont {Pathak}, \citenamefont {Yang}, \citenamefont {Su}, \citenamefont {Galanakis}, \citenamefont {Mikelsons}, \citenamefont {Jarrell},\ and\ \citenamefont {Moreno}}]{PhysRevB.84.245107}%
  \BibitemOpen
  \bibfield  {author} {\bibinfo {author} {\bibfnamefont {K.-S.}\ \bibnamefont {Chen}}, \bibinfo {author} {\bibfnamefont {S.}~\bibnamefont {Pathak}}, \bibinfo {author} {\bibfnamefont {S.-X.}\ \bibnamefont {Yang}}, \bibinfo {author} {\bibfnamefont {S.-Q.}\ \bibnamefont {Su}}, \bibinfo {author} {\bibfnamefont {D.}~\bibnamefont {Galanakis}}, \bibinfo {author} {\bibfnamefont {K.}~\bibnamefont {Mikelsons}}, \bibinfo {author} {\bibfnamefont {M.}~\bibnamefont {Jarrell}}, \ and\ \bibinfo {author} {\bibfnamefont {J.}~\bibnamefont {Moreno}},\ }\bibfield  {title} {\enquote {\bibinfo {title} {{Role of the van Hove singularity in the quantum criticality of the Hubbard model}},}\ }\href {\doibase 10.1103/PhysRevB.84.245107} {\bibfield  {journal} {\bibinfo  {journal} {Phys. Rev. B}\ }\textbf {\bibinfo {volume} {84}},\ \bibinfo {pages} {245107} (\bibinfo {year} {2011})}\BibitemShut {NoStop}%
\bibitem [{\citenamefont {Schmitt}(2010)}]{PhysRevB.82.155126}%
  \BibitemOpen
  \bibfield  {author} {\bibinfo {author} {\bibfnamefont {Sebastian}\ \bibnamefont {Schmitt}},\ }\bibfield  {title} {\enquote {\bibinfo {title} {{Non-Fermi-liquid signatures in the Hubbard model due to van Hove singularities}},}\ }\href {\doibase 10.1103/PhysRevB.82.155126} {\bibfield  {journal} {\bibinfo  {journal} {Phys. Rev. B}\ }\textbf {\bibinfo {volume} {82}},\ \bibinfo {pages} {155126} (\bibinfo {year} {2010})}\BibitemShut {NoStop}%
\bibitem [{\citenamefont {Bl\"ochl}(1994)}]{PhysRevB.50.17953}%
  \BibitemOpen
  \bibfield  {author} {\bibinfo {author} {\bibfnamefont {P.~E.}\ \bibnamefont {Bl\"ochl}},\ }\bibfield  {title} {\enquote {\bibinfo {title} {{Projector augmented-wave method}},}\ }\href {\doibase 10.1103/PhysRevB.50.17953} {\bibfield  {journal} {\bibinfo  {journal} {Phys. Rev. B}\ }\textbf {\bibinfo {volume} {50}},\ \bibinfo {pages} {17953--17979} (\bibinfo {year} {1994})}\BibitemShut {NoStop}%
\bibitem [{\citenamefont {Kresse}\ and\ \citenamefont {Joubert}(1999)}]{PhysRevB.59.1758}%
  \BibitemOpen
  \bibfield  {author} {\bibinfo {author} {\bibfnamefont {G.}~\bibnamefont {Kresse}}\ and\ \bibinfo {author} {\bibfnamefont {D.}~\bibnamefont {Joubert}},\ }\bibfield  {title} {\enquote {\bibinfo {title} {{From ultrasoft pseudopotentials to the projector augmented-wave method}},}\ }\href {\doibase 10.1103/PhysRevB.59.1758} {\bibfield  {journal} {\bibinfo  {journal} {Phys. Rev. B}\ }\textbf {\bibinfo {volume} {59}},\ \bibinfo {pages} {1758--1775} (\bibinfo {year} {1999})}\BibitemShut {NoStop}%
\bibitem [{\citenamefont {Kresse}\ and\ \citenamefont {Furthm{\"u}ller}(1996)}]{KRESSE199615}%
  \BibitemOpen
  \bibfield  {author} {\bibinfo {author} {\bibfnamefont {G.}~\bibnamefont {Kresse}}\ and\ \bibinfo {author} {\bibfnamefont {J.}~\bibnamefont {Furthm{\"u}ller}},\ }\bibfield  {title} {\enquote {\bibinfo {title} {{Efficiency of ab-initio total energy calculations for metals and semiconductors using a plane-wave basis set}},}\ }\href {\doibase https://doi.org/10.1016/0927-0256(96)00008-0} {\bibfield  {journal} {\bibinfo  {journal} {Comput. Mater. Sci.}\ }\textbf {\bibinfo {volume} {6}},\ \bibinfo {pages} {15--50} (\bibinfo {year} {1996})}\BibitemShut {NoStop}%
\bibitem [{\citenamefont {Kresse}\ and\ \citenamefont {Furthm\"uller}(1996)}]{PhysRevB.54.11169}%
  \BibitemOpen
  \bibfield  {author} {\bibinfo {author} {\bibfnamefont {G.}~\bibnamefont {Kresse}}\ and\ \bibinfo {author} {\bibfnamefont {J.}~\bibnamefont {Furthm\"uller}},\ }\bibfield  {title} {\enquote {\bibinfo {title} {{Efficient iterative schemes for ab initio total-energy calculations using a plane-wave basis set}},}\ }\href {\doibase 10.1103/PhysRevB.54.11169} {\bibfield  {journal} {\bibinfo  {journal} {Phys. Rev. B}\ }\textbf {\bibinfo {volume} {54}},\ \bibinfo {pages} {11169--11186} (\bibinfo {year} {1996})}\BibitemShut {NoStop}%
\bibitem [{\citenamefont {Perdew}\ \emph {et~al.}(1996)\citenamefont {Perdew}, \citenamefont {Burke},\ and\ \citenamefont {Ernzerhof}}]{PhysRevLett.77.3865}%
  \BibitemOpen
  \bibfield  {author} {\bibinfo {author} {\bibfnamefont {John~P.}\ \bibnamefont {Perdew}}, \bibinfo {author} {\bibfnamefont {Kieron}\ \bibnamefont {Burke}}, \ and\ \bibinfo {author} {\bibfnamefont {Matthias}\ \bibnamefont {Ernzerhof}},\ }\bibfield  {title} {\enquote {\bibinfo {title} {{Generalized Gradient Approximation Made Simple}},}\ }\href {\doibase 10.1103/PhysRevLett.77.3865} {\bibfield  {journal} {\bibinfo  {journal} {Phys. Rev. Lett.}\ }\textbf {\bibinfo {volume} {77}},\ \bibinfo {pages} {3865--3868} (\bibinfo {year} {1996})}\BibitemShut {NoStop}%
\bibitem [{\citenamefont {Marzari}\ and\ \citenamefont {Vanderbilt}(1997)}]{PhysRevB.56.12847}%
  \BibitemOpen
  \bibfield  {author} {\bibinfo {author} {\bibfnamefont {Nicola}\ \bibnamefont {Marzari}}\ and\ \bibinfo {author} {\bibfnamefont {David}\ \bibnamefont {Vanderbilt}},\ }\bibfield  {title} {\enquote {\bibinfo {title} {{Maximally localized generalized Wannier functions for composite energy bands}},}\ }\href {\doibase 10.1103/PhysRevB.56.12847} {\bibfield  {journal} {\bibinfo  {journal} {Phys. Rev. B}\ }\textbf {\bibinfo {volume} {56}},\ \bibinfo {pages} {12847--12865} (\bibinfo {year} {1997})}\BibitemShut {NoStop}%
\bibitem [{\citenamefont {Marzari}\ \emph {et~al.}(2012)\citenamefont {Marzari}, \citenamefont {Mostofi}, \citenamefont {Yates}, \citenamefont {Souza},\ and\ \citenamefont {Vanderbilt}}]{RevModPhys.84.1419}%
  \BibitemOpen
  \bibfield  {author} {\bibinfo {author} {\bibfnamefont {Nicola}\ \bibnamefont {Marzari}}, \bibinfo {author} {\bibfnamefont {Arash~A.}\ \bibnamefont {Mostofi}}, \bibinfo {author} {\bibfnamefont {Jonathan~R.}\ \bibnamefont {Yates}}, \bibinfo {author} {\bibfnamefont {Ivo}\ \bibnamefont {Souza}}, \ and\ \bibinfo {author} {\bibfnamefont {David}\ \bibnamefont {Vanderbilt}},\ }\bibfield  {title} {\enquote {\bibinfo {title} {{Maximally localized Wannier functions: Theory and applications}},}\ }\href {\doibase 10.1103/RevModPhys.84.1419} {\bibfield  {journal} {\bibinfo  {journal} {Rev. Mod. Phys.}\ }\textbf {\bibinfo {volume} {84}},\ \bibinfo {pages} {1419--1475} (\bibinfo {year} {2012})}\BibitemShut {NoStop}%
\bibitem [{\citenamefont {Mostofi}\ \emph {et~al.}(2008)\citenamefont {Mostofi}, \citenamefont {Yates}, \citenamefont {Lee}, \citenamefont {Souza}, \citenamefont {Vanderbilt},\ and\ \citenamefont {Marzari}}]{MOSTOFI2008685}%
  \BibitemOpen
  \bibfield  {author} {\bibinfo {author} {\bibfnamefont {Arash~A.}\ \bibnamefont {Mostofi}}, \bibinfo {author} {\bibfnamefont {Jonathan~R.}\ \bibnamefont {Yates}}, \bibinfo {author} {\bibfnamefont {Young-Su}\ \bibnamefont {Lee}}, \bibinfo {author} {\bibfnamefont {Ivo}\ \bibnamefont {Souza}}, \bibinfo {author} {\bibfnamefont {David}\ \bibnamefont {Vanderbilt}}, \ and\ \bibinfo {author} {\bibfnamefont {Nicola}\ \bibnamefont {Marzari}},\ }\bibfield  {title} {\enquote {\bibinfo {title} {{wannier90: A tool for obtaining maximally-localised Wannier functions}},}\ }\href {\doibase https://doi.org/10.1016/j.cpc.2007.11.016} {\bibfield  {journal} {\bibinfo  {journal} {Comput. Phys. Commun.}\ }\textbf {\bibinfo {volume} {178}},\ \bibinfo {pages} {685--699} (\bibinfo {year} {2008})}\BibitemShut {NoStop}%
\end{thebibliography}%


\begin{thebibliography}{7}%
\makeatletter
\providecommand \@ifxundefined [1]{%
 \@ifx{#1\undefined}
}%
\providecommand \@ifnum [1]{%
 \ifnum #1\expandafter \@firstoftwo
 \else \expandafter \@secondoftwo
 \fi
}%
\providecommand \@ifx [1]{%
 \ifx #1\expandafter \@firstoftwo
 \else \expandafter \@secondoftwo
 \fi
}%
\providecommand \natexlab [1]{#1}%
\providecommand \enquote  [1]{``#1''}%
\providecommand \bibnamefont  [1]{#1}%
\providecommand \bibfnamefont [1]{#1}%
\providecommand \citenamefont [1]{#1}%
\providecommand \href@noop [0]{\@secondoftwo}%
\providecommand \href [0]{\begingroup \@sanitize@url \@href}%
\providecommand \@href[1]{\@@startlink{#1}\@@href}%
\providecommand \@@href[1]{\endgroup#1\@@endlink}%
\providecommand \@sanitize@url [0]{\catcode `\\12\catcode `\$12\catcode
  `\&12\catcode `\#12\catcode `\^12\catcode `\_12\catcode `\%12\relax}%
\providecommand \@@startlink[1]{}%
\providecommand \@@endlink[0]{}%
\providecommand \url  [0]{\begingroup\@sanitize@url \@url }%
\providecommand \@url [1]{\endgroup\@href {#1}{\urlprefix }}%
\providecommand \urlprefix  [0]{URL }%
\providecommand \Eprint [0]{\href }%
\providecommand \doibase [0]{http://dx.doi.org/}%
\providecommand \selectlanguage [0]{\@gobble}%
\providecommand \bibinfo  [0]{\@secondoftwo}%
\providecommand \bibfield  [0]{\@secondoftwo}%
\providecommand \translation [1]{[#1]}%
\providecommand \BibitemOpen [0]{}%
\providecommand \bibitemStop [0]{}%
\providecommand \bibitemNoStop [0]{.\EOS\space}%
\providecommand \EOS [0]{\spacefactor3000\relax}%
\providecommand \BibitemShut  [1]{\csname bibitem#1\endcsname}%
\let\auto@bib@innerbib\@empty
%</preamble>
\bibitem [{\citenamefont {Strand}\ \emph {et~al.}(2019)\citenamefont {Strand},
  \citenamefont {Zingl}, \citenamefont {Wentzell}, \citenamefont {Parcollet},\
  and\ \citenamefont {Georges}}]{PhysRevB.100.125120}%
  \BibitemOpen
  \bibfield  {author} {\bibinfo {author} {\bibfnamefont {Hugo U.~R.}\
  \bibnamefont {Strand}}, \bibinfo {author} {\bibfnamefont {Manuel}\
  \bibnamefont {Zingl}}, \bibinfo {author} {\bibfnamefont {Nils}\ \bibnamefont
  {Wentzell}}, \bibinfo {author} {\bibfnamefont {Olivier}\ \bibnamefont
  {Parcollet}}, \ and\ \bibinfo {author} {\bibfnamefont {Antoine}\ \bibnamefont
  {Georges}},\ }\bibfield  {title} {\enquote {\bibinfo {title} {{Magnetic
  response of ${\mathrm{Sr}}_{2}{\mathrm{RuO}}_{4}$: Quasi-local spin
  fluctuations due to Hund's coupling}},}\ }\href {\doibase
  10.1103/PhysRevB.100.125120} {\bibfield  {journal} {\bibinfo  {journal}
  {Phys. Rev. B}\ }\textbf {\bibinfo {volume} {100}},\ \bibinfo {pages}
  {125120} (\bibinfo {year} {2019})}\BibitemShut {NoStop}%
\bibitem [{\citenamefont {Vaugier}\ \emph {et~al.}(2012)\citenamefont
  {Vaugier}, \citenamefont {Jiang},\ and\ \citenamefont
  {Biermann}}]{PhysRevB.86.165105}%
  \BibitemOpen
  \bibfield  {author} {\bibinfo {author} {\bibfnamefont {Lo\"{\i}g}\
  \bibnamefont {Vaugier}}, \bibinfo {author} {\bibfnamefont {Hong}\
  \bibnamefont {Jiang}}, \ and\ \bibinfo {author} {\bibfnamefont {Silke}\
  \bibnamefont {Biermann}},\ }\bibfield  {title} {\enquote {\bibinfo {title}
  {{Hubbard $U$ and Hund exchange $J$ in transition metal oxides: Screening
  versus localization trends from constrained random phase approximation}},}\
  }\href {\doibase 10.1103/PhysRevB.86.165105} {\bibfield  {journal} {\bibinfo
  {journal} {Phys. Rev. B}\ }\textbf {\bibinfo {volume} {86}},\ \bibinfo
  {pages} {165105} (\bibinfo {year} {2012})}\BibitemShut {NoStop}%
\bibitem [{\citenamefont {Mravlje}\ \emph {et~al.}(2011)\citenamefont
  {Mravlje}, \citenamefont {Aichhorn}, \citenamefont {Miyake}, \citenamefont
  {Haule}, \citenamefont {Kotliar},\ and\ \citenamefont
  {Georges}}]{PhysRevLett.106.096401}%
  \BibitemOpen
  \bibfield  {author} {\bibinfo {author} {\bibfnamefont {Jernej}\ \bibnamefont
  {Mravlje}}, \bibinfo {author} {\bibfnamefont {Markus}\ \bibnamefont
  {Aichhorn}}, \bibinfo {author} {\bibfnamefont {Takashi}\ \bibnamefont
  {Miyake}}, \bibinfo {author} {\bibfnamefont {Kristjan}\ \bibnamefont
  {Haule}}, \bibinfo {author} {\bibfnamefont {Gabriel}\ \bibnamefont
  {Kotliar}}, \ and\ \bibinfo {author} {\bibfnamefont {Antoine}\ \bibnamefont
  {Georges}},\ }\bibfield  {title} {\enquote {\bibinfo {title}
  {{Coherence-Incoherence Crossover and the Mass-Renormalization Puzzles in
  ${\mathrm{Sr}}_{2}{\mathrm{RuO}}_{4}$}},}\ }\href {\doibase
  10.1103/PhysRevLett.106.096401} {\bibfield  {journal} {\bibinfo  {journal}
  {Phys. Rev. Lett.}\ }\textbf {\bibinfo {volume} {106}},\ \bibinfo {pages}
  {096401} (\bibinfo {year} {2011})}\BibitemShut {NoStop}%
\bibitem [{\citenamefont {Mackenzie}\ \emph {et~al.}(1996)\citenamefont
  {Mackenzie}, \citenamefont {Julian}, \citenamefont {Diver}, \citenamefont
  {McMullan}, \citenamefont {Ray}, \citenamefont {Lonzarich}, \citenamefont
  {Maeno}, \citenamefont {Nishizaki},\ and\ \citenamefont
  {Fujita}}]{PhysRevLett.76.3786}%
  \BibitemOpen
  \bibfield  {author} {\bibinfo {author} {\bibfnamefont {A.~P.}\ \bibnamefont
  {Mackenzie}}, \bibinfo {author} {\bibfnamefont {S.~R.}\ \bibnamefont
  {Julian}}, \bibinfo {author} {\bibfnamefont {A.~J.}\ \bibnamefont {Diver}},
  \bibinfo {author} {\bibfnamefont {G.~J.}\ \bibnamefont {McMullan}}, \bibinfo
  {author} {\bibfnamefont {M.~P.}\ \bibnamefont {Ray}}, \bibinfo {author}
  {\bibfnamefont {G.~G.}\ \bibnamefont {Lonzarich}}, \bibinfo {author}
  {\bibfnamefont {Y.}~\bibnamefont {Maeno}}, \bibinfo {author} {\bibfnamefont
  {S.}~\bibnamefont {Nishizaki}}, \ and\ \bibinfo {author} {\bibfnamefont
  {T.}~\bibnamefont {Fujita}},\ }\bibfield  {title} {\enquote {\bibinfo {title}
  {{Quantum Oscillations in the Layered Perovskite Superconductor
  S${\mathrm{r}}_{2}$Ru${\mathrm{O}}_{4}$}},}\ }\href {\doibase
  10.1103/PhysRevLett.76.3786} {\bibfield  {journal} {\bibinfo  {journal}
  {Phys. Rev. Lett.}\ }\textbf {\bibinfo {volume} {76}},\ \bibinfo {pages}
  {3786--3789} (\bibinfo {year} {1996})}\BibitemShut {NoStop}%
\bibitem [{\citenamefont {Bergemann}\ \emph {et~al.}(2003)\citenamefont
  {Bergemann}, \citenamefont {Mackenzie}, \citenamefont {Julian}, \citenamefont
  {Forsythe},\ and\ \citenamefont
  {Ohmichi}}]{doi:10.1080/00018730310001621737}%
  \BibitemOpen
  \bibfield  {author} {\bibinfo {author} {\bibfnamefont {C.}~\bibnamefont
  {Bergemann}}, \bibinfo {author} {\bibfnamefont {A.~P.}\ \bibnamefont
  {Mackenzie}}, \bibinfo {author} {\bibfnamefont {S.~R.}\ \bibnamefont
  {Julian}}, \bibinfo {author} {\bibfnamefont {D.}~\bibnamefont {Forsythe}}, \
  and\ \bibinfo {author} {\bibfnamefont {E.}~\bibnamefont {Ohmichi}},\
  }\bibfield  {title} {\enquote {\bibinfo {title} {{Quasi-two-dimensional Fermi
  liquid properties of the unconventional superconductor Sr$_2$RuO$_4$}},}\
  }\href {\doibase 10.1080/00018730310001621737} {\bibfield  {journal}
  {\bibinfo  {journal} {Advances in Physics}\ }\textbf {\bibinfo {volume}
  {52}},\ \bibinfo {pages} {639--725} (\bibinfo {year} {2003})}\BibitemShut
  {NoStop}%
\bibitem [{\citenamefont {Tamai}\ \emph {et~al.}(2019)\citenamefont {Tamai},
  \citenamefont {Zingl}, \citenamefont {Rozbicki}, \citenamefont {Cappelli},
  \citenamefont {Ricc\`o}, \citenamefont {de~la Torre}, \citenamefont
  {McKeown~Walker}, \citenamefont {Bruno}, \citenamefont {King}, \citenamefont
  {Meevasana}, \citenamefont {Shi}, \citenamefont
  {Radovi\ifmmode~\acute{c}\else \'{c}\fi{}}, \citenamefont {Plumb},
  \citenamefont {Gibbs}, \citenamefont {Mackenzie}, \citenamefont {Berthod},
  \citenamefont {Strand}, \citenamefont {Kim}, \citenamefont {Georges},\ and\
  \citenamefont {Baumberger}}]{PhysRevX.9.021048}%
  \BibitemOpen
  \bibfield  {author} {\bibinfo {author} {\bibfnamefont {A.}~\bibnamefont
  {Tamai}}, \bibinfo {author} {\bibfnamefont {M.}~\bibnamefont {Zingl}},
  \bibinfo {author} {\bibfnamefont {E.}~\bibnamefont {Rozbicki}}, \bibinfo
  {author} {\bibfnamefont {E.}~\bibnamefont {Cappelli}}, \bibinfo {author}
  {\bibfnamefont {S.}~\bibnamefont {Ricc\`o}}, \bibinfo {author} {\bibfnamefont
  {A.}~\bibnamefont {de~la Torre}}, \bibinfo {author} {\bibfnamefont
  {S.}~\bibnamefont {McKeown~Walker}}, \bibinfo {author} {\bibfnamefont
  {F.~Y.}\ \bibnamefont {Bruno}}, \bibinfo {author} {\bibfnamefont {P.~D.~C.}\
  \bibnamefont {King}}, \bibinfo {author} {\bibfnamefont {W.}~\bibnamefont
  {Meevasana}}, \bibinfo {author} {\bibfnamefont {M.}~\bibnamefont {Shi}},
  \bibinfo {author} {\bibfnamefont {M.}~\bibnamefont
  {Radovi\ifmmode~\acute{c}\else \'{c}\fi{}}}, \bibinfo {author} {\bibfnamefont
  {N.~C.}\ \bibnamefont {Plumb}}, \bibinfo {author} {\bibfnamefont {A.~S.}\
  \bibnamefont {Gibbs}}, \bibinfo {author} {\bibfnamefont {A.~P.}\ \bibnamefont
  {Mackenzie}}, \bibinfo {author} {\bibfnamefont {C.}~\bibnamefont {Berthod}},
  \bibinfo {author} {\bibfnamefont {H.~U.~R.}\ \bibnamefont {Strand}}, \bibinfo
  {author} {\bibfnamefont {M.}~\bibnamefont {Kim}}, \bibinfo {author}
  {\bibfnamefont {A.}~\bibnamefont {Georges}}, \ and\ \bibinfo {author}
  {\bibfnamefont {F.}~\bibnamefont {Baumberger}},\ }\bibfield  {title}
  {\enquote {\bibinfo {title} {{High-Resolution Photoemission on
  ${\mathrm{Sr}}_{2}{\mathrm{RuO}}_{4}$ Reveals Correlation-Enhanced Effective
  Spin-Orbit Coupling and Dominantly Local Self-Energies}},}\ }\href {\doibase
  10.1103/PhysRevX.9.021048} {\bibfield  {journal} {\bibinfo  {journal} {Phys.
  Rev. X}\ }\textbf {\bibinfo {volume} {9}},\ \bibinfo {pages} {021048}
  (\bibinfo {year} {2019})}\BibitemShut {NoStop}%
\bibitem [{\citenamefont {Mackenzie}\ and\ \citenamefont
  {Maeno}(2003)}]{RevModPhys.75.657}%
  \BibitemOpen
  \bibfield  {author} {\bibinfo {author} {\bibfnamefont {Andrew~Peter}\
  \bibnamefont {Mackenzie}}\ and\ \bibinfo {author} {\bibfnamefont {Yoshiteru}\
  \bibnamefont {Maeno}},\ }\bibfield  {title} {\enquote {\bibinfo {title} {{The
  superconductivity of ${\mathrm{Sr}}_{2}{\mathrm{RuO}}_{4}$ and the physics of
  spin-triplet pairing}},}\ }\href {\doibase 10.1103/RevModPhys.75.657}
  {\bibfield  {journal} {\bibinfo  {journal} {Rev. Mod. Phys.}\ }\textbf
  {\bibinfo {volume} {75}},\ \bibinfo {pages} {657--712} (\bibinfo {year}
  {2003})}\BibitemShut {NoStop}%
\end{thebibliography}%

\end{document}

\title{Supplemental Material\\[0.4cm]
Orbital-Selective Diffuse Magnetic Fluctuations in Sr$_2$RuO$_4$: a Unified Theoretical Picture}

\author{Maria Chatzieleftheriou}
\affiliation{CPHT, CNRS, {\'E}cole polytechnique, Institut Polytechnique de Paris, 91120 Palaiseau, France}

\author{Alexander N. Rudenko}
\affiliation{\mbox{Radboud University, Institute for Molecules and Materials, Heijendaalseweg 135, 6525AJ Nijmegen, The Netherlands}}

\author{Yvan Sidis}
\affiliation{Universit{\'e} Paris-Saclay, CNRS, CEA, Laboratoire L{\'e}on Brillouin, 91191, Gif-sur-Yvette, France}

\author{Silke Biermann}
\affiliation{CPHT, CNRS, {\'E}cole polytechnique, Institut Polytechnique de Paris, 91120 Palaiseau, France}
\affiliation{Coll{\`e}ge de France, Universit{\'e} PSL, 11 place Marcelin Berthelot, 75005 Paris, France}
\affiliation{European Theoretical Spectroscopy Facility, 91128 Palaiseau, France}

\author{Evgeny A. Stepanov}
\affiliation{CPHT, CNRS, {\'E}cole polytechnique, Institut Polytechnique de Paris, 91120 Palaiseau, France}
\affiliation{Coll{\`e}ge de France, Universit{\'e} PSL, 11 place Marcelin Berthelot, 75005 Paris, France}

\maketitle

\section{Model}

\noindent
In this work we study the Hubbard-Kanamori Hamiltonian: 
\begin{align}
\hat{H} =& \sum_{ij,mm',\sigma}t_{ij}^{mm'}\hat{c}_{im\sigma}^{\dagger}\hat{c}^{\phantom{\dagger}}_{jm'\sigma} + U\sum_{i,m}\hat{n}_{im\uparrow}\hat{n}_{im\downarrow} + \notag\\
&+U'\sum_{i,m\neq m}\hat{n}_{im\uparrow}\hat{n}_{im'\downarrow} + (U'-J)\sum_{i,m<m',\sigma}\hat{n}_{im\sigma}\hat{n}_{im'\sigma} + \notag\\
&+J\sum_{i,m\neq m'}(\hat{c}_{im\uparrow}^{\dagger}\hat{c}_{im\downarrow}^{\dagger}\hat{c}^{\phantom{\dagger}}_{im'\downarrow}\hat{c}^{\phantom{\dagger}}_{im'\uparrow} - \hat{c}_{im\uparrow}^{\dagger}\hat{c}^{\phantom{\dagger}}_{im\downarrow}\hat{c}_{im'\downarrow}^{\dagger}\hat{c}^{\phantom{\dagger}}_{im'\uparrow}),
\end{align}
that models the electronic behavior in Sr$_2$RuO$_4$. 
The first term corresponds to the kinetic energy, where $\hat{c}^{(\dagger)}_{i\sigma}$ are annihilation (creation) operators, with lattice site index $i$, orbital index $m$ and spin polarization $\sigma$, and $t_{ij}^{mm'}$ is the hopping amplitude. 
All other terms correspond to the interaction energy; 
$U$ the intra- and ${U'=U-2J}$ the inter-orbital local (on-site) Coulomb repulsion and $J$ the Hund's exchange coupling, favoring high-spin states. $\hat{n}_{im\sigma}=\hat{c}^{\dagger}_{im\sigma}\hat{c}_{im\sigma}$ is the density operator for spin $\sigma$ and orbital character $m$ on site $i$. 
The model cannot, in general, be solved exactly and approximate methods are typically used. We employ the DMFT and \mbox{D-TRILEX} many-body techniques, discussed in the main text. 

\section{Effect of Hund's coupling}

\begin{figure}[t!]
\includegraphics[width=1\linewidth]{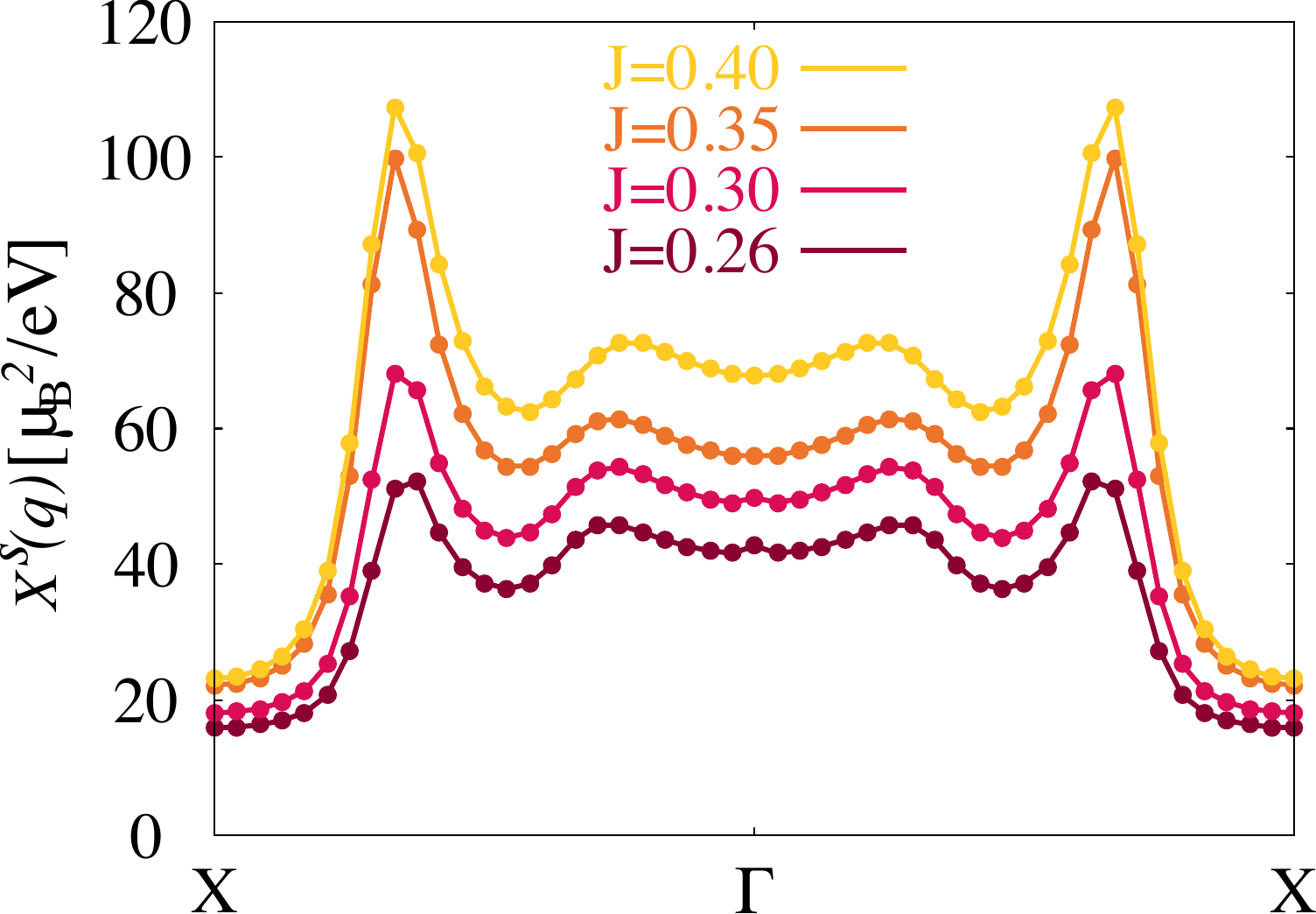} 
\\[15pt]
\includegraphics[width=1\linewidth]{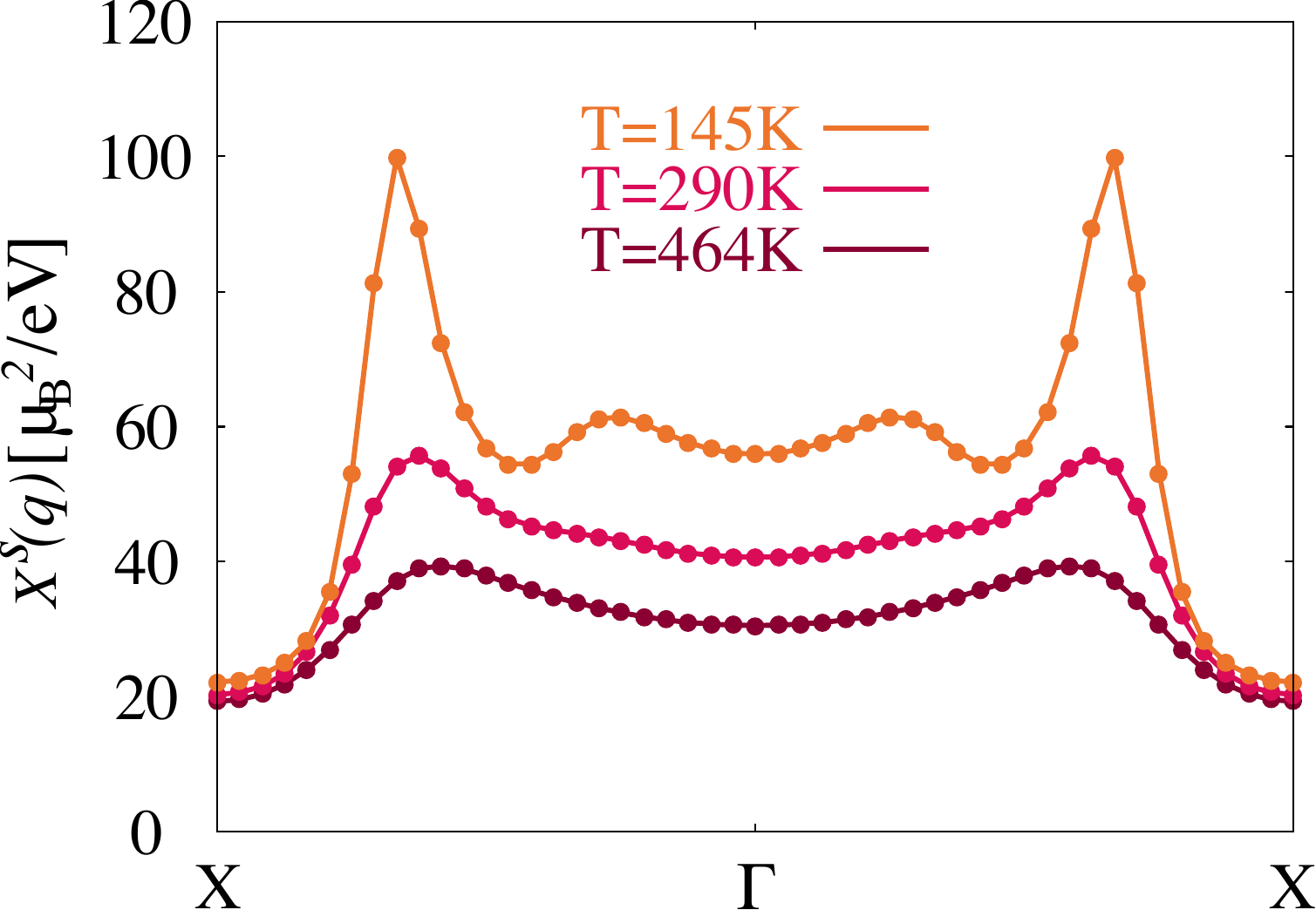}
\caption{Total spin susceptibility across the high symmetry path $X-\Gamma-X$ of the first Brillouin Zone ($X=(\pi,\pi)$, $\Gamma=(0,0)$), for \textbf{Left panel}: different values of the Hund's exchange coupling $J$. An increased $J$ leads to an overall shift of $X^{s}$ to larger values. Comparing the center ($\Gamma$) and corner ($X$) of the BZ, the increase of $X^{s}$ at the center is much more pronounced, as a direct result of the ferromagnetic fluctuations promoted by $J$.
\textbf{Right panel:} $X^{s}$ for different values of the temperature. Decreasing the temperature leads to an overall larger magnetic susceptibility, with the effect being more pronounced at the incommensurate SDW vector and at the center of the BZ (at the $\Gamma$ point). 
\label{fig:SM1}}
\end{figure}

In this section we investigate the effect of Hund's exchange coupling $J$ on properties of Sr$_2$RuO$_4$. The results for the magnetic susceptibility shown in the left panel of Fig.~\ref{fig:SM1} reveal that an increased value of $J$ leads to an overall enhancement of the spin susceptibility at all $k$-points. However, it becomes immediately clear that the $X^{s}(q)$ value at the center of the Brillouin Zone (at the $\Gamma=(0,0,0)$ point) is significantly more affected by $J$ compared to the one at the edges of the zone (at the $X=(\pi,\pi,0)$ point). 
This result confirms the DMFT conclusions and it is not surprising as the main mechanism of Hund's rule is to favor the high-spin states in the system, therefore to increase the ferromagnetic fluctuations. 
Nevertheless, what is interesting about our analysis is that we find the dependence of the susceptibility on $J$ to be more restricted compared to the DMFT findings. As already shown in previous studies~\cite{PhysRevB.100.125120}, in DMFT a value of $J$ within the range suggested by cRPA analysis leads to an increased signal at the AFM $X$ ordering vector, compared to $\Gamma$, and only upon strong increase of the value of $J$, the suppression of this unphysical signal is obtained. To the contrary, the inclusion of non-local electronic correlations in the description appears to limit the effect of Hund's coupling and the change of the spin susceptibility as a function of $J$ is only quantitative and mainly manifested in the increase of the response at the $\Gamma$ point. 

As discussed in the main text, the Coulomb interaction value ${U=2.56}$\,eV has been chosen following the cRPA study of ~\cite{PhysRevB.86.165105}. In that work, the Hund's coupling had been estimated to be ${J=0.26}$\,eV, however in our analysis we chose a slightly larger value ${J=0.35}$\,eV. Different theoretical studies have revealed that in order to capture accurately the orbital-dependent mass enhancement of the system an increased value of $J$ is necessary~\cite{PhysRevLett.106.096401,PhysRevB.100.125120}.

\begin{table}[b!]
\begin{tabular}{ |c|c|c|c|c|c|}  
 \hline
 ~~~~$J$~~~~ & ~~$m^{*}/m$ $(xz/yz)$~~ & ~~$m^{*}/m$ $(xy)$~~ & ~~$X^{s}_{\Gamma}$~~ & ~~$X^{s}_{X}$~~ & ~~$X^{s}_{SDW}$~~\\  
 \hline
 0.26 & 1.75 & 2.38 & 42 & 16 & 53 \\ 
 \hline
 0.35 & 2.44 & 3.13 & 56 & 22 & 100 \\ 
 \hline
 0.40 & 3.03 & 3.85 & 68 & 23 & 108 \\  
 \hline
\end{tabular}
\caption{Dependence of different quantities on the value of the Hund's coupling $J$ (first column), in units of [eV]; calculations performed at $T=145K$. The mass enhancements per orbital (second and third columns) consistently increase with $J$ and the one corresponding to the 2D $xy$ orbital (third column) is always larger. The majority of experimental findings suggest values for the mass enhancement that are between those found in this work for $J=0.35$\,eV and $J=0.40$\,eV. The spin susceptibility at the $\Gamma=(0,0,0)$, $X=(\pi,\pi,0)$ and $SDW=(3\pi/5,3\pi/5,0)$ points are shown in columns four, five and six respectively, in units of [$\mu_B^2/$eV]. They suggest that a small $J$ gives an overall small spin signal, while by increasing $J$ we retrieve the values that best agree with experiments (between $J=0.30$\,eV and $J=0.35$\,eV). The main effect of $J$ at this larger $J$ regime is to increase the ferromagnetic response, i.e., the susceptibility at the $\Gamma$ point.} 
\label{table}
\end{table}

We find that within our method indeed a realistic $m^*/m$ is obtained for ${J\sim0.35-0.40}$\,eV, while for the spin susceptibility the best agreement with experiments is seen for ${J\sim0.30-0.35}$\,eV. 
In Table~\ref{table} we summarize the dependence of the mass enhancement on the value of Hund's coupling $J$ - for the $xz/yz$ and $xy$ orbitals - and the dependence of the spin susceptibility at the center ($\Gamma$) and edge (X) of the BZ, as well as on the SDW ordering vector. 
The mass enhancement is calculated as the slope of $Im\Sigma$ at zero frequency.
In the temperature regime of our study, the necessary extrapolation from the lowest Matsubara frequency leads to a relatively large error bar.  
However, we find that $m^{*}/m$ is consistently larger for the 2D $xy$ orbital, in agreement with quantum oscillations and ARPES experiments~\cite{PhysRevLett.76.3786,doi:10.1080/00018730310001621737,PhysRevX.9.021048} and the values obtained for $J$ between $0.35$\,eV and $0.4$\,eV are at a regime close to the experimental expectations~\cite{RevModPhys.75.657,doi:10.1080/00018730310001621737}. 
The momentum-resolved spin susceptibility results, as discussed earlier, reveal that the main effect of Hund's coupling is the control over the ferromagnetic behavior, reflected on the $\Gamma$ point, while they also show that a small $J$ value results in an overall suppressed magnetic susceptibility.

However, we are at a temperature regime where the calculation of $m^*/m$ cannot be expected to be extremely accurate, and at the same time we know that $X^{s}(q)$ is not directly measured experimentally, it is rather estimated indirectly, rending a quantitative comparison with our calculations not adequate. Therefore, we choose the value ${J=0.35}$\,eV which gives the best compromise between mass enhancement and spin susceptibility and an overall good qualitative agreement with the experiments. 

\section{Effect of temperature}

In this section we discuss the effect of temperature on the magnetic susceptibility of Sr$_2$RuO$_4$. The right panel of Fig.~\ref{fig:SM1} demonstrates that, as expected, by increasing the temperature there is an overall decrease of the magnetic susceptibility, as the spin fluctuations become weaker. In particular, the SDW peaks appear significantly diminished, as is the value of $X^{s}(q)$ at the center of the BZ (at $\Gamma$), while for the edges of the zone (at X) the already smaller value, is not particularly affected. 
In Fig.2 of the main text, the inverse of $X^{s}$ at the SDW vector is plotted as a function of temperature and from a linear extrapolation of the data, one observes that no magnetically ordered state is found at finite temperatures. Therefore, our calculations accounting for non-local electronic correlations, resolve the strongly enhanceed magnetic fluctuations found within DMFT, which predict a finite temperature magnetic transition. 

\section{Momentum dependence of self-energy}

\begin{figure*}[t!]
\includegraphics[width=0.24\linewidth]{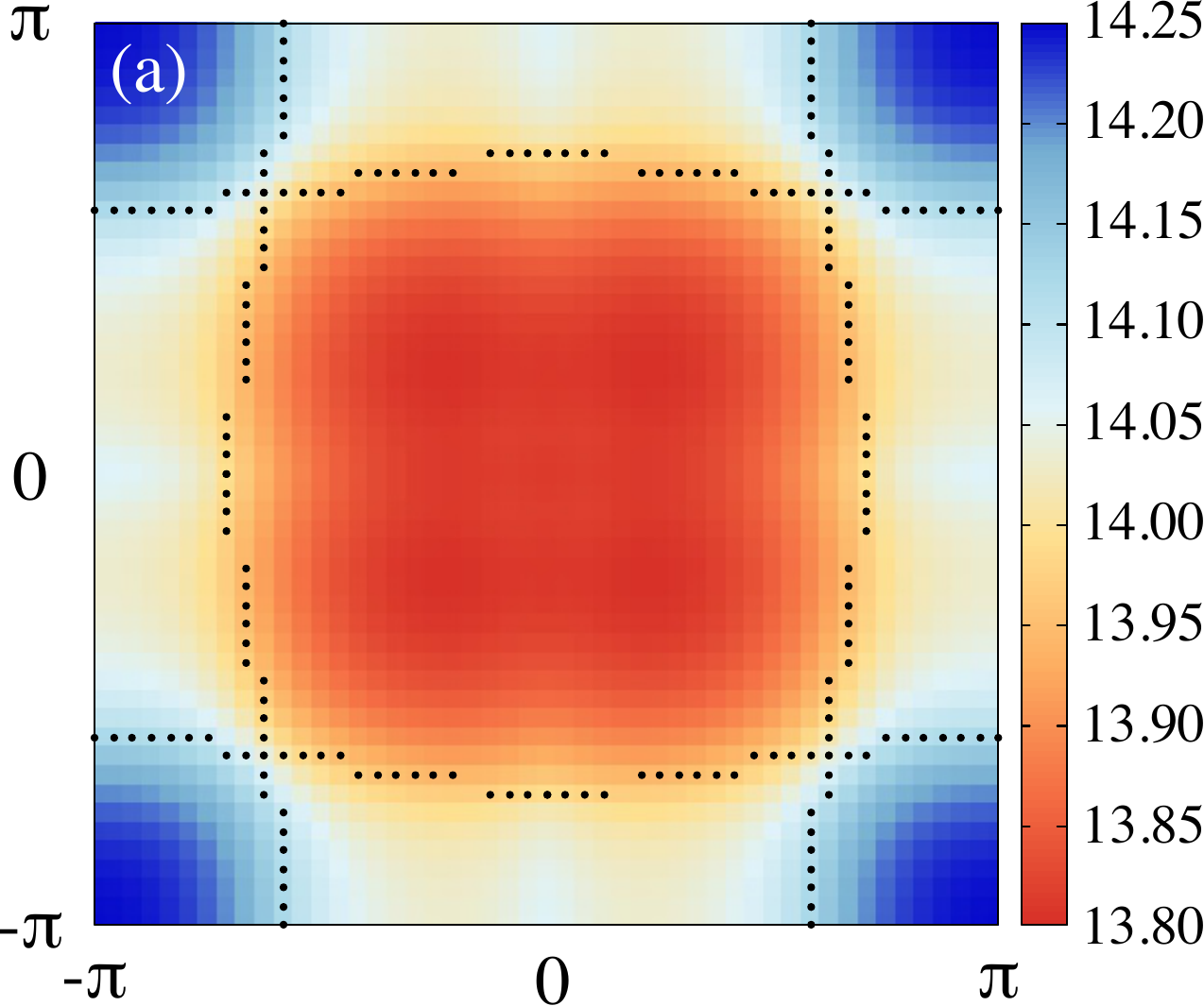} \hspace{0.08cm}
\includegraphics[width=0.24\linewidth]{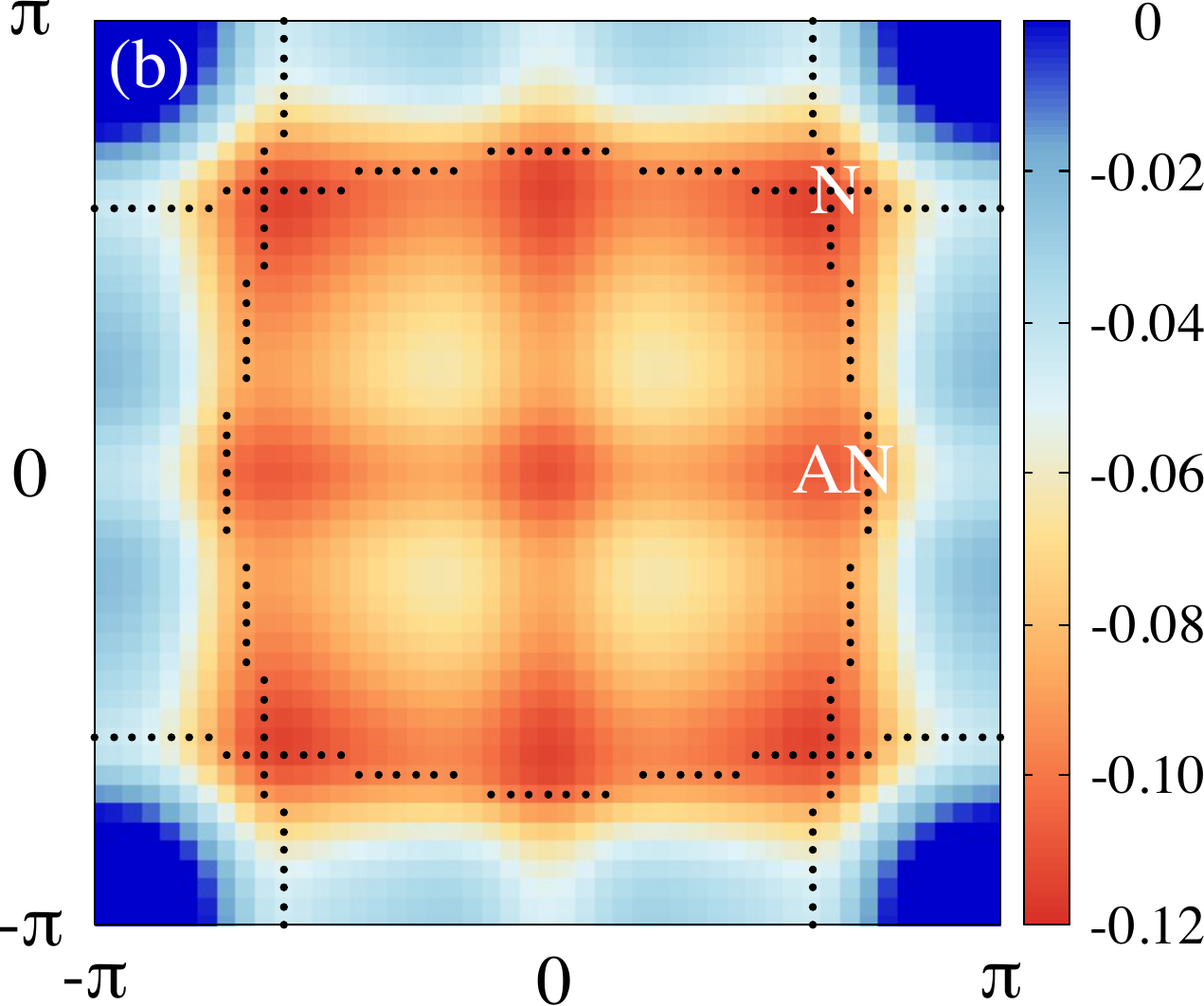} 
\hspace{0.08cm}
\includegraphics[width=0.24\linewidth]{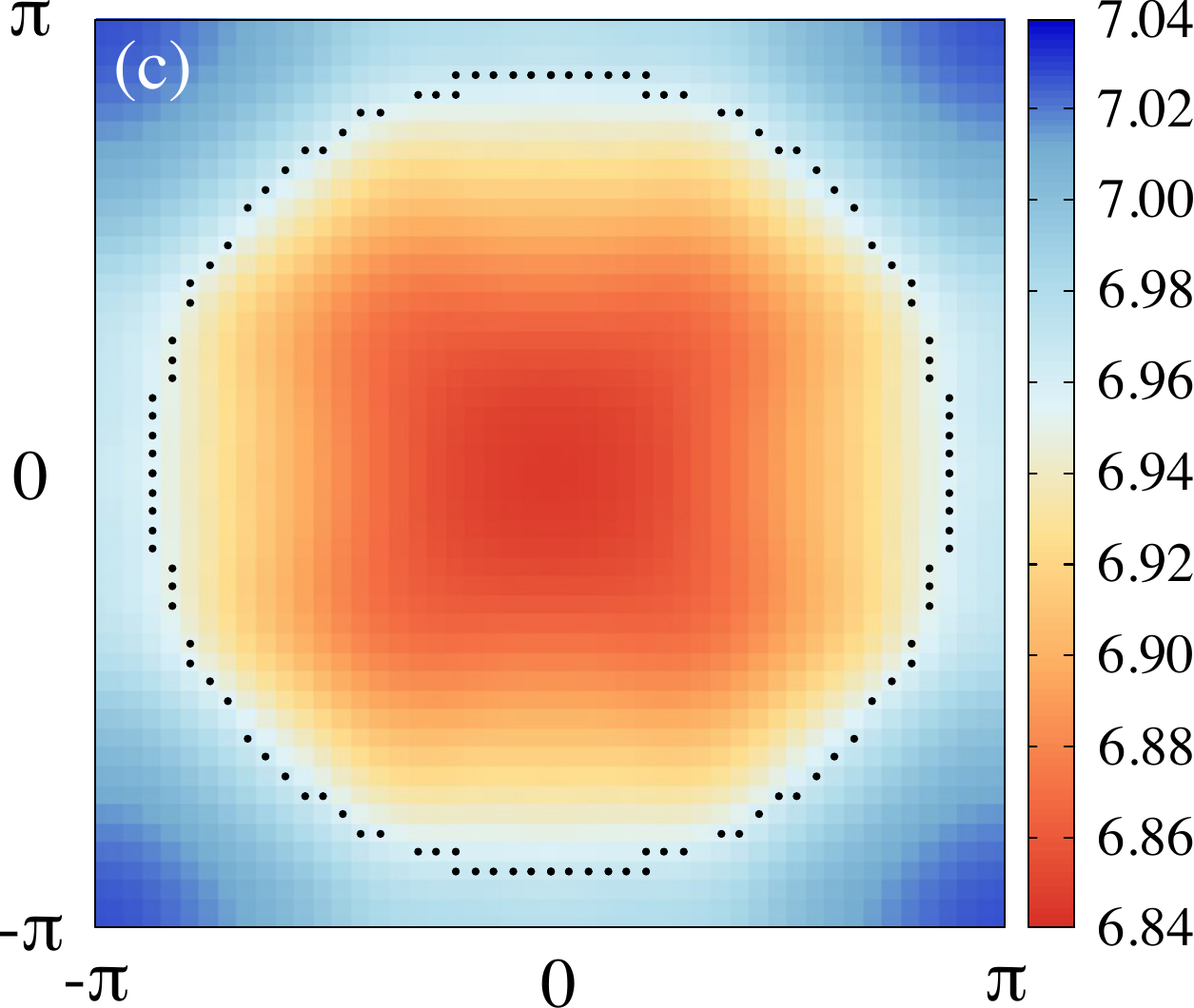} \hspace{0.08cm}
\includegraphics[width=0.24\linewidth]{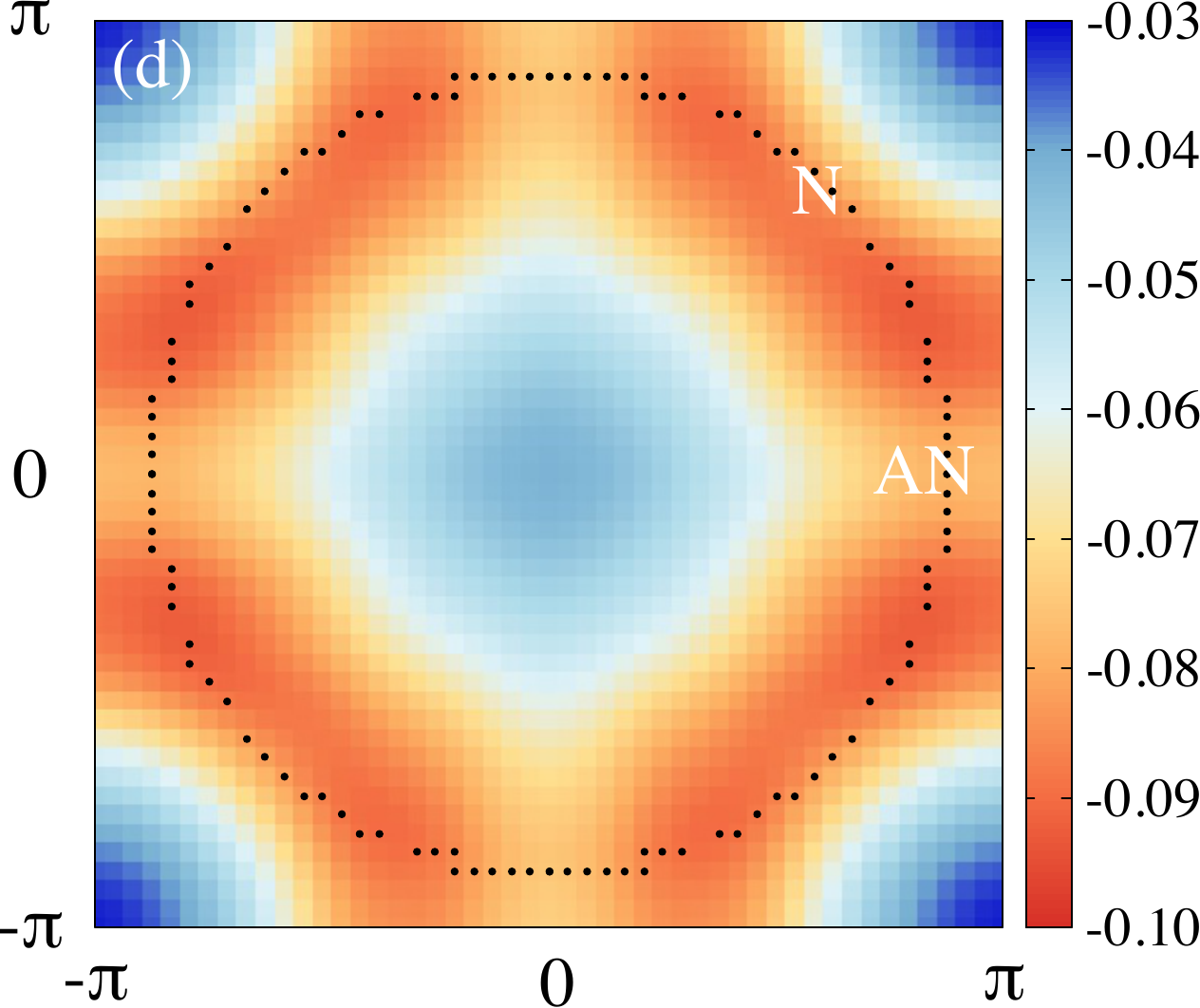} 
\caption{
\textbf{Two left panels:} 
Real \textbf{(a)} and imaginary \textbf{(b)} part of the self-energy at the first Matsubara frequency - very close to the Fermi energy - for the sum of the 1D $xz$ and $yz$ orbitals and similarly in \textbf{(c)} and \textbf{(d)} for the 2D $xy$ orbital. The FS sheets corresponding to each orbital are plotted with black dots. For the $xz+yz$ case, Im$\Sigma$ appears to be quasi-local on the FS, with slightly larger values around the nodal points and suppressed ones at the corners of the BZ - around the $X$ point. For the $xy$ orbital's Im$\Sigma$, however, a more extended momentum-dependence is observed along the FS, revealing that the quasi-local picture of the material's self-energy is of limited validity. The Re$\Sigma$ reveals that for the $xy$ orbital along the FS there is no momentum-dependence, for the $xz+yz$ case on the other hand there is a finite differentiation with $k$. This last finding justifies the suppression of the local spectral function of the $xz/yz$ orbitals, leading to the suppressed magnetic fluctuations.
\label{fig:selfen_FS}}
\end{figure*}

The importance of non-local electronic correlations can be more directly seen through the analysis of the system's self-energy $\Sigma(i\omega)$.
To this end we plot in Fig.\ref{fig:selfen_FS}, the real and imaginary part of the self-energy on the whole BZ at the first Matsubara frequency $i\omega_0$, so essentially on the Fermi energy. 
Panels (a) and (b) correspond to the Re$\Sigma$ and Im$\Sigma$ of the summed $xz$,$yz$ 1D orbitals respectively and panels (c) and (d) to Re$\Sigma$ and Im$\Sigma$ of the $xy$ 2D orbital. 
One immediately observes that there is a much more pronounced spatial dependence of Im$\Sigma$ across the FS, which is illustrated with the black dots, for the wide $xy$ orbital compared to the narrow $xz/yz$ ones. 
In particular, we find that for $xy$ there is a quasi-constant self-energy across the arcs located around the N point, while around the AN point $Im\Sigma(i\omega_0)$ is evidently diminished. 
The qualitative behavior is quite similar for the $xz/yz$ orbitals, however the range of values is much more restricted rending the effect almost unimportant. 
Concerning the Re$\Sigma$ the situation is reversed, with the $xy$ orbital being almost $k-$independent and the $xz/yz$ ones exhibiting a finite momentum-dependence along the FS. 
In conclusion, we find that the conjecture of a quasi-local self-energy is not completely accurate, however its finite momentum-dependence is not particularly strong along the Fermi surface. 

\bibliography{Ref_sup}